\documentclass[english,tightenlines,eqsecnum,amsmath,amssymb,aps,prd,superscriptaddress,nofootinbib,showpacs,floats]{article}
\usepackage{amsmath,amsfonts,amssymb,multicol}
\usepackage{graphicx,caption,subcaption}
\usepackage[usenames]{xcolor}
\definecolor{AHZ}{rgb}{0.0,1,0.0}
\usepackage[colorlinks,linkcolor=AHZ,citecolor=red,bookmarks]{hyperref}
\usepackage{rotating}
\usepackage[mathscr]{eucal}
\usepackage{graphicx}
\usepackage[T1]{fontenc}
\usepackage[utf8]{inputenc}
\usepackage{babel}
\usepackage{authblk}
\usepackage{epsfig}
\usepackage{cite}
\usepackage{color}
\usepackage{amssymb}
\usepackage{fancyhdr}

\def\nn{\nonumber\\}
\newcommand{\f}[2]{\frac{#1}{#2}}
\def\be{\begin{equation}}
	\def\ee{\end{equation}}
\def\bea{\begin{eqnarray}}
	\def\eea{\end{eqnarray}}

\def\bwt{\begin{widetext}}
	\def\ewt{\end{widetext}}

\usepackage{amsmath}

\usepackage{amssymb}

\begin{document}
\title{Novel Casimir wormholes in Einstein gravity}
\author{$^{1}$Mohammad Reza Mehdizadeh\thanks{mehdizadeh.mr@uk.ac.ir}}
\author{${^2}$Amir Hadi Ziaie\thanks{ah.ziaie@maragheh.ac.ir}}
\affil{${^1}${\rm Department~of~ Physics,~ Shahid~ Bahonar~ University, P.~ O.~ Box~ 76175, Kerman, Iran}}
\affil{{\rm ${^2}$Research~Institute~for~Astronomy~and~Astrophysics~of~ Maragha~(RIAAM), University of Maragheh,  P.~O.~Box~55136-553,~Maragheh, Iran}}
\renewcommand\Authands{ and }
\maketitle
\begin{abstract}
In the context of General Relativity (GR), violation of the null energy condition (NEC) is necessary for existence of static spherically symmetric wormhole solutions. Also, it is a well-known fact that the energy conditions are violated by certain quantum fields, such as the Casimir effect. The magnitude and sign of the Casimir energy depend on Dirichlet or Neumann boundary conditions and geometrical configuration of the objects involved in a Casimir setup. The Casimir energy may act as an ideal candidate for the  matter that supports the wormhole geometry. In the present work, we firstly find traversable wormhole solutions supported by a general form for the Casimir energy density assuming a constant redshift function. As well, in this framework, assuming that the radial pressure and energy density obey a linear equation of state, we derive for the first time Casimir traversable wormhole solutions admitting suitable shape function. Then, we consider three geometric configurations of the Casimir effect such as (i) two parallel plates, (ii) two parallel cylindrical shells, and (iii) two spheres. We study wormhole solutions for each case and their property in detail. We also check the weak and strong energy conditions in the spacetime for the obtained wormhole solutions. The stability of the Casimir traversable wormhole solutions are investigated using the Tolman-Oppenheimer-Volkoff (TOV) equation. Finally, we study trajectory of null as well as timelike particles {along with quasi-normal modes (QNMs) of a scalar field} in the wormhole spacetime.
\end{abstract}
%\maketitle
\section{Introduction}
%\DeclareUnicodeCharacter{2009}{\,}
Wormholes are theoretical passages between two different universes, or sometimes between two distant parts of the same universe. The concept of wormhole  was first put forward by Einstein and Rosen in their famous Einstein-Rosen bridge in 1935~\cite{ERB,ERB1}. Later in 1957, Misner and Wheeler coined the term \lq{}\lq{}{\it wormhole}\rq{}\rq{} in their seminal works, as an attempt to present a mechanism for having \lq{}\lq{}{\it charge without charge}\rq{}\rq{}~\cite{mwhee,mwhee1,mwhee2}. They found that wormholes that connect two asymptotically flat spacetimes may render non-trivial solutions to Einstein-Maxwell equations, where, the lines of electric field as observed in one part of the universe could thread the throat and emerge in other part. Traversable wormhole structures were first studied by Morris and Thorne in 1988~\cite{mothorn,mothorn1}. They found exact static spherically symmetric solutions and discussed the required conditions for physically meaningful Lorentzian traversable wormholes. In the framework of GR, the Morris-Thorne wormholes allow a two way communication between two regions of the spacetime through a minimal surface called the wormhole throat. Such a two way travel requires the fulfillment of the fundamental flare-out condition at the wormhole throat. However, the EMT components for such a wormhole configuration always violate the NEC~\cite{book visser}. The matter distribution responsible for such a situation is the so-called \lq{}\lq{}{\it exotic matter}\rq{}\rq{} which has negative energy density~\cite{kar1,kar2}. Therefore, the issue of exotic matter and the establishment of standard energy conditions have been one of the most important challenges in wormhole physics until today. In this respect, there have been many attempts in literature in order to avoid or at least to minimize the usage of exotic matter in the wormhole geometry~\cite{pow1}. For instance, dynamical wormhole geometries which satisfy the energy conditions during a time period on a timelike or null geodesic have been studied in~\cite{dynami1,dynami2,dynami3}. Visser and Poisson have studied construction of thin-shell wormholes which are constructed by the cut-and-paste technique and their supporting matter is concentrated on the wormhole's throat~\cite{pvis}. Wormhole configurations with phantom or quintom-type energy as the supporting matter have been investigated in~\cite{phantworm,phantworm1,phantworm2} and wormholes supported by nonminimal interaction between dark matter and dark energy has been explored in~\cite{intdarksec}, see also~\cite{hisnotewo} for historical notes and~\cite{loboreview} for a comprehensive review. In GR, the thin-shell wormholes do not respect the standard energy conditions at the throat. However, in the context of modified theories of gravity, the presence of higher-order terms in curvature would allow for building thin-shell wormholes supported by ordinary matter~\cite{highthin,highthin1,highthin2,highthin3}. As a matter of fact, the correction terms or additional degrees of freedom not present in GR can provide a setting for traversable wormhole solutions. Studies in this arena have been performed in the context of different modified gravity theories and under various circumstances, for example: traversable wormholes in Einstein-Gauss-Bonnet gravity~\cite{EGBW,EGBW1,EGBW2,EGBW3}, higher-dimensional GR~\cite{highdgr,highdgr1,highdgr2}, nonsymmetric gravitational theory~\cite{nonsymgr}, Lovelock~\cite{loveworm,loveworm1,loveworm2} and $f(R)$ gravity theories~\cite{frworm,frworm1,frworm2,frworm3,frworm4,frworm5}, modified gravities with curvature-matter coupling~\cite{frtworm,frtworm1,frtworm2,frtworm3,frtworm4,frtworm5}, Brans-Dicke~\cite{bdw,bdw1,bdw2,bdw3,bdw4,bdw5,bdw6,bdw7,bdw8,bdw9,bdw10,bdw11,bdw12,bdw13} and other modified gravity theories~\cite{othmodw,othmodw1,othmodw2,othmodw3,othmodw4,othmodw5,othmodw6,othmodw7,othmodw8,parssah}.
\par
It is now well known that the theoretical existence of traversable Lorentzian wormholes in GR is accompanied by the violation of NEC and consequently the existence of exotic types of matter. However, the quest for finding the promising candidates of exotic matter is not a routine and easy task, and the footprints of such type of matter have been recognized only in a small area, such as
the experimentally verified Casimir effect~\cite{HC1948,casmirexp,casmirexp1} and semi-classical Hawking radiation~\cite{klin}. The Casimir effect is a famous quantum field theoretical phenomenon
that appears as an attractive force between neutral parallel conducting plates in a vacuum. The associated negative energy density to this effect is a manifestation of the quantum fluctuations of the vacuum of the electromagnetic field confined between the two plates~\cite{bordag,miltonbook}. In view of the exotic nature of Casimir energy, Morris and Thorne~\cite{mothorn} and some time later Visser~\cite{book visser} argued that this type of exotic energy can be considered as an appropriate source for modeling traversable wormholes. However, wormhole solutions in GR with Casimir energy as the supporting matter have been found only very recently by Garattini~\cite{gara}, where the author studied negative energy density due the Casimir effect and explored the consequences of quantum weak
energy conditions on the traversability of the wormhole. Subsequently, the study of traversable wormholes in the presence of Casimir energy has been carried out in different frameworks and under various circumstances among which we can quote, the case of three~\cite{casi3} and D-dimensions~\cite{casid}, alternative gravity theories~\cite{altcas,altcas1,altcas2}, GUP corrections~\cite{casgup,casgup2,casgup3,casgup4}, Casimir source modified by a Yukawa term~\cite{yukcas,yukcas1,yukcas2} and other frameworks~\cite{othercas,othercas2,othercas3,othercas4}. The Casimir effect has a strong dependence on the type of the quantum field under investigation, shape of the objects and the boundary conditions imposed on them~\cite{bordag,miltonbook}. Our aim in the present paper is then to study wormhole configurations with Casimir energy as the supporting matter for different Casimir setups. The organization of the this paper is as follows: In Sec.~(\ref{Gb}) we give a brief review on Morris-Thorne wormholes. In Sec.~(\ref{WHS}) we proceed to find traversable wormhole solutions assuming a power-law form for the Casimir energy density. The zero tidal force solutions are given in Subsec.~(\ref{WHS1}) and those of non-constant redshift function are presented in Subsec.~(\ref{WHS2}). Sec.~(\ref{Eqcon}) is devoted to the equilibrium conditions on wormhole structure. In Sec.~(\ref{traject}) we investigate trajectory of null and timelike particles in the wormhole spacetime. {Finally, in Sec.~(\ref{QNMO}) we study QNMs and their physical properties in the wormhole spacetime.} Our conclusions are presented in Sec.~(\ref{concluding}). Throughout the present work we set the units so that $\hbar=C=G=1$.
%%%%%%%%%%%%%%%%%%%%%%%%%%%%%%%%%%%%%%%%%%%%%%%%%%%%%
\section{Morris-Thorne wormholes: a brief review}\label{Gb}
In their seminal work, Morris and Thorne introduced the following spherically symmetric line element
\begin{equation}\label{evw}
ds^{2}=-e^{2\phi (r)}dt^{2}+\left[\frac{dr^{2}}{1-b(r)/r} +r^{2}d\Omega^{2}\right],
\end{equation}
as a possible solution to obtain viable wormhole structure. In the above metric, $\phi(r)$ is the redshift function as it is related to the gravitational redshift and $b(r)$ is the wormhole shape function. The shape function must satisfy the flare-out condition at the throat, i.e., we must have $b^{\prime}(r_0)<1$ and $b(r)<r$ for $r>r_0$ in the whole spacetime, where $r_0$ is the throat radius. To obtain the components of Einstein field equation we utilize a set of orthonormal basis vectors. These vector fields are defined as the proper reference frame of a set of observers who remain at rest in the coordinate system $(t,r, \theta, \phi)$, with $(r,\theta, \phi)$ fixed. If we denote the basis vectors in this coordinate system as
$({\rm e}_t,{\rm e}_r,{\rm e}_\theta,{\rm e}_\phi)=\left(\partial/\partial_t,\partial/\partial_r,\partial/\partial_\theta,\partial/\partial_\phi\right)$, then the orthonormal basis vectors are given by
\be\label{orthoframe}
{\rm e}_{\hat{t}}={\rm e}^{-\phi}{\rm e}_{t},~~~~~{\rm e}_{\hat{r}}=\left(1-\f{b}{r}\right)^{\f{1}{2}}{\rm e}_{r},~~~~{\rm e}_{\hat{\theta}}=\f{{\rm e}_{\theta}}{r},~~~~{\rm e}_{\hat{\phi}}=\f{{\rm e}_{\phi}}{r\sin\theta},
\ee
by the virtue of which, the components of spacetime metric (\ref{evw}) take on their standard, special relativity forms as $g_{\alpha\beta}={\rm diag}\left[-1,1,1,1\right]$. Working in this basis simplifies the mathematical analysis and physical interpretation~\cite{mothorn,mothorn1}. The non-vanishing components of Einstein tensor in this orthonormal reference frame are then found as
\bea\label{gortho}
&&G_{\hat{t}\hat{t}}=\f{b^\prime}{r^2},~~~~G_{\hat{r}\hat{r}}=-\f{b}{r^3}+2\left[1-\f{b}{r}\right]\f{\phi^\prime}{r},\nn
&&G_{\hat{\theta}\hat{\theta}}=G_{\hat{\phi}\hat{\phi}}=\left(1-\f{b}{r}\right)\left[\phi^{\prime\prime}+(\phi^{\prime})^2-\f{rb^\prime-b}{2r(r-b)}\phi^\prime-\f{rb^\prime-b}{2r^2(r-b)}+\f{\phi^\prime}{r}\right],
\eea
where a prime denotes differentiation with respect to $r$. Also, the nonzero components of stress-energy tensor (SET) in the orthonormal reference frame are given by
\begin{equation}\label{setortho}
T_{\hat{t}\hat{t}}=\rho(r),~~~~~~~~~~T_{\hat{r}\hat{r}}=P_r(r),~~~~~~~~~~T_{\hat{\theta}\hat{\theta}}=T_{\hat{\phi}\hat{\phi}}=P_t(r),
\end{equation}
where $\rho(r)$ is the energy density and $P_{r}(r)$ and $P_{t}(r)$ are the radial and transverse pressures, respectively. Thus, the Einstein field equation $G_{\hat{\alpha}\hat{\beta}}=8\pi T_{\hat{\alpha}\hat{\beta}}$ provides us with the following expressions
\bea
8\pi r^2\rho(r)&=&1-g^{\prime}(r)r-g(r),\label{feild1}\\
8\pi r^2P_{r}(r)&=&rg(r)\frac{f^{\prime}(r)}{f(r)}+g(r)-1,\label{feild2}\\
32\pi r^2P_{t}(r)&=&2rg^\prime(r)-g\left[\f{rf^\prime(r)}{f(r)}\right]^2+r^2g^\prime(r)\f{f^\prime(r)}{f(r)}+\f{2rg(r)}{f(r)}\left[f^\prime(r)+rf^{\prime\prime}(r)\right],\label{feild3}
\eea
where a prime denotes differentiation with respect to radial coordinate $r$ and we have defined the positive functions $f(r)$ and $g(r)$ as
\bea\label{fgfuncs}
g(r)\equiv\frac{r-b}{r},~~~~~~~~~f(r)\equiv e^{2\phi (r)}.
\eea  
The flare-out condition at the wormhole throat leads to the conditions $g^{\prime}(r_0)>0$ and $g(r_0)=0$. It is well-known that existence of traversable Lorentzian wormholes in four dimensions as solutions to the Einstein equations requires some kind of the so-called \lq{}\lq{}exotic matter\rq{}\rq{}, i.e., matter that violates the NEC~\cite{kar1,kar2}. This is due to the fulfillment of flaring-out condition at the throat of the wormhole and for $r>r_0$. Physically, the flare-out condition at the throat is responsible for holding back the wormhole throat from collapsing and is crucial for its traversability. Hence, in classical GR, a traversable wormhole configuration requires exotic matter at or in the neighborhood of the wormhole’s throat. We note that NEC is a part of the weak energy condition (WEC) whose physical meaning is that the energy density is non-negative in any reference frame. In other words, WEC requires that $T_{\mu \nu }U^{\mu }U^{\nu}\geq 0$, where  $U^{\mu}$ is a timelike vector field\footnote{The WEC is utilized in the proof of Penrose singularity theorem~\cite{penbh}.}. For the SET (\ref{setortho}), the WEC leads to the following inequalities
\begin{equation}\label{E11}
\rho \geq0 ,~~~~~~~~~~~~\rho+P_{r}\geq0,~~~~~~~~~~~~~~~\rho+P_{t}\geq0.
\end{equation} 
Note that the last two inequalities are defined as the NEC. In addition, the strong energy condition (SEC) is satisfied through the following inequalities~\cite{book visser}
\begin{equation}\label{E11}
\rho+P_{r}+2P_{t} \geq0,~~~~~~~~~~~\rho+P_{r}\geq0,~~~~~~~~~~~\rho+P_{t}\geq0.
\end{equation}
Using Eqs.~(\ref{feild1})-(\ref{feild3}), one finds the following relationships
\bea
&&8\pi(\rho +P_{r})=\f{1}{r}\left[g(r)\f{f^{\prime}(r)}{f(r)}-g^{\prime}(r)\right],\label{EGBNEC}\\
&&32\pi(\rho +P_{t})=2g(r)\f{f^{\prime\prime}(r)}{f(r)}+\f{f^\prime(r)}{rf(r)}\left[2g(r)+rg^\prime(r)\right]-g\left[\f{f^\prime(r)}{f(r)}\right]^2\!\!-\f{2}{r^2}\left[rg^\prime(r)+2g(r)-2\right],\label{EGBNEC1}\\
&&16\pi(\rho+P_{r}+2P_{t})=4g(r)\f{f^\prime(r)}{rf(r)}+g^\prime(r)\f{f^\prime(r)}{f(r)}+2g(r)\f{f^{\prime\prime}(r)}{f(r)}-g(r)\left[\f{f^\prime(r)}{f(r)}\right]^2.\label{EGBNEC2}
\eea
The above expressions at the throat take the following forms
\begin{eqnarray}
\rho + P_r\Big|_{r=r_0}&=&-\f{g^\prime(r_0)}{8\pi r_0},\nn32\pi(\rho + P_t)\Big|_{r=r_0}&=&\f{g^\prime(r_0)f^\prime(r_0)}{f(r_0)}-\f{2}{r_0}g^\prime(r_0)+\f{4}{r_0^2},~~~~~~\rho + P_r +2 P_t\Big|_{r=r_0}=\f{g^\prime(r_0)f^\prime(r_0)}{16\pi f(r_0)},
\end{eqnarray}
whence we observe that the NEC in radial direction is violated as a consequence of flare-out condition. However, in tangential direction the fulfillment of NEC depends on the values of metric components and their derivatives at the throat. Also, for $f^\prime(r_0)<0$ the SEC is violated at the throat, due to the flare-out condition. Also, for the line element (\ref{evw}), the Ricci $\mathcal{R}=g_{\mu\nu}\mathcal{R}^{\mu\nu}$ and the Kretschmann $\mathcal{K}=\mathcal{R}_{\mu \nu \gamma \beta}\mathcal{R}^{\mu \nu \gamma \beta}$ scalars are given as
\begin{align}
\mathcal{R}(r)=g(r)\left[\f{f^\prime(r)}{f(r)}\right]^2-\left[4g(r)+rg^\prime(r)\right]\f{f^\prime(r)}{rf(r)}-2g(r)\f{f^{\prime\prime}(r)}{f(r)}-\f{4}{r^2}\left[g(r)+rg^\prime(r)-1\right],
\label{ric1}
\end{align}
\begin{eqnarray}
	\mathcal{K}(r)\!\!&=&\!\! \left[\f{g(r)f^{\prime\prime}(r)}{f(r)}\right]^2-\f{f^{\prime\prime}(r)}{f(r)}\left[\f{g(r)f^{\prime}(r)}{f(r)}\right]^2+g(r)g^{\prime}(r)\f{f^{\prime}(r)f^{\prime\prime}(r)}{f^2(r)}+\f{g(r)^2}{4}\left[\f{f^{\prime}(r)}{f(r)}\right]^4\nn&-&\f{1}{2}g(r)g^\prime(r)\left[\f{f^\prime(r)}{f(r)}\right]^3+\left[\f{f^\prime(r)g^\prime(r)}{2f(r)}\right]^2+2\left[\f{g(r)f^\prime(r)}{rf(r)}\right]^2+\f{4}{r^4}\left[\f{1}{2}\left(rg^\prime(r)\right)^2+g(r)^2-2g(r)+1\right].\nn
	\label{kr11}
\end{eqnarray}

%\begin{eqnarray}
%\mathcal{K}(r)& =& \frac{4\,{r}^{4}{{\it f''}}^{2}{f}^{2}{g}^{2}-4\,{r}^{4}{\it f''}\,f{g}^{2}{{\it f'}}^{2}+4\,{r}^{4}{\it f''}\,{f}^{2}g{\it f'}\,{\it g'}+{r}^{4}{{\it f'}}^{4}{g}^{2}-2\,{r}^{4}{{\it f'}}^{3}g{\it g'}\,f
%}{4r^4 f^4}\nn
%&&+\frac{{r}^{4}{{\it f'}}^{2}{{\it g'}}^{2}{f}^{2}+8\,{g}^{2}{{\it f'}}^{2}{f}
%	^{2}{r}^{2}+8\,{{\it g'}}^{2}{f}^{4}{r}^{2}+16\,{f}^{4}-32\,{f}^{4}g+16\,{f}^{4}{g}^{2}}{4r^4 f^4}
%\label{kr11}
%\end{eqnarray}
The above two expressions are useful in determining the possible occurrence (or absence) of spacetime singularities through their divergent (regular) behaviors~\cite{RicKretch,RicKretch1}.
%%%%%%%%%%%%%%%%%%%%%%%%%%%%%%%%%%%%%%%%%%%%%%%%%%%%%%%%%%%%%%%%%%%%%%%%%%%%%%%%%%%%%%%%%%%%%%%%%%%%%%%%%%%%%%%%%%%%%%%%%%%%%%%%%%%%%%%%%%%%%%%%%%
\section{Casimir wormhole solutions}\label{WHS}
\subsection{Specific case: constant redshift function}\label{WHS1}
\begin{figure}
 	\begin{center}
\includegraphics[scale=0.35]{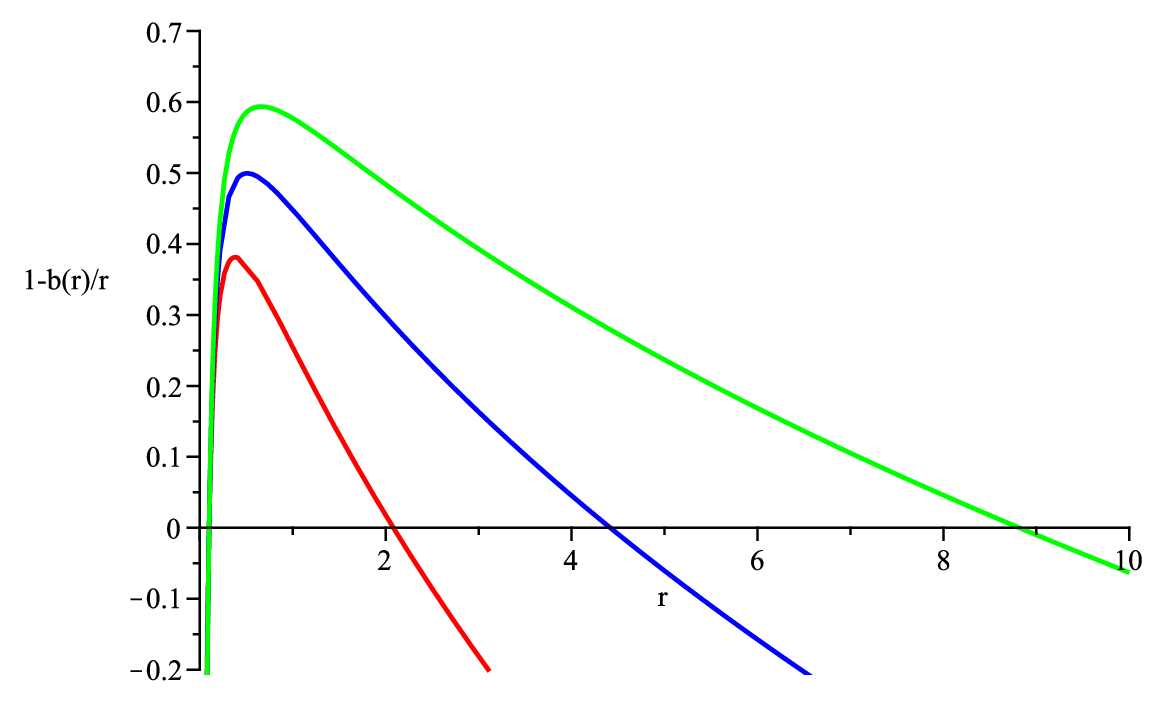}
\caption{The behavior of  $1-b(r)/r$  with respect to $r$ for $r_0=0.1$, $m=1.5$ and $\lambda=1, 0.7, 0.5$ from down to up, respectively.}\label{figbr1}
 	\end{center}
 \end{figure}

The Casimir effect which was discovered more than 70 years by Dutch physicist Hendrik Casimir, is one of the most direct manifestations of the existence of zero-point vacuum oscillations~\cite{HC1948}. This is a purely quantum effect which, in its simplest form, is the attraction of
a pair of neutral, parallel conducting plates resulting from the distortion of the electromagnetic vacuum by the boundaries. According to Casimir's
prediction, the energy per unit area of two infinitely large, neutral parallel planes made of an ideal metal at zero temperature, is given by \cite{bordag}
\begin{align}
 E(d)=-\frac{{\pi}^2}{720 d^3},
 \label{Eqca}
 \end{align}
which depends on the separation distance between the planes. The Casimir pressure can be obtained as
\begin{align}
P(d)=-\frac{\partial E}{\partial d}=-\frac{{\pi}^2}{240 d^4}.
\end{align}
From the above expressions we can recognize the following form for energy density as
\begin{align}
\rho=-\frac{{\pi}^2}{720 d^4}.
\label{rhca1}
\end{align}
It is obvious that for the given pressure and energy density we may assume the linear equation of state (EoS) $P = w\rho$ with EoS parameter $w = 3$. The force associated to Casimir energy Eq.~(\ref{Eqca}) is attractive, i.e, under the effect of this force, the plates tend to move toward each other. We further note that the vacuum energy of different quantum fields (e.g., scalar, spinor, electromagnetic and etc.) depends on the boundary conditions imposed on the bodies that compose a Casimir setup~\cite{bordag,miltonbook}. The ideal-metal planes are idealized thin plates made of a material with an infinitely large magnitude of the dielectric permittivity. For such arrangement, the imposed boundary conditions that lead to Eq.~(\ref{Eqca}) imply that the transverse component of the electric field and the normal component of the magnetic field on the surface of each plate be zero. The fulfillment of these conditions signal that an electromagnetic field can exist only outside an ideal conductor~\cite{bordag}. The problem of Casimir interaction between an ideal-metal plane and an infinitely permeable plane was considered by Boyer~\cite{Boyer1974}. An infinitely permeable plane is characterized by an infinitely large magnetic permeability on which, the tangential component of the magnetic induction vanishes. The Casimir energy density for such a set up then reads
\begin{align}
\rho=\frac{7}{8}\frac{{\pi}^2}{720 d^4}.
\label{rhca4}
\end{align}  
The above result is equal to a factor of $-7/8$ times the of ideal-metal planes. We note that the corresponding Casimir force for this case is repulsive in the manner that the plates tend to move away from each other. This change from attraction to repulsion occurs due to the
mixed boundary conditions~\cite{bordag}. Also, the Casimir energy depends on the geometry and shape of the objects, temperature and the interplay between geometry and material properties~\cite{bordag,miltonbook,Rob2007}. For instance, the case of a sphere in front a plane has been discussed in~\cite{hgies}, a plate and a cylinder has been studied in~\cite{temi}, eccentric cylinders in~\cite{dalvi,dalvi1}, a hyperboloid opposite to a plane in~\cite{sch} and, a flat and a corrugated plane in~\cite{hang,goles}. The calculation of the Casimir energy for nontrivial geometries is a complicated task. Because of this, several approximate methods have been developed so far among which, a powerful one for calculating the Casimir force between bodies of arbitrary shape is the proximity force approximation (PFA) method which was suggested by Derjaguin~\cite{Der1934}. In this method, the Casimir energy can be computed as an integral over infinitesimal parallel surface elements at their local distance $L$ as measured perpendicular to a surface $\Sigma$ that can be one of the two surfaces of the objects or an auxiliary surface placed between them. The PFA approximation for the energy is then given by~\cite{PFAEMIG}
\be\label{PFAAPP}
E_{\rm PFA}=\f{1}{A}\int_{\Sigma}E_{||}(L)dS,
\ee
where $E_{||}(L)/A$ is the energy per unit area for two parallel plates at the distance $L$. Using PFA method one can show that the dependence of the Casimir force on distance for a Cylindrical shell in front of a conducting plane is $d^{-7/2}$, which is an intermediate between the plane-spherical ($d^{-3}$) and the parallel plate configuration ($d^{-4}$). In addition, Casimir energy between two concentric cylinders (using PFA method) is proportional $d^{-2}$~\cite{mazit} and in the case of two parallel cylinders outside each other, to $d^{-5/2}$~\cite{dalvi1,dalvit1,dalvi}. Regarding the above considerations, we may assume the replacement $r$ instead of $d$ and obtain the Casimir energy density in the general form 
 \begin{align}
\rho=\frac{ \lambda}{8\pi r^m},
\label{rh10}
 \end{align}
where $m>0$ and $\lambda=8\pi\lambda_0 $ is a constant which depends on the type of quantum field, shape of the objects and boundary conditions. It may assume positive or negative values for different combinations of Dirichlet (D) and Neumann (N) boundary conditions, e.g. in the case of two parallel cylinders outside each other one gets $\lambda_0<0$ for DD and NN boundary conditions and $\lambda_0>0$ For DN and ND boundary conditions~\cite{dalvi1}. From equations (\ref{rh10}) and (\ref{feild1}), we obtain the following form for the shape function as
 \begin{align}
 b(r)=r\left[1-g(r)\right]=-\frac{\lambda r^{3-m}+{C}_0\left( m-3 \right) }{m-3},
 \label{shp1}
 \end{align}
where the integration constant can be determined through the condition $b(r_0)=r_0$ as
 \begin{align}
 {C}_0=-\frac{r_0(m-3)+\lambda r_0^{3-m}}{m-3}.
 \end{align}
Also, the flare-out condition at the throat results in the following inequality
\begin{align}
r_0-\lambda r_0^{3-m}>0.
\end{align}
The simplest case is a model with $ \phi=constant$, namely a spacetime with no tidal forces. As is clear from Eq.~(\ref{shp1}) for $m<2$, we obtain anti-de Sitter-like or de Sitter-like wormhole solutions for $\lambda<0$ and $\lambda>0$, respectively. However, the spatial extension of de Sitter-like wormhole solutions cannot be arbitrarily large. Fig.~(\ref{figbr1}) shows that a decrease in the value of parameter $\lambda$ enlarges the wormhole spatial extension. 
\par 
Next, we proceed to find the expressions for radial and tangential pressure profiles in the case of zero tidal forces. This can be achieved by using Eqs.~(\ref{feild2}) and (\ref{feild3}) together with the shape function (\ref{shp1}) 
\begin{eqnarray}
P_{r}&=&\f{\lambda}{8\pi(m-3)r^m}\left[1-\left(\f{r_0}{r}\right)^{3-m}\right]-\f{r_0}{8\pi r^3},\label{EGBNEC3}\\
P_{t}&=&\f{r_0}{16\pi r^3}-\f{\lambda}{16\pi(m-3)r^m}\left[m-2-\left(\f{r_0}{r}\right)^{3-m}\right].\label{EGBNE2}
\end{eqnarray}
Also, using equations (\ref{EGBNEC3}) and (\ref{EGBNE2}), we get the radial ($w_r=P_r/\rho$) and tangential ($w_t=P_t/\rho$) EoS parameters respectively, as
\begin{eqnarray}
w_{r}&=&\f{1}{m-3}\left[1-\left(\f{r_0}{r}\right)^{3-m}\right]-\f{r_0}{\lambda}r^{m-3},\label{EGBNEC4}\\
w_{t}&=&\f{r_0r^{m-3}}{2\lambda}-\f{1}{2(m-3)}\left[m-2-\left(\f{r_0}{r}\right)^{3-m}\right].
\end{eqnarray}
Furthermore, at the throat, we can see that $w_{r}^0=-r_0^{m-2}/\lambda$ and $w_{t}^0=-(w_{r}^0+1)/2$. We note that for $m<3$, in the limit $r\rightarrow\infty$ we have $w_{r}\rightarrow-1/(m-3)$ and $w_{t}\rightarrow-(m-2)/2(m-3)$. From Eqs.~(\ref{EGBNEC3}) and (\ref{EGBNE2}) we also get
\begin{eqnarray}
\rho +P_{r}&=&{\frac{ \left( m-2 \right) \lambda}{8\pi r^{m}\left( m-3\right)}}-\f{\left[\left(m-3 \right)r_0^{m}+\lambda r_0^{2}\right]r_0}{8\pi \left( m-3 \right)r^3r_0^{m}},\label{EGBNEC5}\\
\rho +P_{t}&=&{\frac{ \left(m-4\right) \lambda}{16\pi{r}^{m} \left( m-3\right) }}+{\f{\left[\left( m-3 \right)r_0^{m}+\lambda r_0^{2} \right] r_0}{16\pi \left( m-3 \right)r^3 r_0^{m}}},\label{EGBNEC6}\\
&&\rho +P_{r}+2P_{t}=0.
\label{EGBNEC7}
\end{eqnarray}%
Using Eq.~(\ref{kr11}) we can calculate the Kretschmann scalar for the solution (\ref{shp1}) as 
\begin{eqnarray}
\mathcal{K}(r)&=&6\f{r_0^{2}}{{r}^{6}}+\f{\lambda}{(m-3)r^6}\left[12 r_0^{4-m}-4 r_0m{r}^{3-m}\right]\nn
 &&+\f{\lambda^{2}}{ \left( m-3 \right)^{2}{r}^{6}}\left[6r_0^{6-2m}-4mr_0^{3-m}{r}^{3-m}+\left(2{m}^{2}-8m+12\right){r}^{6-2m}\right],
 \end{eqnarray}
whereby we find that the Kretschmann scalar is finite in the whole range $r>r_0$ and approaches zero asymptotically. 
\par
\begin{figure}
 	\begin{center}
 		\includegraphics[width=7cm]{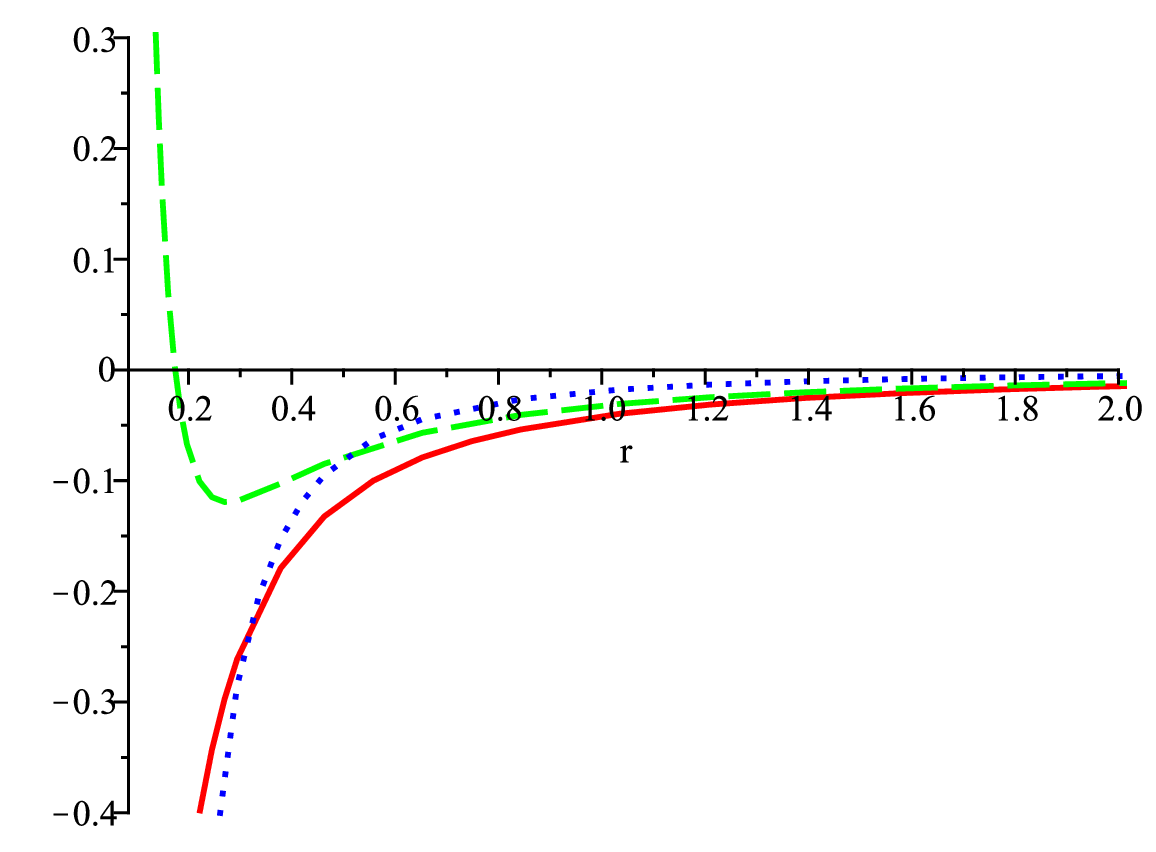}
 		\includegraphics[width=7cm]{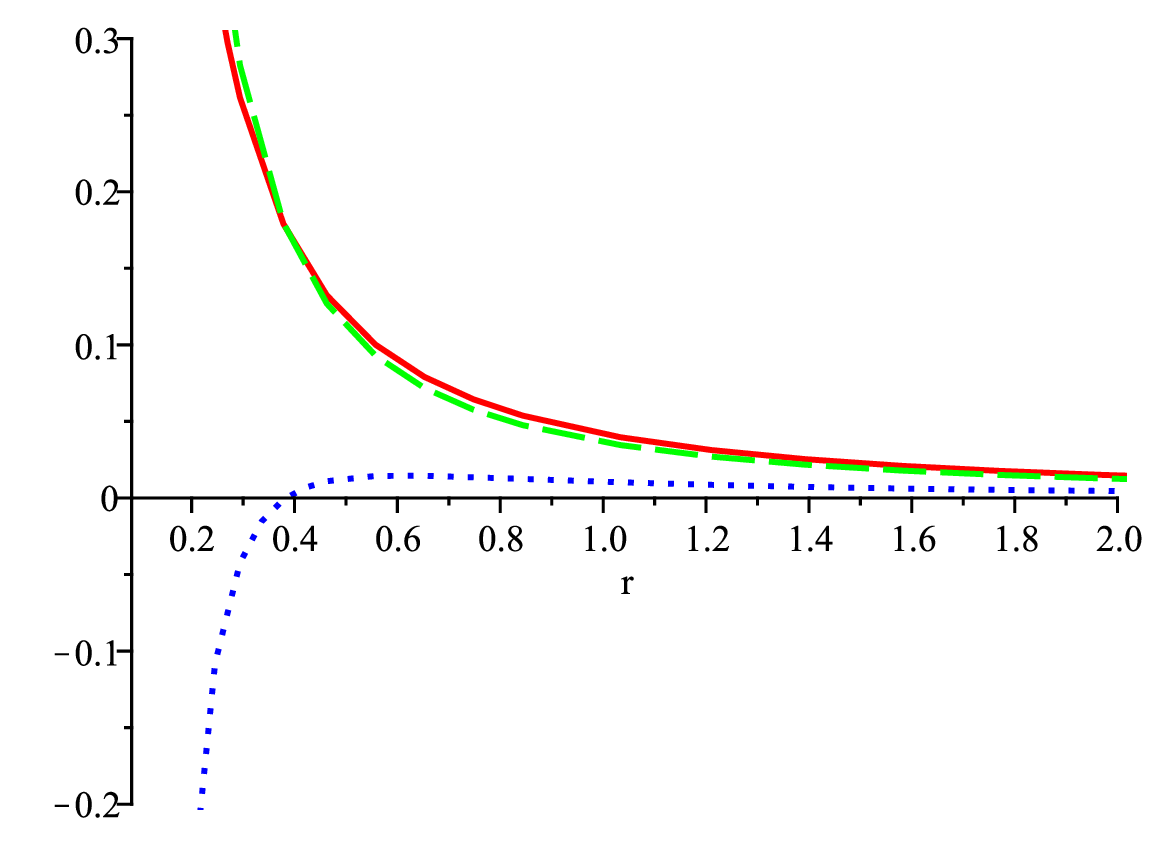}
 		\caption{The behavior of $\protect\rho $ (solid curve), $\protect\rho +p_{r}$ (dotted curve) and $\protect\rho +p_{t}$ (dashed curve) versus $r$. The model parameters have been set as, $m=1.5$ and $r_0=0.1$, $\lambda=-1$ (left panel) and $\lambda=1$ (right panel).}\label{wor2}
 	\end{center}
 \end{figure}

\begin{figure}
 	\begin{center}
 		\includegraphics[width=7cm]{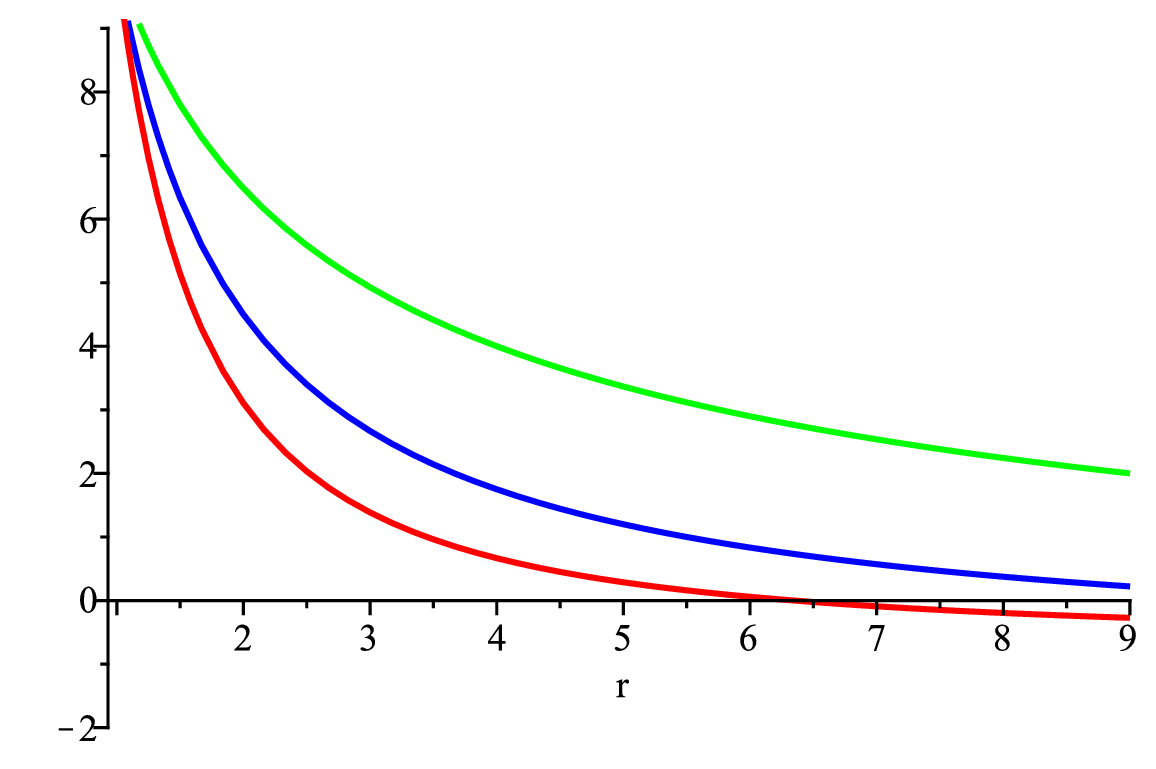}
 		\includegraphics[width=7cm]{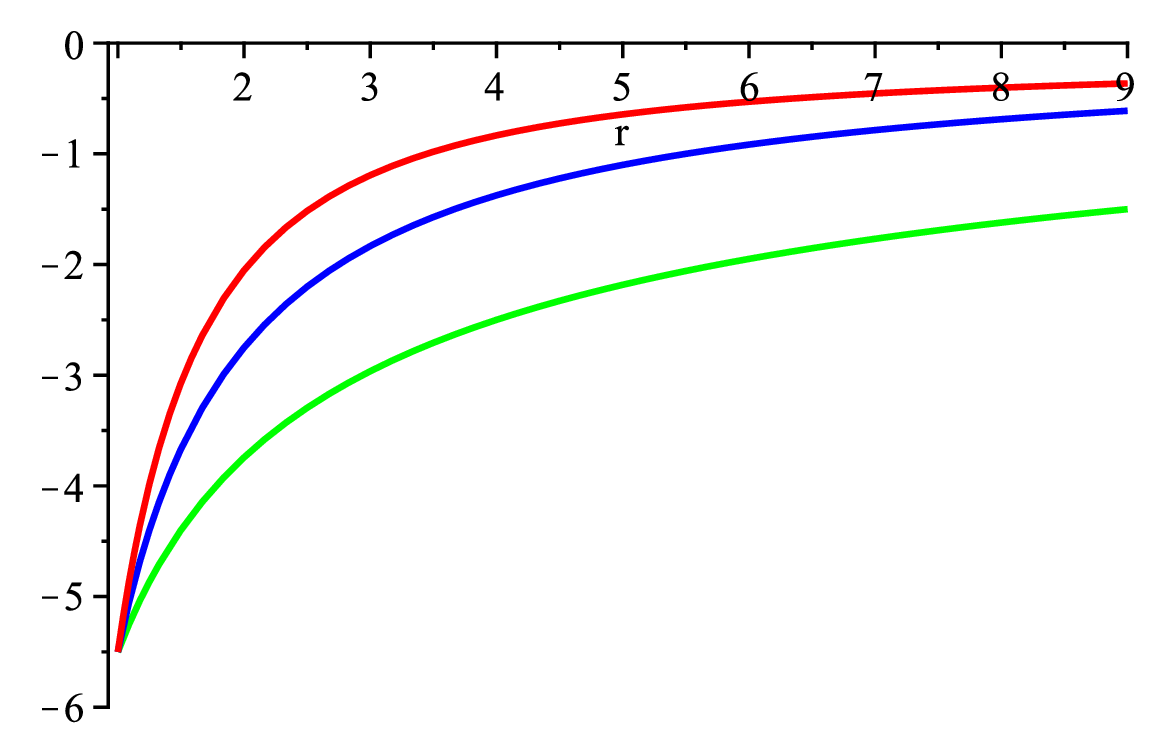}
 		\caption{The behavior radial (left panel) and tangential (right panel) EoS parameters against radial coordinate. The model parameters have been set as $\lambda=-0.1 $, $r_0=1$, $m=1.5$ (red curve), $m=2$ (blue curve) and $m=2.5$ (green curve).}\label{wor3}
 	\end{center}
 \end{figure}

\begin{figure}
 	\begin{center}
 		\includegraphics[width=7cm]{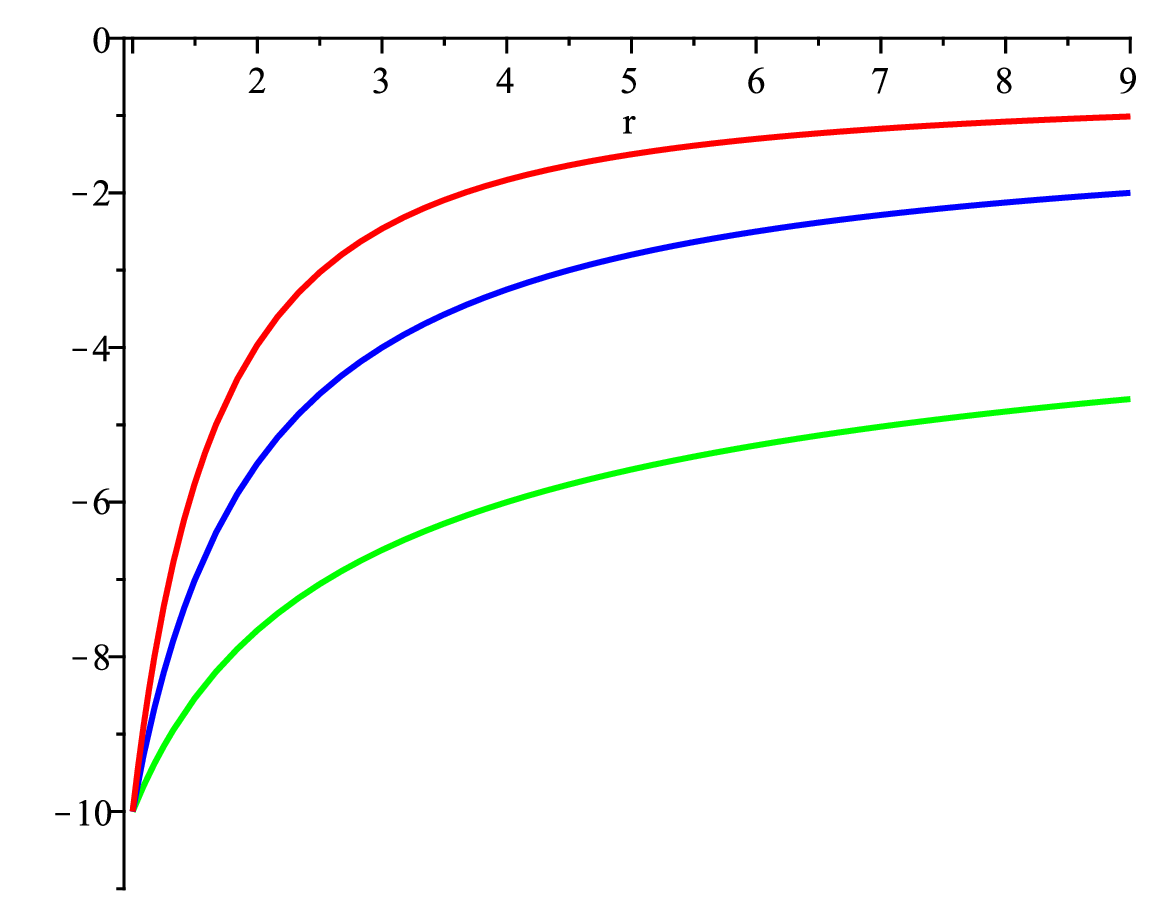}
 		\includegraphics[width=7cm]{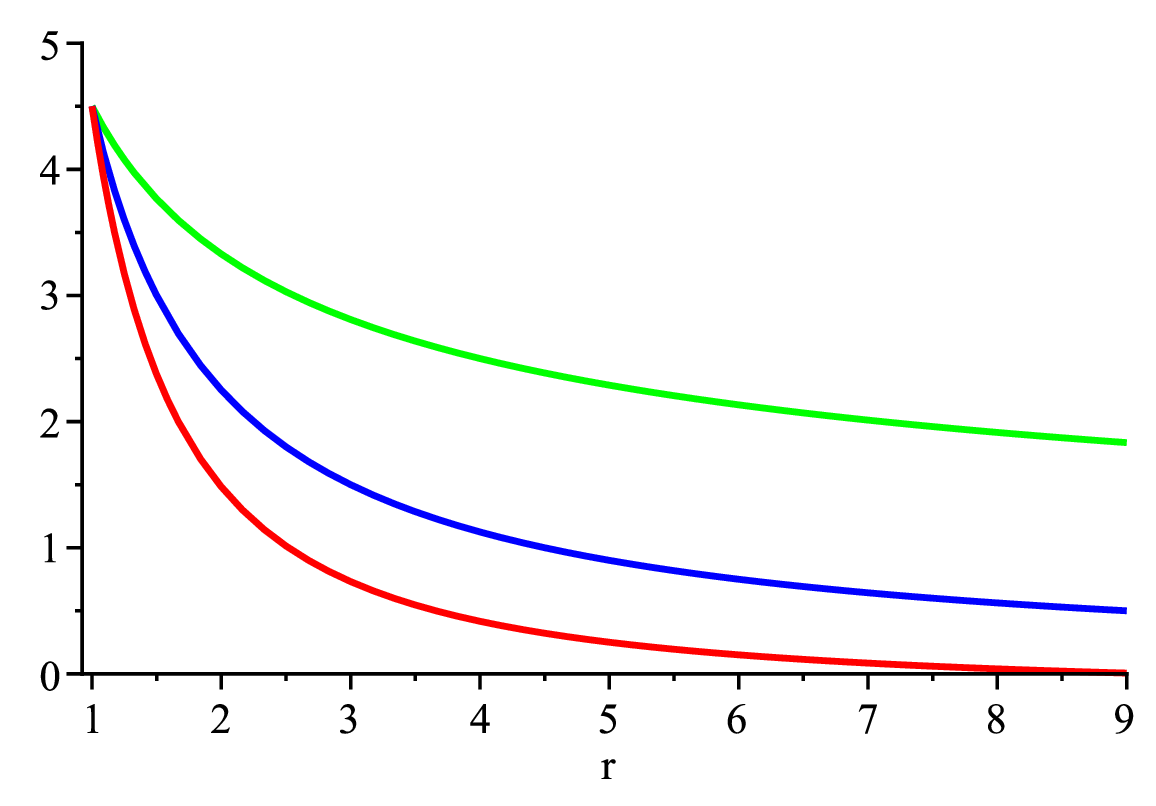}
 		\caption{The behavior radial (left panel) and tangential (right panel) EoS parameters against radial coordinate. The model parameters have been set as $\lambda=0.1 $, $r_0=1$, $m=1.5$ (red curve), $m=2$ (blue curve) and $m=2.5$ (green curve).}\label{wor4}
 	\end{center}
 \end{figure}

 
In what follows, we consider the energy density for the configuration of a sphere (or a spherical lens) situated above a large disk. The closest separation between the sphere and disk points is taken as $a\ll R$, where R is a sphere (lens) radius. In~\cite{lifsh}, it is shown that for a configuration of a perfectly conducting disk and lens with large separations, the energy density is given by
\begin{align}
	\rho=\frac{\lambda}{8\pi r^3},\label{lens}
\end{align}
in which $\lambda=-\pi^3/90$. We consider the distance between the lens and disk $a$ as the radial coordinate $r$ in wormhole spacetime. Thus, for the special case of $m=3$ and by using equation (\ref{feild1}) the shape function is obtained as
\begin{align}
b(r)=\lambda\ln\left( \f{r}{r_0} \right)+r_0,\label{bl3}
\end{align}
where the integration constant has been set according to the condition $b(r_0)=r_0$.
% \begin{align}
% C_0=\lambda \ln (r_0)-r_0
% \end{align}
The flare-out condition at the throat implies the inequality $\lambda<r_0$. It is then clear that these solutions are asymptotically flat, i.e. $b(r)/r$ tends to zero as $r\rightarrow\infty$. For this case the radial and tangential profiles of NEC along with the SEC are given as
\begin{eqnarray}
\rho +P_{r}&=&\frac{1}{8\pi r^3}\left[\lambda\ln\left(\frac{r_0}{r}\right)+\lambda-r_0\right],\label{EG7}\\
\rho +P_{t}&=&\frac{1}{16 \pi r^3}\left[-\lambda\ln\left(\frac{r_0}{r}\right)+\lambda+r_0\right],\label{EG8}
\end{eqnarray}%
\begin{eqnarray}
\rho +P_{r}+2P_{t}&=&0.\label{EG11}
\end{eqnarray}%
We then observe that the flare-out condition, i.e., $\lambda<r_0$, leads to $\rho(r_0)+P_r(r_0)<0$. Hence, the radial profile of NEC is violated at the throat. However, the tangential profile of NEC is satisfied at $r=r_0$. Moreover, for this case the Kretschmann scalar reads
\begin{eqnarray}
\mathcal{K}(r)=\frac{\lambda^{2}}{r^6}\left[6\ln(r_0)^2+6\ln(r)^2+4\ln\left(\frac{r_0}{r}\right)-12\ln(r)\ln(r_0) +2\right]+\frac{-4\lambda r_0}{r^6}\left[3\ln\left(\frac{r_0}{r}\right)+1\right]+\frac{6 r_0^2}{{r}^{6}},
\end{eqnarray}
whence we find that the Kretschmann scalar is finite in the whole range $r>r_0$. In Fig.~(\ref{wor2}) we have shown the behavior  $\rho+P_{t}$, $\rho+P_{r}$ and SEC versus radial coordinate for negative and positive values of $\lambda$ parameter in the left and right panels respectively. We see that we can choose suitable values for the parameter $\lambda$ in order to have normal matter at the throat and at spatial infinity. Fig.~(\ref{wor3}) shows the behavior of $w_r$ and $w_t$ against $r$ for $\lambda<0$, where we observe that $w_r$ gets larger values as we increase the value of parameter $m$, whereas $w_t$ takes larger values in negative direction as $m$ increases. Fig.~(\ref{wor4}) shows a plot of EoS parameters in radial and tangential directions for $\lambda>0$, where, we see that the behavior of $w_r$ ($w_t$) resembles that of $w_t$ ($w_r$) for $\lambda<0$.
 
\subsection{Specific case: non-constant redshift function}\label{WHS2}
In this section, we seek for spacetimes admitting wormhole structures with non zero tidal force. To this aim we must adopt a strategy for specifying the redshift function. Here, we consider a linear EoS, which provides a relation between the EMT components, namely, $P_{r}=w\rho$. Using then Eqs.~(\ref{feild1}) and (\ref{feild2}), we arrive at the following differential equation
\begin{align}
rg(r)f^\prime(r)+f(r)\left[rwg^\prime(r)+\left(g(r)-1\right)\left(1+w\right)\right]=0.
\label{dereqs}
\end{align}
Now, substituting for the shape function from Eq.~(\ref{shp1}) into Eq.~(\ref{dereqs}), we find the redshift function in general form as
\begin{equation}
f(r)={\rm exp}\left[\int^{r}_{r_{0}}X(r)dr+f_0\right],\label{redshift}
\end{equation}
 where 
 \begin{align}
 	 X(r)=\frac{-C_0\left(m-3 \right){r}^{m}+\lambda{r}^{3} \left[w(m-3)-1\right] }{r\left[ \left( m-3 \right)  \left( r+C_0 \right) {r}^{m}+\lambda r^{3} \right]},
 \end{align}
and $f_0$  is an integration constant. In order to check the energy conditions, we use the field equations and the shape function Eq.~(\ref{shp1}) to get the following expressions 
\begin{eqnarray}
\rho +P_{r}&=&(1+w)\rho,\label{EGBNEC}
\end{eqnarray}%
 
\begin{eqnarray}
\rho +P_{t}&=& \f{\Sigma_1}{32\pi r^mr_0^m}\big[(m-3)(r_0-r)-\lambda(r^{3-m}-r_0^{3-m})\big]^{-1},\label{wec2m}
\end{eqnarray}
where
  \begin{eqnarray}
  \Sigma_1&=&\lambda^2\left[(3-(m-3)w)(w-1)r_0^{m}{r}^{3-m}+\left(3-\left(2m-3\right)w\right)r_0^{3}\right]\nn
  &&+\left[\left( -2m^{2}+9m-9 \right) w-9+3m \right]r_0^{1+m}+2\left( m-3 \right)  \left( -2+ \left( m-2 \right) w \right)rr_0^{m}
  \end{eqnarray}
 
 \begin{eqnarray}
 \rho+P_{r}+2P_{t}&=&\f{\Sigma_2}{16\pi r^mr_0^m}\big[(m-3)(r_0-r)-\lambda(r^{3-m}-r_0^{3-m})\big]^{-1},\label{sec2m}
 \end{eqnarray}
where
\begin{eqnarray}
\Sigma_2&=&\Sigma_1-2\lambda\left[{r}^{3-m}r_0^{m}\lambda+r_0^{1+m}\left(3-m\right)+r\left(m-3\right)r_0^{m}-\lambda r_0^{3}\right] \left(w-1\right). 
\end{eqnarray}
In all the above equations, we have employed the solution given in Eq.~(\ref{redshift}) and its derivative with respect to $r$. For $m>3$, the quantities $\rho +P_{t}$ and $\rho+P_{r}+2P_{t}$ in the limit of large values of radial coordinate take the following forms
\begin{eqnarray}
\rho +P_{t}&=&-\frac{\lambda}{16\pi r^m}[w(m-2)-2]+{\mathcal O}\left(\frac{1}{r^{m+1}}\right),
\end{eqnarray}
and
\begin{eqnarray}
\rho +P_{r}+2P_{t}&=&-\frac{\lambda}{8\pi r^m}[w(m-3)-1]+{\mathcal O}\left(\frac{1}{r^{m+1}}\right).
\end{eqnarray}  
It is therefore seen that both of the above quantities tend to zero as $r\rightarrow\infty$. Moreover, at throat we have
\begin{eqnarray}
\rho +P_{t}\Big|_{r=r_0}=\frac{m-1+3\lambda r_0^{2-m}}{32\pi r0^{2}},~~~~	\rho +P_{r}+2P_{t}\Big|_{r=r_0}=\frac{m-3+\lambda r_0^{2-m}}{16 \pi r_0^{2}}.
\end{eqnarray}
whence, we recognize that a suitable choice of $\lambda$ parameter and throat radius can lead to satisfaction of $\rho+P_{r}+2P_{t}$ and NEC in tangential direction throughout the spacetime. We now proceed to find traversable Casimir wormholes for which $i)$ the redshift function is finite everywhere (absence of horizon), $ii)$ any spacetime singularity is avoided at or near the wormhole throat. This can be achieved through suitable values of EoS parameter $w$ at the throat. We note that the second condition comes from this argument that the presence of spacetime singularities in GR signals the break down of the theory in the singular region~\cite{penbh,joshibook}. Also, a singular spacetime contains incomplete paths which means that, any particle or observer traveling through this path would experience only a finite interval of existence that in principle cannot be continued any longer. Hence, the existence of spacetime singularity and consequently path incompleteness could have undesirable effects on traversability of the wormhole~\cite{mothorn,RicKretch}. To investigate this issue we find Ricci scalar given in Eq.~(\ref{ric1}) using the shape function (\ref{shp1}) and redshift (\ref{redshift})
\begin{align}
\mathcal{R}(r) =\f{\xi}{r^{m+1}r_0^m}\big[(m-3)(r_0-r)-\lambda(r^{3-m}-r_0^{3-m})\big]^{-1},\label{ri2ch2}
\end{align}
 where
 \begin{eqnarray}
 \xi&=&{\lambda}^{2}\left[\f{r_0^{m}}{2}\left(3-\left( m-3 \right) {w}^{2}-\left( 2-m \right) w \right) {r}^{4-m}-r\left(\f{3}{2}+ \left(m-\f{5}{2}\right) w \right)r_0^{3}\right]\nn
 &+&\lambda r\left[ - \left(\f{3}{2}+ \left( m-\f{5}{2}\right) w \right) \left( m-3\right)r_0^{1+m}+ r\left(1+\left( m-3 \right) w \right) \left( m-3 \right)r_0^{m} \right].
 \end{eqnarray}
In view of the above expression we realize that the Ricci scalar diverges in the limit of approach to the throat. This occurs due to divergence terms within the denominator of Eq.~(\ref{ri2ch2}). However, it is still possible to find suitable values of EoS parameter in order to avoid divergence in Ricci scalar. An investigation with more scrutiny reveals that if we choose the EoS parameter as $w=-r_0^{m-2}/\lambda$ then $\xi\rightarrow0$ in the limit $r\rightarrow r_0$, hence, the singularity at the wormhole throat can be removed, using the L'Hopital's rule. We therefore get the Ricci scalar at the throat as
% \begin{align}
% 	w=-\frac{{{\it r_0}}^{m-2}}{\lambda}
% 	\label{trawh}
% \end{align}
\begin{align}
\mathcal{R}(r_0)=\frac{3\lambda r_0^{2-m}+3-m}{2 r_0^2}.\label{ri2ch22}
\end{align}
By the same argument, the Kretschmann scalar at the throat assumes the following form
\begin{eqnarray}
 \mathcal{K}(r_0)& =&\frac{9\lambda^{2} r_0^{4-2m}+2\lambda(m-11)r_0^{2-m}+m^2-6m+33}{4r_0^4}. \label{krmr0}
 \end{eqnarray}
As expected, these scalar curvatures are finite at the throat signaling the absence of spacetime singularity. It can be shown that these quantities behave regularly in the regions with $r>r_0$ hence, there is no singular region in wormhole spacetime to affect its traversability. In what follows we study some specific wormhole solutions and their physical properties in more detail.
\subsubsection{Case-I: Parallel plates}\label{2pp}
\begin{figure}
\begin{center}
\includegraphics[scale=0.35]{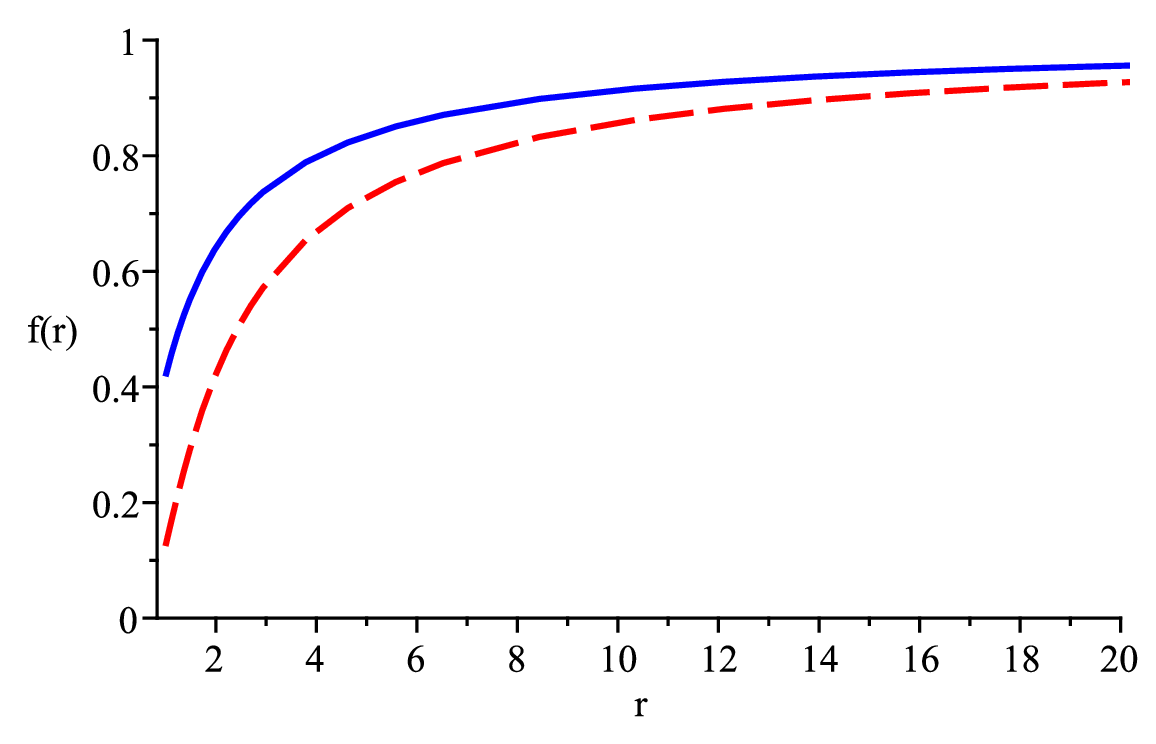}
\caption{The behavior of the metric function $f(r)$ for $r_0=1$, $m=4$ and $\lambda=0.5,-0.09$ from down to up, respectively.}\label{figbr4}
 		\end{center}
 	\end{figure}

\begin{figure}
 	\begin{center}
 		\includegraphics[scale=0.35]{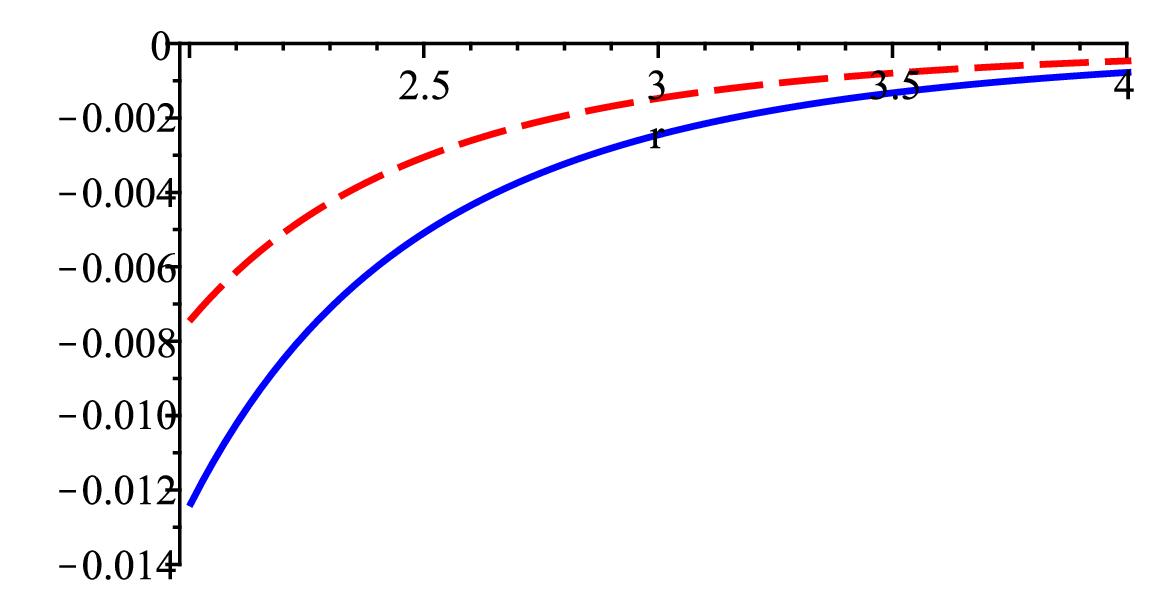}
 		\includegraphics[scale=0.35]{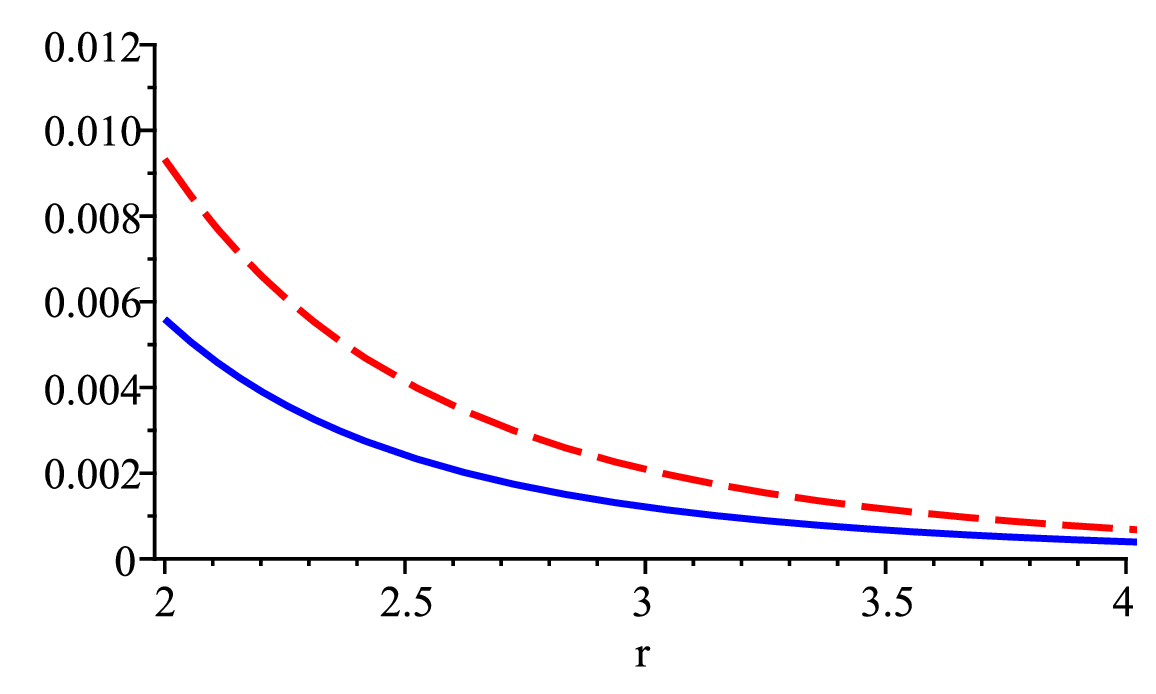}
 		\caption{The behavior of  $\protect\rho +p_{r}$ (left panel) and $\protect\rho +p_{t}$ (right panel) versus $r$. The model parameters have been set as $m=4$ and $r_0=2$, $\lambda=-1$ (solid curve) and $\lambda=1$ (dashed curve).}\label{figbrmw4}
 	\end{center}
 \end{figure}

\begin{figure}
\begin{center}
\includegraphics[scale=0.4]{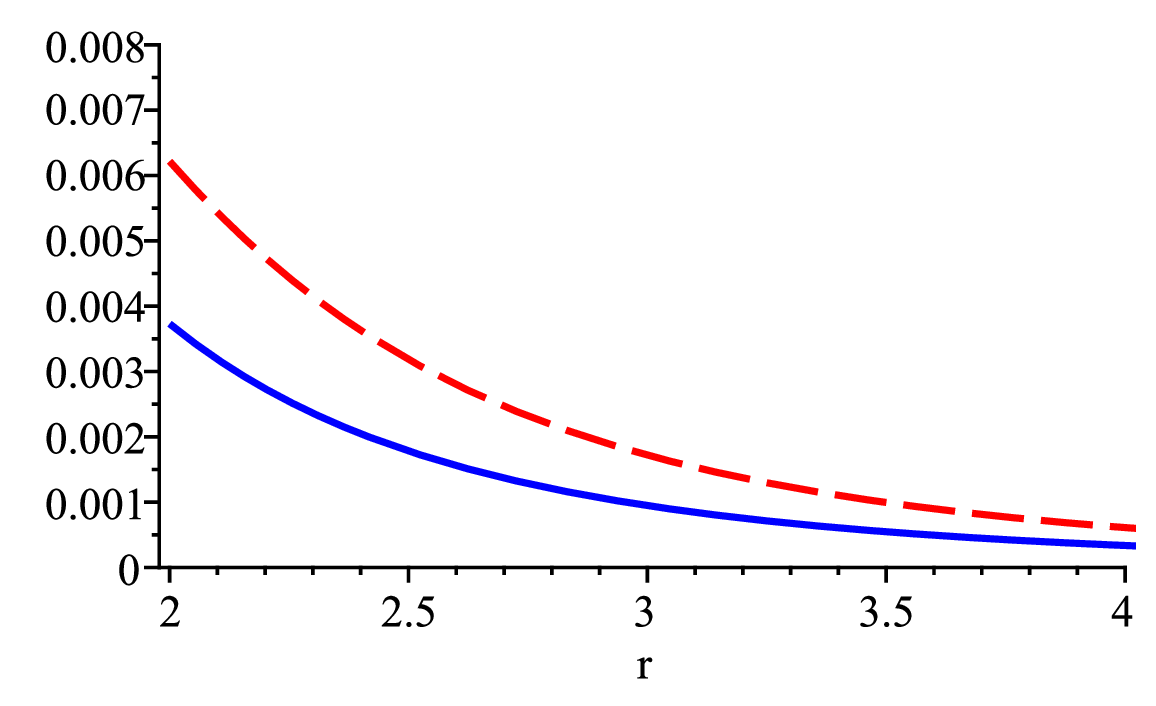}
\caption{The behavior of $\protect\rho +p_{r}+2p_{t}$ versus $r$. The model parameters have been set as $m=4$ and $r_0=2$, $\lambda=-1$ (solid curve) and $\lambda=1$ (dashed curve).}\label{figbrm4}
\end{center}
\end{figure}

\begin{figure}
\begin{center}
\includegraphics[scale=0.35]{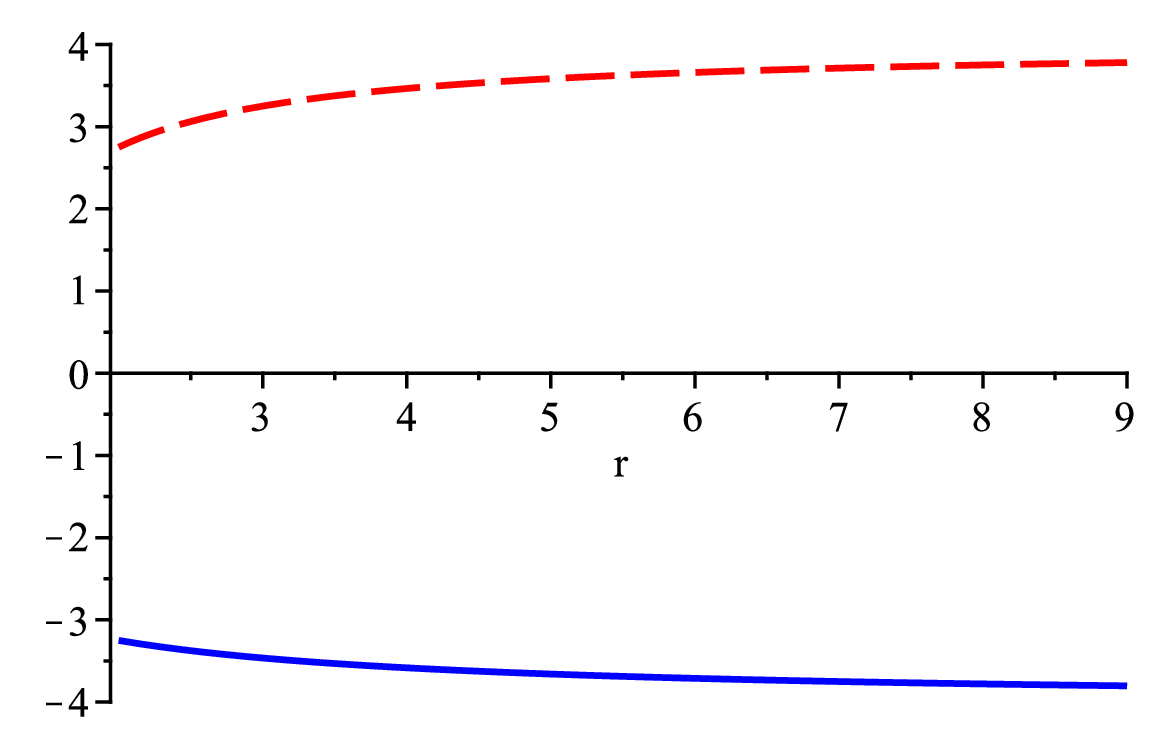}
\caption{The behavior of $w_t$ with respect to $r$ for $r_0=2$, $m=4$ and $\lambda=-1,1$ from down to up, respectively.}\label{figbr}
\end{center}
\end{figure}

As we have discussed earlier, this case has been studied in various wormhole configurations. The Casimir energy density for two parallel plates is given by equations~(\ref{rhca1}) and (\ref{rhca4}). So, the energy density $\rho$ is given by
\begin{align}
\rho=\frac{\lambda}{8\pi r^4},
\end{align}
where $\lambda=-{8\pi^3}/{720}$ for parallel planes made of ideal metal at zero temperature and, $\lambda=7\pi^3/720$ in the case of Casimir interaction between an ideal metal plane and an infinitely permeable plane characterized by an infinitely large magnetic permeability. Using then equations~(\ref{dereqs}) and (\ref{shp1}) for $m=4$, we obtain the shape function and red-shift function as
\begin{eqnarray}
b(r)=\frac{(r-r_0)\lambda+r_0^2 r}{r_0 r},\label{b4}
\end{eqnarray}
and
\begin{eqnarray}
f(r)={C_1}\left(r_0r-\lambda \right)^{-{\frac{wr_0^{2}+\lambda}{r_0^{2}-\lambda}}}{r}^{w-1} \left( r-r_0\right) ^{{\frac{w\lambda+r_0^{2}}{r_0^{2}-\lambda}}},\label{fr4}
\end{eqnarray}
where ${C_1}$ is an integration constant. Also, we see that for all the values of the parameter $\lambda$, the quantity $b(r)/r$ tends to zero at spatial infinity. The flare-out condition at the throat leads to the inequality $\lambda<r_0^2$ which violates the NEC at the throat. From Eq.~(\ref{fr4}), it is clear that $f(r_0)=0$ and consequently the Ricci and Kretschmann scalars diverge as we approach the throat. However, we can remove the spacetime singularity at the throat of wormhole by choosing $w=-r_0^2/\lambda$ in the red shift function which gives
\begin{align}\label{red4}
f(r)=\left(1-\frac{\lambda}{r_0 r} \right)^{\frac{r_0^{2}+\lambda}{\lambda}}.
\end{align}
We therefore find out that the Ricci and Kretschmann scalars assume the following finite values at the throat
\be\label{RicKret4}
{\mathcal R}(r_0)=\f{3\lambda-r_0^2}{2r_0^4},~~~~~~~~~~\mathcal{K}(r_0)=\f{25r_0^4-14r_0^2\lambda+9{\lambda}^2}{4r_0^8}.
\ee
In Fig.~(\ref{figbr4}) the behavior of redshift function is shown where we observe that $f(r)$ is finite everywhere and thus there is no horizon in the wormhole spacetime. This along with the shape function (\ref{b4}) provide us with asymptotically flat traversable wormhole solutions. Using equations~(\ref{wec2m}) and (\ref{sec2m}) for $m=4$ and $w=-r_0^2/\lambda$ we get
\bea
\rho +P_{t}&=&{\frac{\left( 4r_0r-r_0^{2}-3\lambda\right)\left(r_0^{2}+\lambda\right)}{32\pi\left(r_0r-\lambda \right)r^{4}}},\\
\rho+P_r+2P_{t}&=&\frac{\left(2r_0r-r_0^{2}-\lambda\right)\left(r_0^{2}+\lambda \right)}{16\pi\left(r_0r-\lambda\right){r}^{4}}.
\eea
In Fig.~(\ref{figbrmw4}) the behavior of radial (left panel) and tangential (right panel) profiles of NEC is depicted. Fig.~(\ref{figbrm4}) shows the behavior of SEC versus radial coordinate. We see that the radial profile of NEC is negative everywhere ($\rho+P_{r}=\frac{(\lambda-r_0^2)}{\lambda}\rho<0$), although one can choose suitable values for the parameters $\lambda$ and $r_0$ so that both $\rho +P_{t}$ and $\rho+P_{r}+2P_{t}$ take positive values throughout the spacetime. Also, using the field equations we get the tangential EoS for $m=4$ as
\begin{eqnarray}
w_{t}=\frac{\lambda^{2}-4r_0^{2}\lambda-r_0^{4}+4rr_0^{3}}{4\left(r_0r-\lambda\right)\lambda}.\label{EG1C4}
\end{eqnarray}
Fig.~(\ref{figbr}) shows the behavior of $w_t$, where, it is seen that the tangential EoS parameter is a positive (negative) function of radial coordinate and increases in positive (negative) direction for $\lambda>0~(<0)$.
 
\subsubsection{Case-II: Two parallel eccentric cylinders}
\begin{figure}
  	\begin{center}
  		\includegraphics[width=7cm]{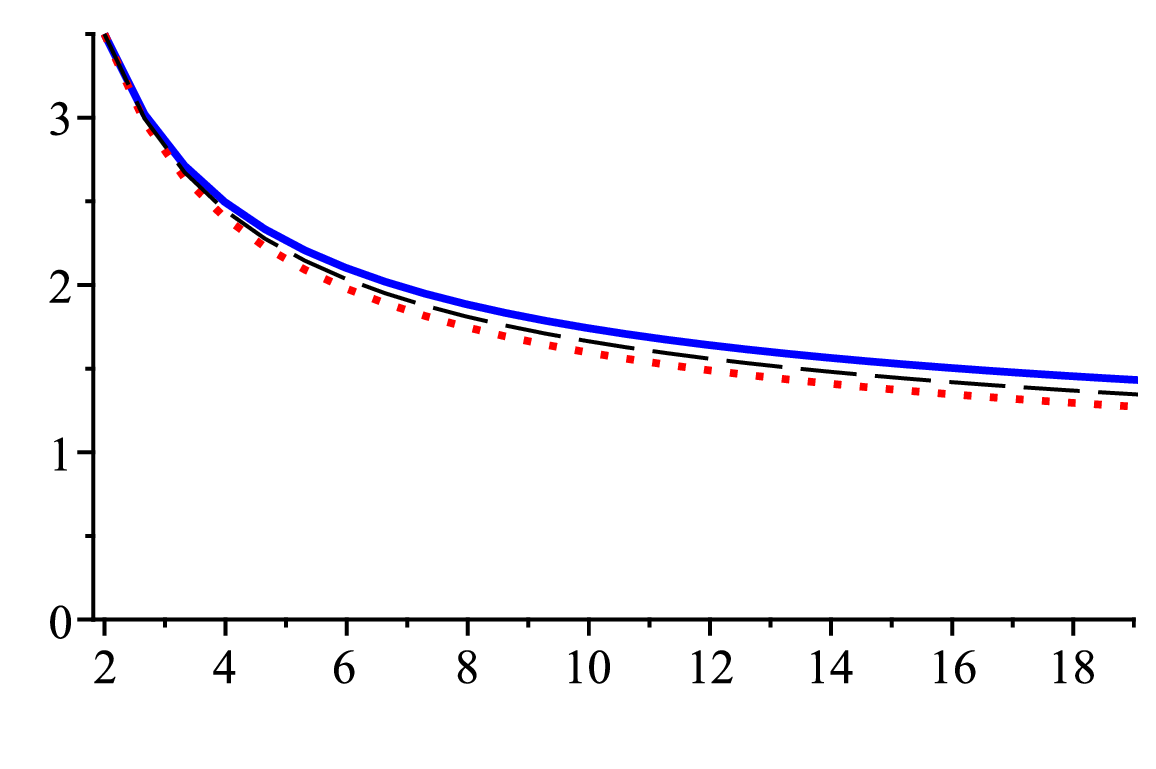}
  		\includegraphics[width=7cm]{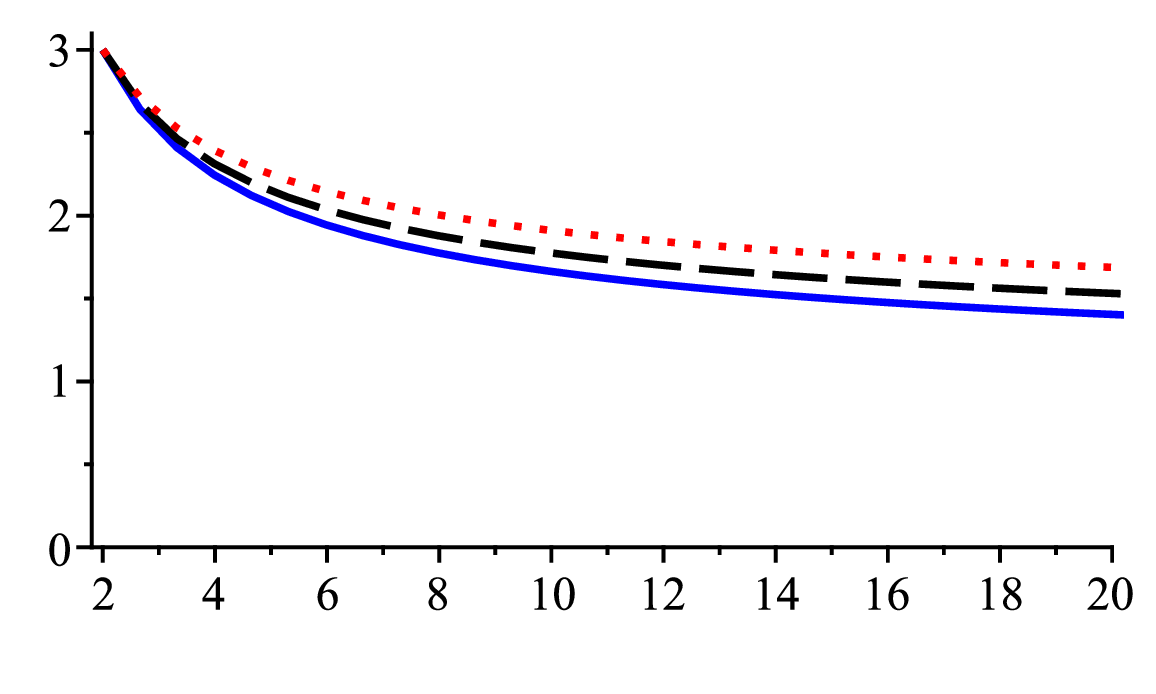}
  		\caption{Behavior of $f(r)$ with respect to $r$ for $r_0=2$ and $m=5/2$. In the left panel we have set $\lambda=-0.1,-0.2,-0.3$ from up to down and in right panel we have set $\lambda=0.3,0.2,0.1$ from up to down.}\label{figwor}
  	\end{center}
  \end{figure}

\begin{figure}
  	\begin{center}
  		\includegraphics[width=7cm]{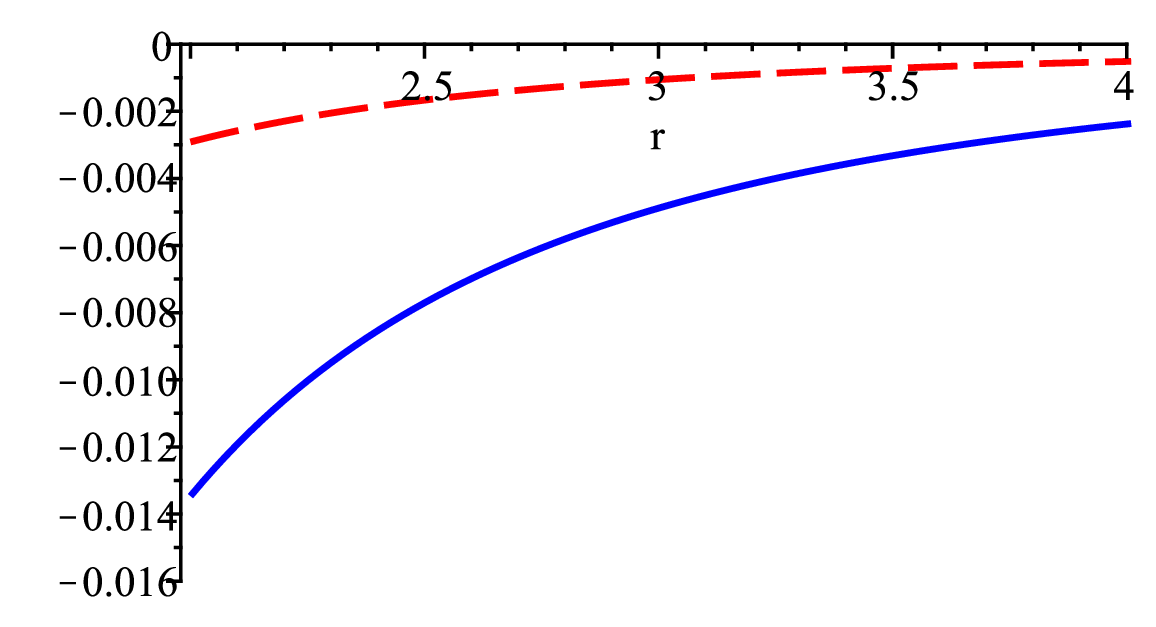}
  		\includegraphics[width=7cm]{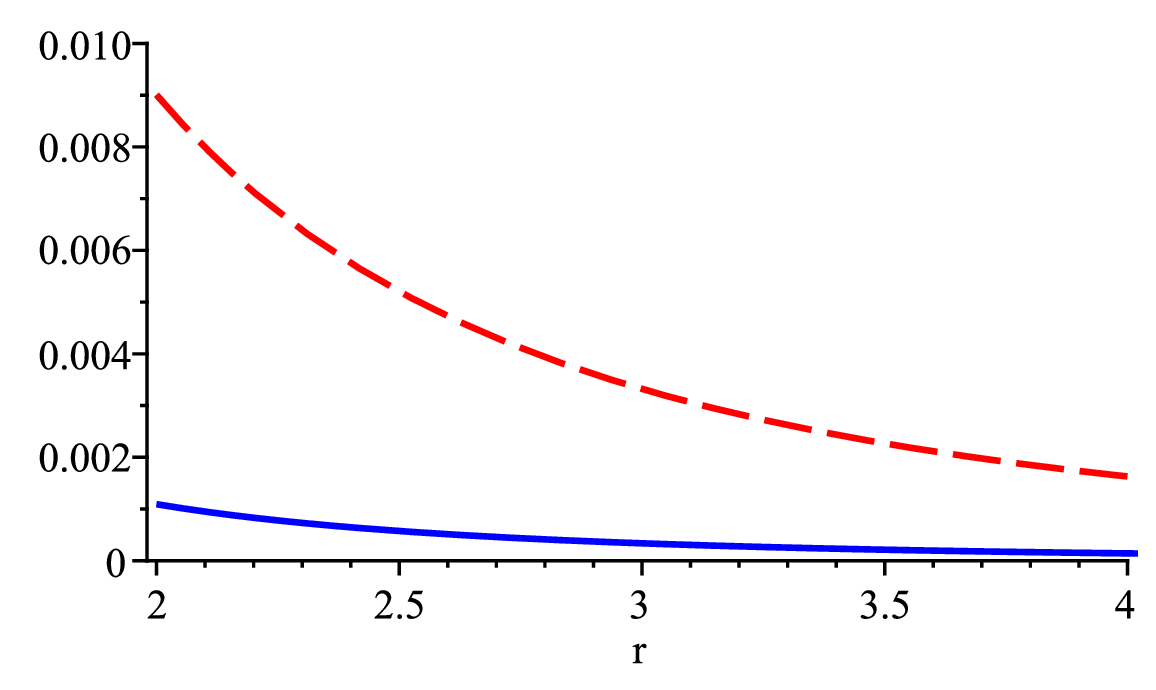}
  		\caption{The behavior of $\protect\rho +p_{r}$ (left panel) and $\protect\rho +p_{t}$ (right panel) versus $r$. The model parameters have been set as $\lambda=1$ (dashed curve), $\lambda=-0.5$ (solid curve), $m=5/2$ and $r_0=2$.}\label{figwor15}
  	\end{center}
  \end{figure}

\begin{figure}
 	  	\begin{center}
 	  		\includegraphics[scale=0.35]{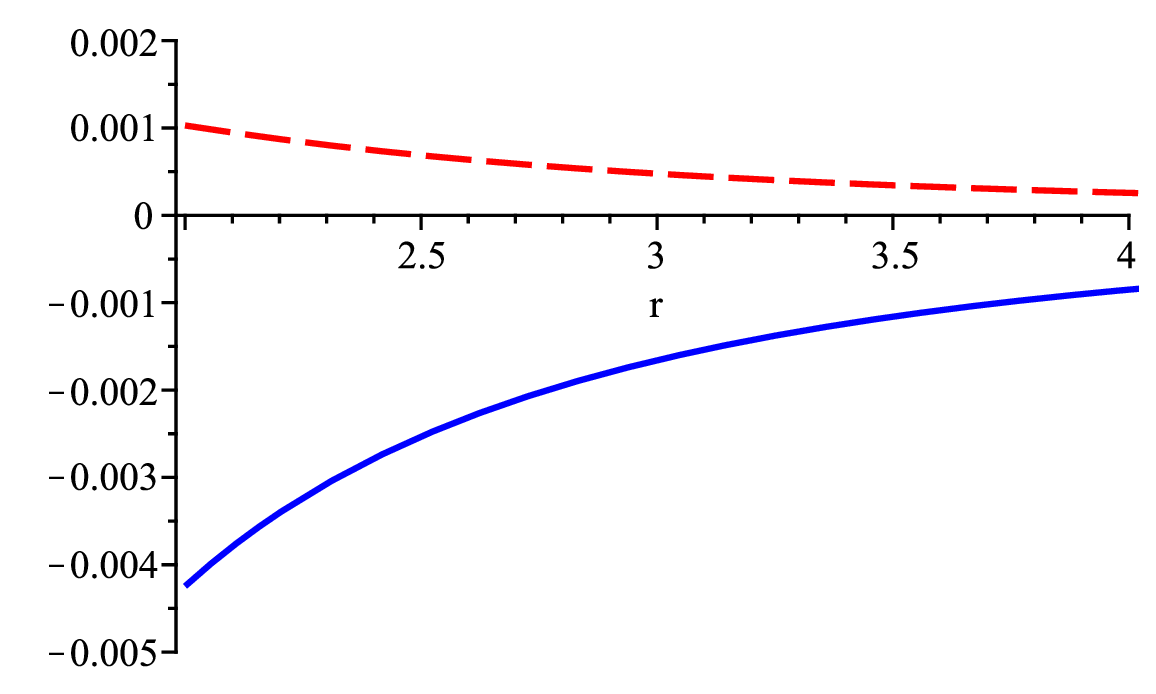}
 	  		\includegraphics[scale=0.35]{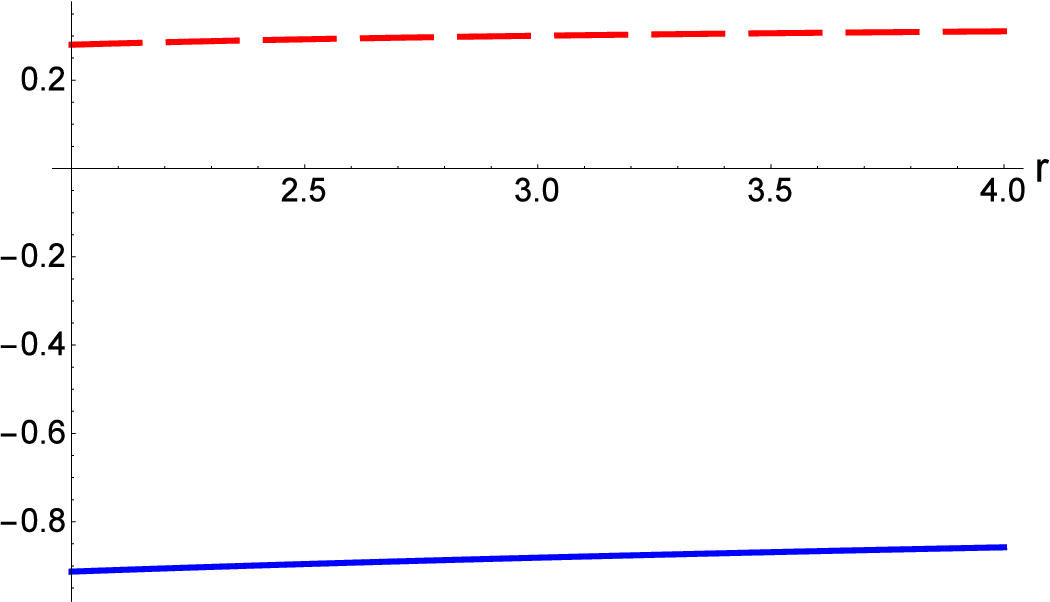}
 	  		\caption{(Left panel) The behavior of $\protect\rho +p_{r}+2p_{t}$ versus $r$. The model parameters have been chosen as $\lambda=1$ (dashed curve), $\lambda=-0.5$ (solid curve), $m=5/2$ and $r_0=2$. (Right panel) The behavior of EoS in tangential direction for the same values of parameters as of the left panel.}\label{figbr5}
 	  	\end{center}
 	  \end{figure}

In this subsection, we consider a Casimir setup consisting of two parallel cylinders of length $L$ and radii $a$ and $b$ respectively, so that, the cylinder of radius $a$ lies inside that of radius $b$. Let us denote by $\delta$ the separation between the centers of the cylinders, and $d$ the (varying) distance between them, hence we have $\delta= b-a-d$. An exact result for the Casimir energy for such a configuration has been
presented in~\cite{dalvi1}. In the case of asymptotic behavior of Casimir interaction, i.e, in the limit when $d\ll(b-a)$, the associated Casimir energy for Dirichlet (D) or Neumann (N) boundary conditions on both cylinders is given by~\cite{Teo2011}
\begin{align}
E^{cc}=-\frac{\pi^3 \sqrt{a b}L}{1920 d^{\frac{5}{2}} \sqrt{2(b-a)}}.
\end{align}
For the case of DN (Dirichlet on one cylinder and Neumann on the other) or ND boundary conditions the Casimir energy reads
\begin{align}
E^{cc}=\frac{7 \pi^3 \sqrt{a b}L}{15360 d^{\frac{5}{2}} \sqrt{2(b-a)}}.
\end{align}
The Casimir energy density can then be obtained using the volume between the two cylinders $V=\pi(b^2-a^2)L$, as
\begin{align}
\rho=\frac{E^{cc}}{\pi(b^2-a^2)}.
\end{align}
Here, we consider the distance between the cylinders $d$ as the radial coordinate $r$ of the wormhole. Then, energy density is found as $\rho=\lambda/8\pi r^{\f{5}{2}}$ where
\bea\label{lambds}
\lambda=-\frac{\pi^3 \sqrt{ab} L}{240\sqrt{2(b-a)^3}(a+b)},~~~~~~~~~\lambda=\frac{7\pi^3 \sqrt{ab} L}{1920\sqrt{2(b-a)^3}(a+b)}.
\eea
The first part of the above expression refers to DD or NN boundary conditions and the second one to DN or ND boundary conditions. Using now equation (\ref{shp1}) for $m=5/2$, we get the shape function as
\begin{eqnarray}
b(r)=2\lambda\left(\sqrt{r}-\sqrt{r_0}\right)+r_0.\label{b5}
\end{eqnarray}
For this solution, we find that the quantity $b(r)/r$ tends to zero at spatial infinity, so these solutions correspond to an asymptotically flat spacetime. The flare-out condition at the throat leads to $\lambda< \sqrt{r_0}$. Also, substituting for the shape function (\ref{b5}) into equation (\ref{dereqs}) and setting $w=-\sqrt{r_0}/\lambda$ we arrive at the following differential equation for the redshift function
\begin{align}
r^{\f{3}{2}}f^\prime(r)+\left(\sqrt{r_0}-2\lambda\right)\left[f(r)+rf^\prime(r)\right]=0.\label{f5ov2}
\end{align}
The above equation admits an exact solution in the form 
\be\label{exactsol}
f(r)=\f{f_0r_0\left[\sqrt{r}+\sqrt{r_0}-2\lambda\right]^2}{4r\left(\sqrt{r_0}-\lambda\right)^2},
\ee
where the integration constant has been set in such a way that the redshift function assumes a finite value at the throat, $f(r_0)=f_0$. Fig.~(\ref{figwor}) shows the behavior of redshift function where we observe that this function is finite for $r>r_0$. Using equations (\ref{wec2m}) and (\ref{sec2m}) for $m=5/2$ and $w=-\sqrt{r_0}/\lambda$, we get 
\begin{eqnarray}
\rho +P_{t}&=&\f{2r_0-\lambda\sqrt{r_0}-6\lambda^2+\sqrt{r}\left(4\lambda+\sqrt{r_0}\right)}{32\pi r^{\f{5}{2}}\left(\sqrt{r}+\sqrt{r_0}-2\lambda\right)},\\
\rho +P_{r}+2P_{t}&=&\f{(\sqrt{r_0}-2\lambda)(\lambda-\sqrt{r})}{16\pi r^{\f{5}{2}}\left(\sqrt{r}+\sqrt{r_0}-2\lambda\right)}.
\end{eqnarray} 
The behavior of NEC is shown in Fig.~(\ref{figwor15}) where we observe its radial profile (left panel) gets violated for both positive and negative values of $\lambda$ parameter. However, the NEC in tangential direction (right panel) is satisfied. Also the left panel in Fig.~(\ref{figbr5}) shows the behavior of SEC against radial coordinate where we see that for $m<3$ and $\lambda<0$ this quantity is negative at the throat and throughout the spacetime. Using the field equations we get the EoS parameter in tangential direction as 
\begin{eqnarray}
w_{t}=\f{\sqrt{rr_0}+2r_0-5\sqrt{r_0}\lambda+2\lambda^2}{4\lambda\left(\sqrt{r}+\sqrt{r_0}-2\lambda\right)}. \label{EG1C52}
\end{eqnarray}
The right panel in Fig.~(\ref{figbr5}) presents the behavior of the above expression where we observe that the NEC in tangential direction is fulfilled, see also the right panel in Fig.~(\ref{figwor15}). Also, from Eq.~(\ref{krmr0}) the Kretschmann scalar for $m=5/2$ is obtained as
\begin{eqnarray}
\mathcal{K}(r_0)=\f{97r_0-68\lambda\sqrt{r_0}+36\lambda^2}{16r_0^5}.
\end{eqnarray}
As it is expected, the Kretschmann scalar is finite at the throat for these traversable wormhole solutions.
  
 	\subsubsection{Case-III: Two concentric spherical shells} 
\begin{figure}
 	\begin{center}
 		\includegraphics[width=7cm]{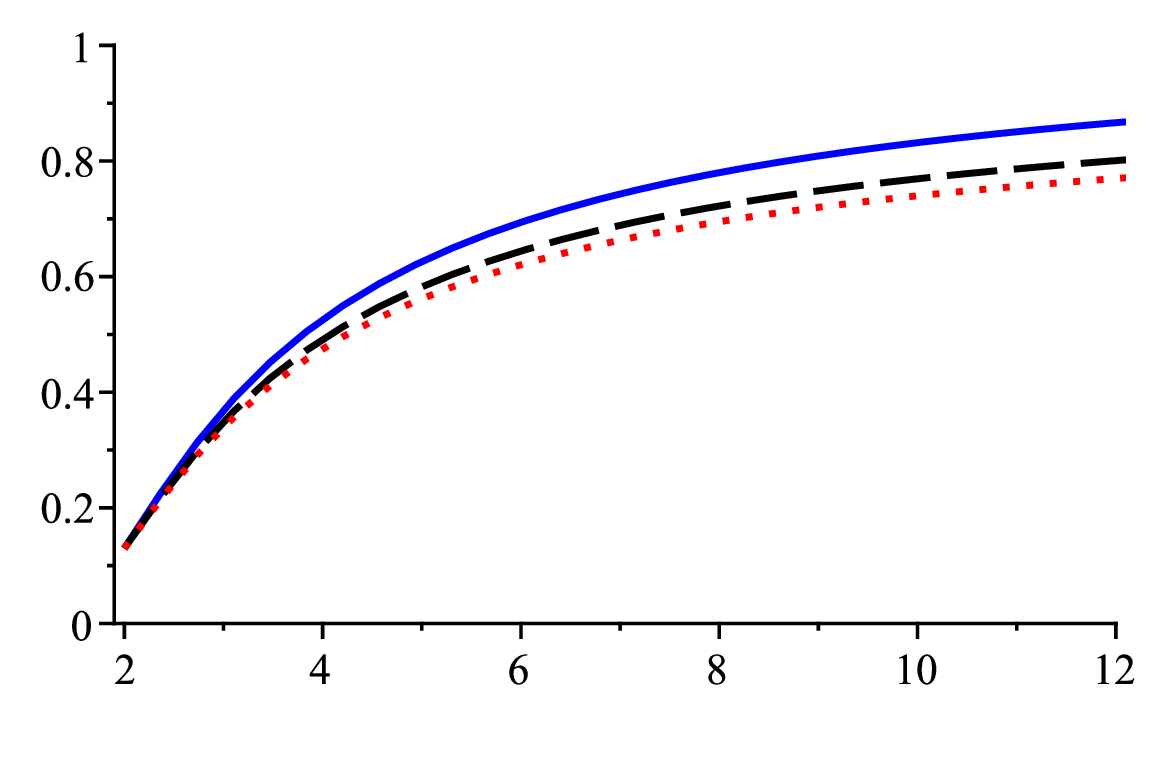}
 		\includegraphics[width=7cm]{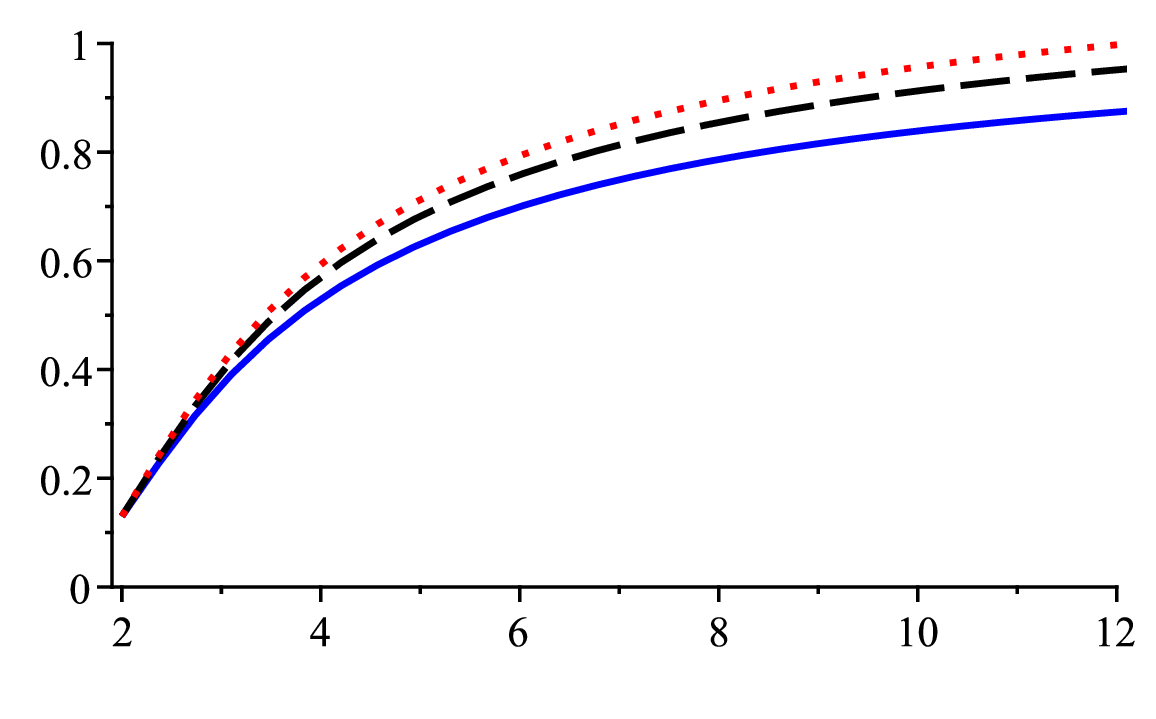}
 		\caption{The behavior of $f(r)$ with respect to $r$ for $r_0=2$, $m=7$ and $\lambda=-0.1,-2,-3$ (left panel) and $\lambda=3,2,0.1$ (right panel) from up to down.}\label{figwor7}
 	\end{center}
 \end{figure}

\begin{figure}
	\begin{center}
		\includegraphics[width=7cm]{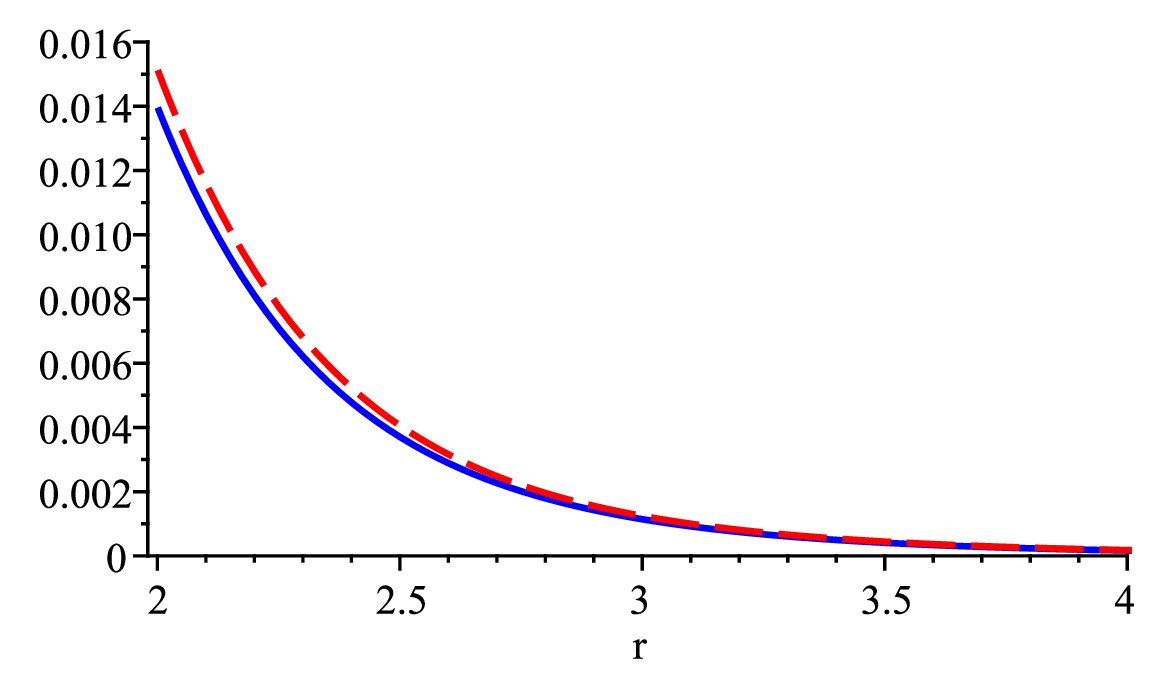}
		\includegraphics[width=7cm]{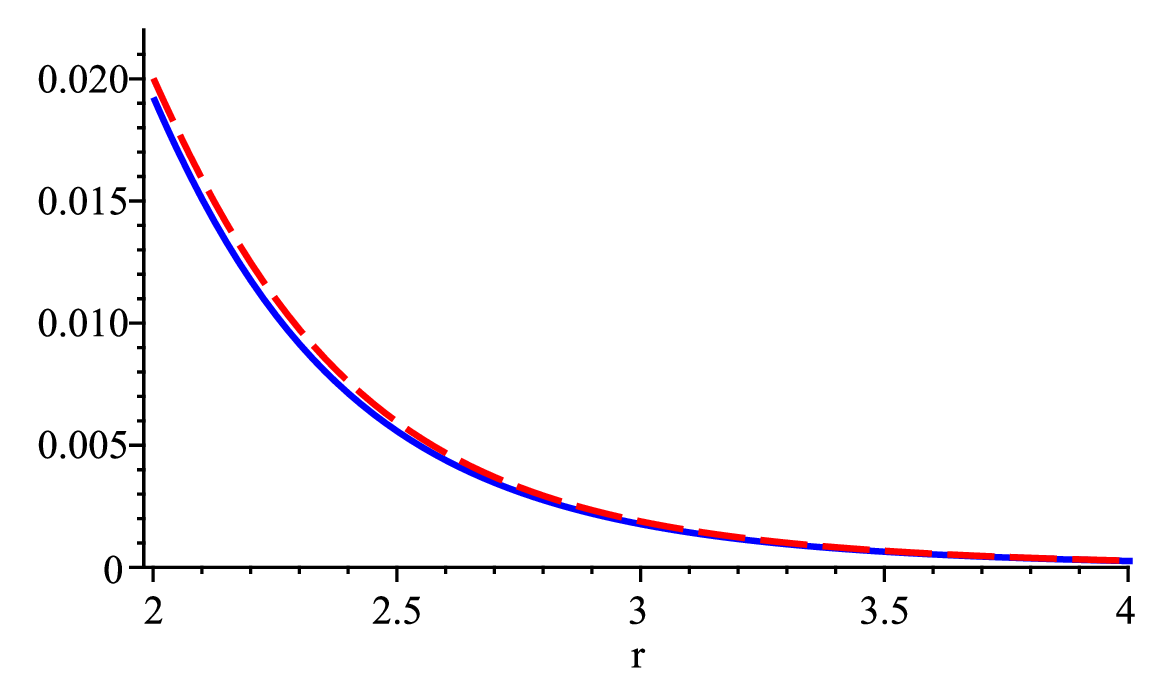}
		\caption {The behavior of  $\protect\rho +p_{t}$ (left panel) and $\protect\rho +2p_{t}+p_{r}$ (right panel) versus $r$. The model parameters have been set as $m=7$ and $r_0=2$, $\lambda=1$ (dashed curve) and $\lambda=-4$ (solid curve).}\label{fig1}
	\end{center}
\end{figure}

\begin{figure}
	\begin{center}
		\includegraphics[width=7cm]{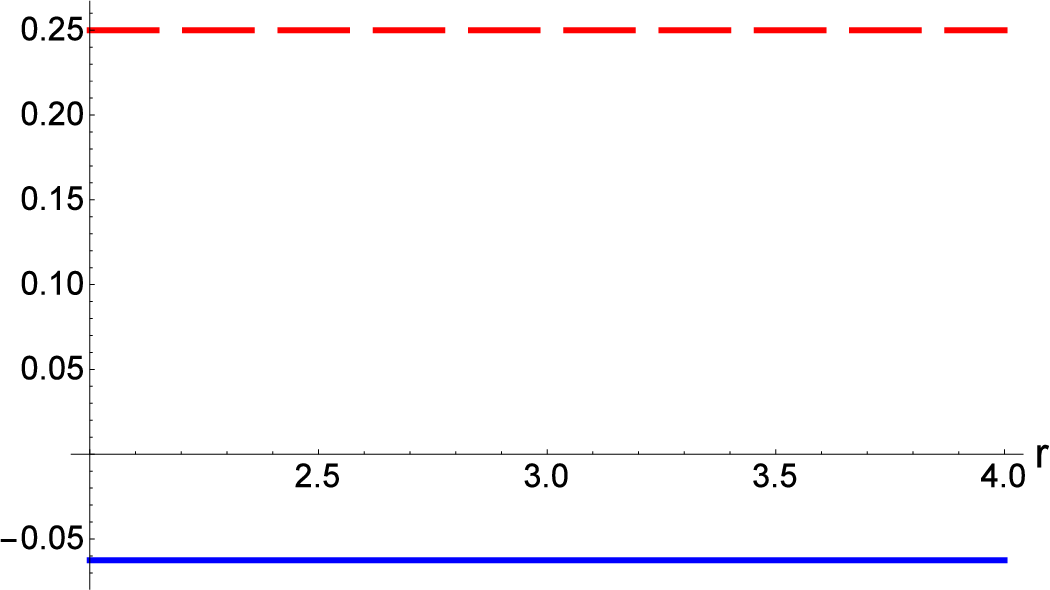}
		\caption {The behavior of tangential EoS parameter against $r$. The model parameters have been set as of Fig.~(\ref{fig1}).}\label{figwt7}
	\end{center}
\end{figure}

Let us  now consider two concentric spherical shells of radii $a$ and $b$ with $\alpha=b/a>1$. The exact form for Casimir energy has been discussed in~\cite{sahn,sahn1}. However, Using PFA method, the Casimir interaction energy can be computed for two limiting situations. In the short distance limit, i.e., $\alpha\rightarrow1$, the interaction energy is given as~\cite{Mazzitelliconf}
 \begin{align}
 E=-\frac{\pi^3 a^2}{180(b-a)^3}\left[1+(\alpha-1)+O(\alpha-1)^2\right].
 \end{align}
 In this case we can find the associated energy density using the fact that the volume between two concentric spheres is $V=4\pi(b^3-a^3)/3$. Hence, the energy density is found as
 \begin{align}
 \rho=\frac{3E}{4\pi(b^3-a^3)}.
 \end{align}
Considering the distance between the spheres $a$ as the radial coordinate, we find energy density as $\rho={\lambda}/{8\pi r^4}$, with $\lambda=-{\pi^3}/{360}$ (for both DD or NN boundary conditions). The wormhole configuration for this type of Casimir energy is similar to the case of two parallel plates which was studied in detail in sub Sec.~\ref{2pp}. We therefore proceed to study the opposite limit for which $\alpha\gg1$. The interaction energy is then written in the form~\cite{Mazzitelliconf}
\begin{align}
E=-\frac{2 \pi^3 }{35 a \alpha^4},
\end{align}
whence the Casimir energy density is found as
\begin{align}
\rho=-\frac{\lambda}{8\pi r^7}
\end{align}
where $\lambda=-{12\pi^3 a^3}/{35}$ and we have taken the outer radius of the sphere i.e., $b$, as the radial coordinate in our model. Using equation (\ref{shp1}) for $m=7$ we obtain the shape function as %we can choose $w=-\frac{r_0^5}{\lambda}$ or ($m=7$ in the eq (\ref{sec2m}))  so that the redshift function $f(r)$ finity at  the wormhole throat and all wormhole spacetime.
\begin{eqnarray}
b(r)=\frac{\left(r^4-r_0^4\right) \lambda}{4 r^4 r_0^4}+r_0.
\label{bob7}
\end{eqnarray}
The above solution corresponds to an asymptotically flat spacetime and the flare-out condition at the throat leads to $\lambda<r_0^5$. Substituting the shape function (\ref{bob7}) into Eq.~(\ref{dereqs}) along with setting $w=-r_0^5/\lambda$ we arrive at following differential equation for the redshift function  
\begin{align}
%\left( 4\,{{\it r_0}}^{9}-4\,{{\it r_0}}^{5}{r}^{4}+\lambda\,{{\it r_0}}
%^{4}-\lambda\,{r}^{4} \right) f \left( r \right) + \left( 4\,{r}^{6}{{
%		\it r_0}}^{4}+\lambda\,r{{\it r_0}}^{4}-4\,{{\it r_0}}^{5}{r}^{5}-\lambda
%\,{r}^{5} \right) {\frac{d}{dr}}f \left( r \right)=0
rf^\prime(r)\left(4r^5 r_0^4-r^4 \left(\lambda +4 r_0^5\right)+\lambda r_0^4\right)-f(r)\left(r^4-r_0^4\right)\left(\lambda+4 r_0^5\right)=0.\label{ff7}
\end{align}
For the above differential equation, an analytic solution can not be found in terms of elementary standard functions. We then resort to numerical techniques. Fig.~(\ref{figwor7}) shows the behavior $f(r)$ for $r_0=2$, $m=7$, where it is seen that the redshift function is finite everywhere and that we have no singularity in the wormhole spacetime. We may also examine the NEC and SEC for these solutions. Using then equations (\ref{wec2m}) and (\ref{sec2m}) for $m=7$ with $w=-r_0^5/\lambda$ we get
 \begin{align}
\rho+P_t\,\,&=&\!\!\!\!\!\!\frac{r_0^4 \left[r_0^5 \left(7 \lambda +40 r^5\right)+\lambda  \left(3 \lambda +16 r^5\right)-23 \lambda  r^4 r_0-44 r^4 r_0^6+4 r_0^{10}\right]-3 \lambda ^2 r^4}{32 \pi  r^7 \left(r_0^4 \left(\lambda +4 r^5-4 r_0 r^4\right)-\lambda  r^4\right)},\\
\rho+P_r+2P_t\,\,&=&\!\!\!\!\!\!\!\!\frac{r_0^4 \left[r_0^5 \left(5 \lambda +32 r^5\right)+\lambda  \left(\lambda +8 r^5\right)-13 \lambda  r^4 r_0-36 r^4 r_0^6+4 r_0^{10}\right]-\lambda ^2 r^4}{16 \pi  r^7 \left(r_0^4 \left(\lambda +4 r^5-4 r_0 r^4\right)-\lambda  r^4\right)}.
\end{align}
The left and right panels in Fig.~(\ref{fig1}) show the behavior tangential component of NEC and SEC where we see that one can choose suitable values of $\lambda$ and $r_0$ parameters so that both $\rho+P_{t}$ and $\rho+P_{r}+2P_{t}$ are satisfied at all space. Moreover, using the field equations, one can find the EoS in tangential direction as
\begin{align}
w_t=\frac{\lambda  r^4 (11 w+1)+r_0^4 \left[-40 r^5 w+4 r_0 r^4 (11 w+1)+\lambda  (w (4 w-7)-1)\right]}{4 \left(4 r^5 r_0^4-\lambda  r^4-4 r_0^5 r^4+\lambda r_0^4\right)}.
\end{align}
Fig.~(\ref{figwt7}) shows the behavior of the above quantity for the same values of model parameters as of Fig.~(\ref{fig1}). Also, from Eq.~(\ref{krmr0}), one may realize that the Kretschmann scalar is finite at the throat of wormhole.


 \section{Equilibrium conditions}\label{Eqcon}
\begin{figure}
 	\begin{center}
 		\includegraphics[width=7cm]{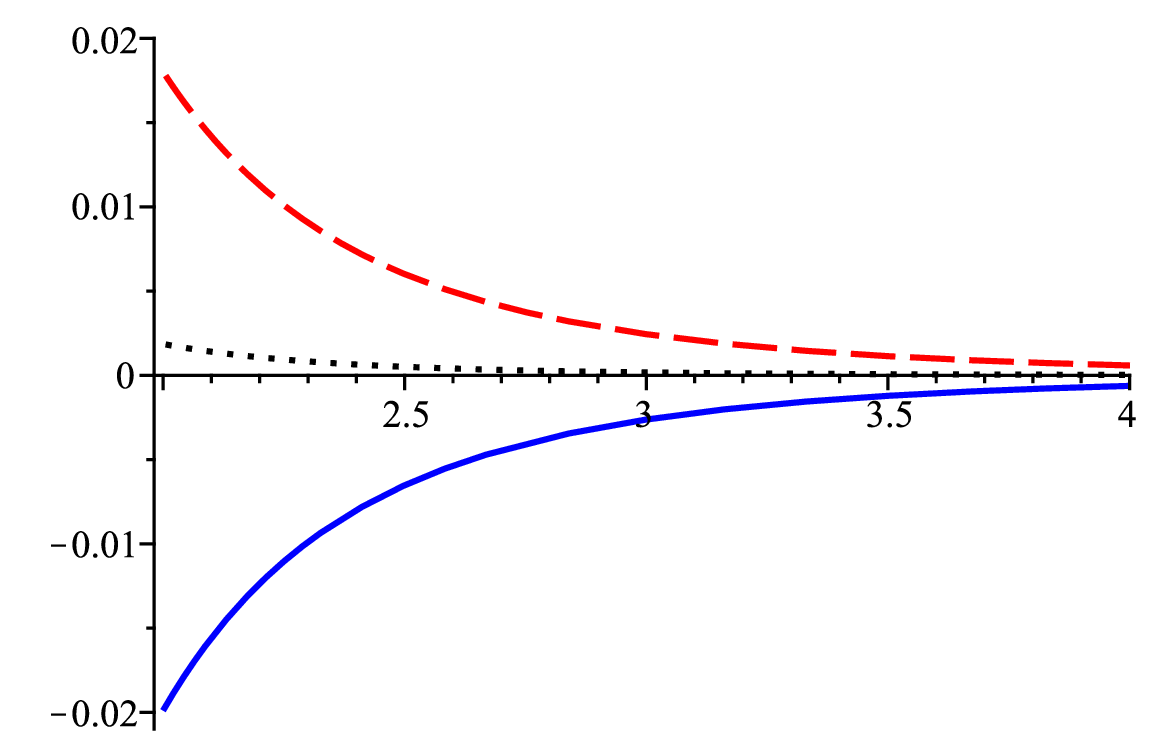}
 		\includegraphics[width=7cm]{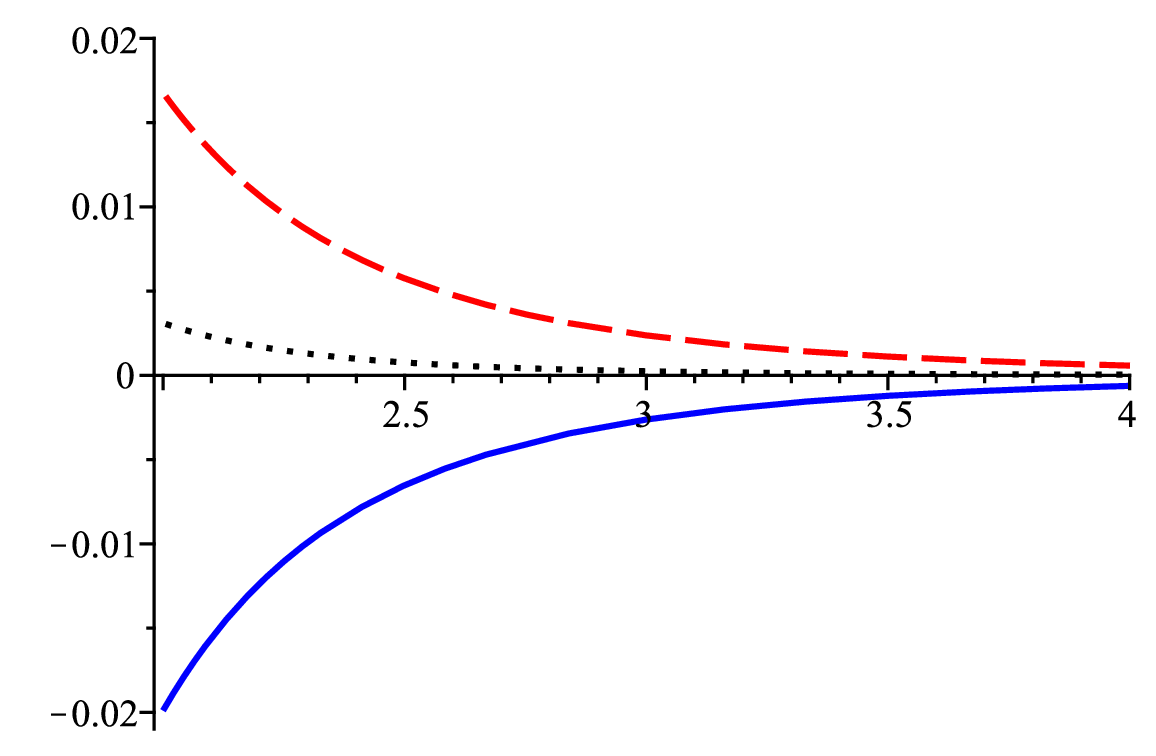}
 		\caption {The behavior of  $F_h$ (solid curve), $F_a$ (dashed curve) and $F_g$ (dotted curve) against $r$. The model parameters have been set as $m=4$ and $r_0=2$, $\lambda=-1$ (left panel) and $\lambda=1$ (right panel).}\label{figw}
 	\end{center}
 \end{figure}

\begin{figure}
 	\begin{center}
 		\includegraphics[width=7cm]{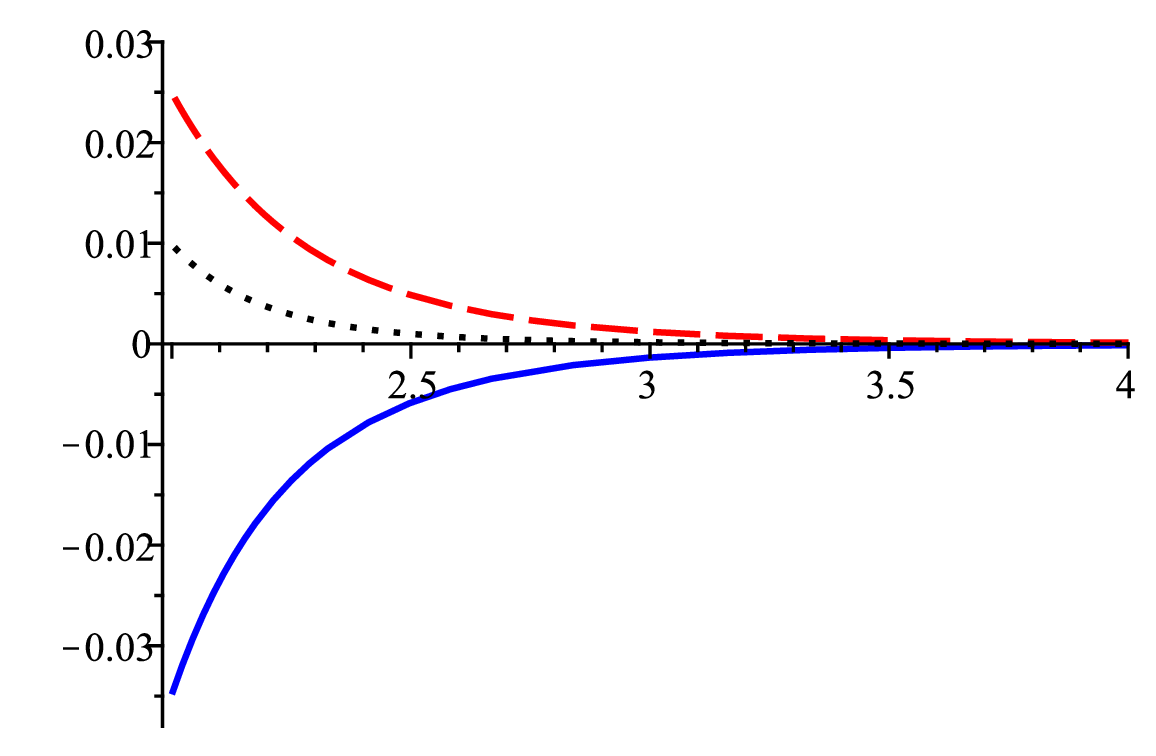}
 		\includegraphics[width=7cm]{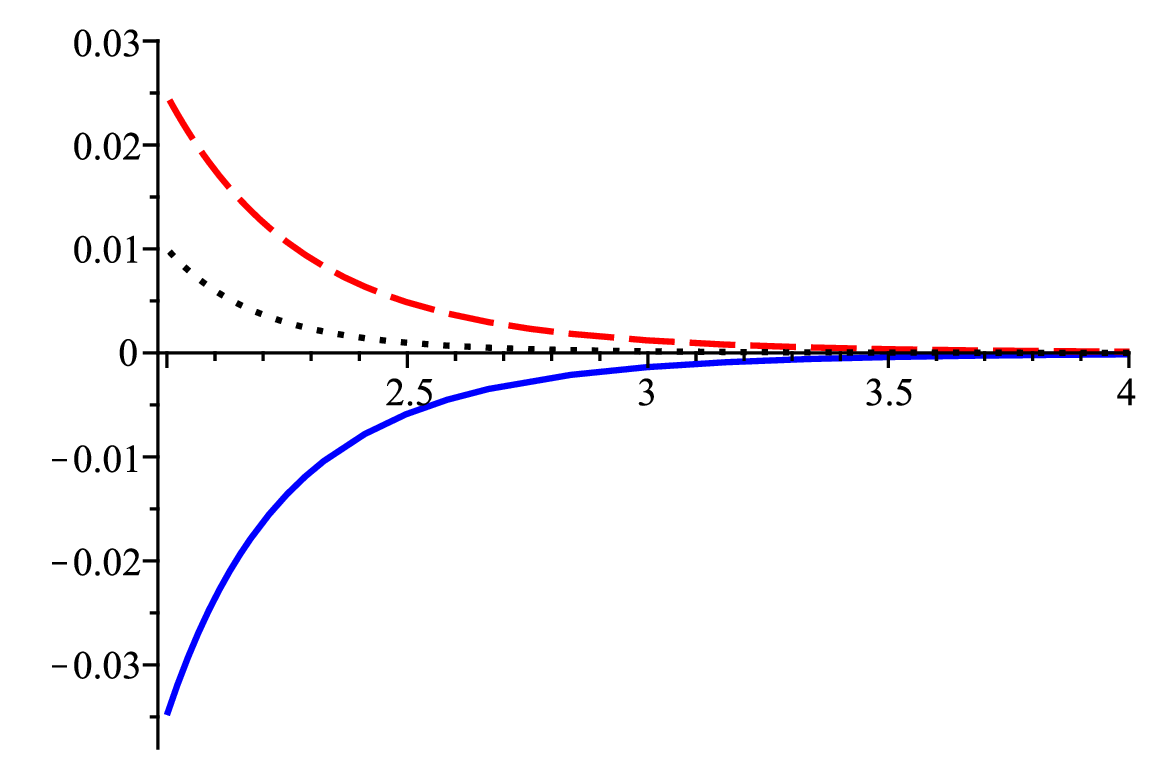}
 		\caption {The behavior of  $F_h$ (solid curve), $F_a$ (dashed curve) and $F_g$ (dotted curve) versus $r$. The model parameters have been set as $m=7$, $r_0=2$, $\lambda=-1$ (left panel) and $\lambda=1$ (right panel).}\label{figw7}
 	\end{center}
 \end{figure}

In the present section we examine the stability of the obtained wormhole solutions by employing the equilibrium condition. In the context of GR, this condition can be deduced using the famous Tolman-Oppenheimer-Volkov (TOV) equation~\cite{rakp,rakp1,rakp2,rakp3,rakp4},\cite{armen,armen1}. For an isotropic EMT fluid the TOV equation is given as
 \begin{align}
 -{\frac{dp_r}{dr}} +2\frac{({\it P_t}-P_r)}{r}-\frac{\phi^\prime(r)}{2}(\rho+P_r)=0.
 \end{align}
Given the above equation one can determine the equilibrium state of a wormhole configuration by taking the gravitational ($F_g$), hydrostatic ($F_h$) as well as the anisotropic ($F_a$) forces (arising due to anisotropy of matter) into account. These forces are defined through the following relations
\begin{eqnarray}
F_g=-\frac{\phi^\prime(r)}{2}(\rho+P_r),~~~~~~~~~~F_h=-{\frac{dP_r}{dr}},~~~~~~~~~F_a=\frac{2}{r}(P_t-P_r).\label{3forces}
\end{eqnarray}
In terms of the above forces the TOV equation can be rewritten as
\begin{align}
F_g+F_h+F_a=0.
\end{align}
Now, using equations (\ref{feild1})-(\ref{feild3}) and considering $P_{r}=w\rho$, we get the corresponding relations for the forces as
\bea
F_g&=&\f{1}{16\pi g(r)r^3}\left[g^\prime(r)r+g(r)-1\right]\left(w+1 \right)\left[w g^\prime(r)r+ \left(g(r)-1\right) \left(w+1\right) 			\right],\\\label{fgs}
F_h&=&\f{w}{8\pi r^3}\left[g^{\prime\prime}(r)r^2-2g(r)+2\right],\\\label{fhs}
F_a&=&\f{1}{16\pi r^3}\Bigg\{r^2g(r)w\left(w+1\right)\left[\f{g^{\prime}(r)}{g(r)}\right]^2-2wr^2g^{\prime\prime}(r)+r\left(1+2w\right)\left(w+1\right)\left(g(r)-1\right)\f{g^\prime(r)}{g(r)}\nn&+& \left[\left(w^{2}+6w+1\right)g(r)-\left(w+1\right)^{2}\right]\left(\f{g(r)-1}{g(r)}\right)\Bigg\}.\label{fas}
\eea
Substituting for the shape function (\ref{shp1}) we finally get
\bea
F_g&=&\frac{\Sigma_3}{16\pi r_0^mr^{m+1}}\left[(m-3)(r_0-r)-\lambda\left(r^{3-m}-r_0^{3-m}\right)\right]^{-1},~~~~~F_h=-\frac{mr_0^{m-2}}{8\pi {r}^{m+1}},\nn
F_a&=&\frac{\Sigma_4}{8\pi r_0^mr^{m+1}}\left[(m-3)(r_0-r)-\lambda\left(r^{3-m}-r_0^{3-m}\right)\right]^{-1},\label{fa1}
\eea
where
\bea
 \Sigma_3&=& \left(r_0^{m}{r}^{3-m}-r_0^{3}\right){\lambda}^{2} + \left[\left(m-4\right)r_0^{2m-2}{r}^{3-m}-\left(m-4\right)r_0^{1+m}\right]\lambda \nn
 &+&\left(3-m\right)r_0^{3m-4}{r}^{3-m}+\left(m-3\right)r_0^{2m-1},\nn
 \Sigma_4&=&m \left[r\left(\lambda{r}^{2-m}+m-3 \right)r_0^{2m-2}+\left(3-m\right)r_0^{2m-1}-r_0^{1+m}\lambda\right]-\frac{\Sigma_3}{2}.
 \eea
Figs.~(\ref{figw}) and (\ref{figw7}) show the graph of gravitational, hydrostatic and anisotropic forces given in Eq.~(\ref{fa1}) for each case. It is therefore seen that these forces cancel the effects of each other leaving thus a stable wormhole configuration. Also, in the case for which $m=3$, we can substitute the shape function (\ref{bl3}) into equations (\ref{feild1})-(\ref{feild3}) to find the EMT components. Then, from Eq.~(\ref{3forces}), a simple calculation gives
\begin{eqnarray}
F_g=0,~~~~~~~~~F_h=\f{1}{8\pi r^4}\left[\lambda-3r_0+3\lambda\ln\left(\f{r_0}{r}\right)\right]=-F_a,
\end{eqnarray}
whence we readily find that for this class of solutions, the gravitational force becomes zero as a result of constant redshift function. Also, the hydrostatic and anisotropic forces are exactly equal and opposite to each other. Thus, the equilibrium of the three forces is achieved due to the combined effect of them, and hence this supports the stability of the wormhole configuration. 
 
\section{Particle trajectories around the wormhole}\label{traject}
\begin{figure}
	\begin{center}
		\includegraphics[scale=0.38]{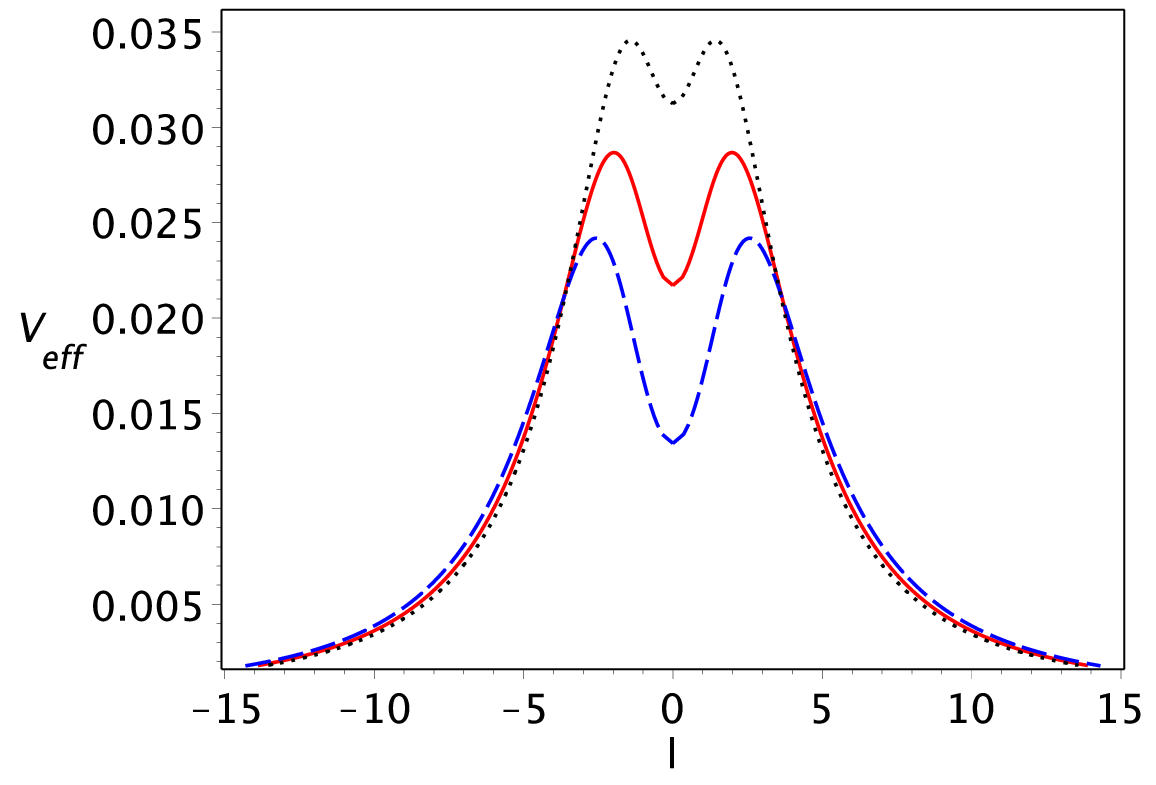}
		\includegraphics[scale=0.4]{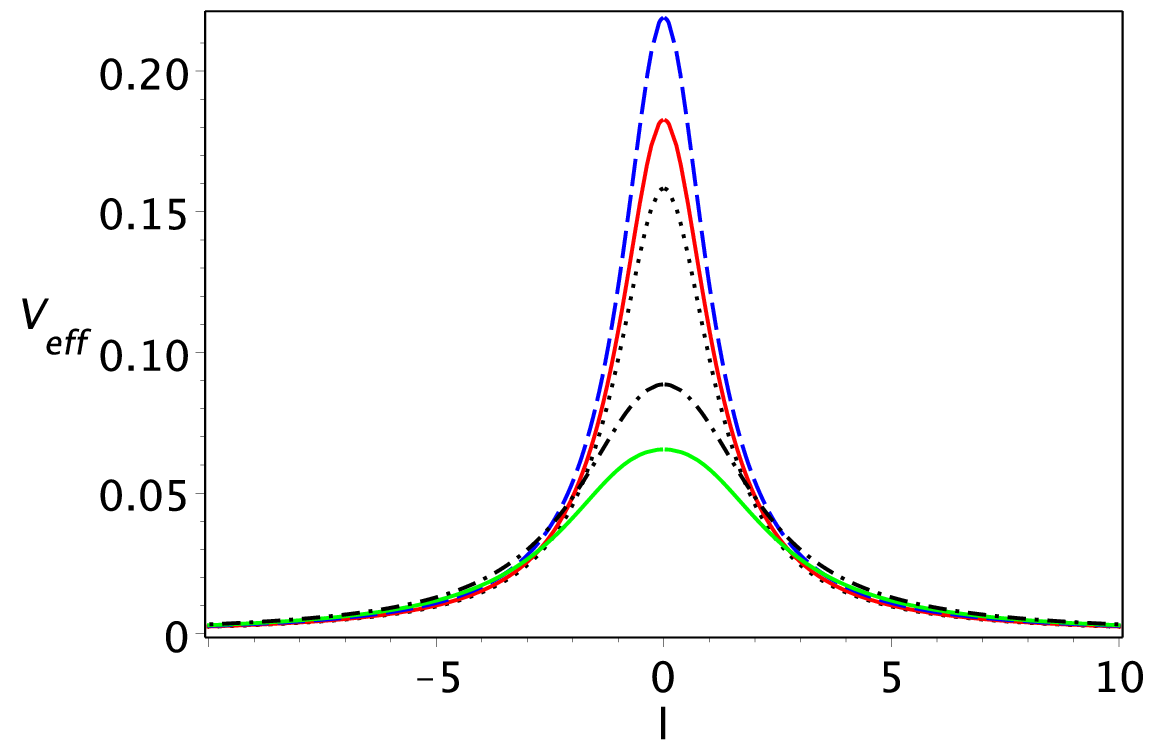}
		\caption{Left panel: The behavior of effective potential for null geodesics against proper radial distance. For the model parameters we have set $\lambda=0.7$ (blue dashed curve), $\lambda=0.6$ (red solid curve) and $\lambda=0.5$ (black dotted curve). Right panel: The behavior of effective potential for $\lambda=-0.7$ (blue dashed curve), $\lambda=-0.6$ (red solid curve), $\lambda=-0.5$ (black dotted curve), $\lambda=0.2$ (green curve) and $\lambda=0.1$ (dot-dashed curve). The throat radius and angular momentum have been taken as $r_0=1$ and $L=0.5$, respectively.}\label{nuleff1}
	\end{center}
\end{figure}

\begin{figure}
	\begin{center}
		\includegraphics[scale=0.348]{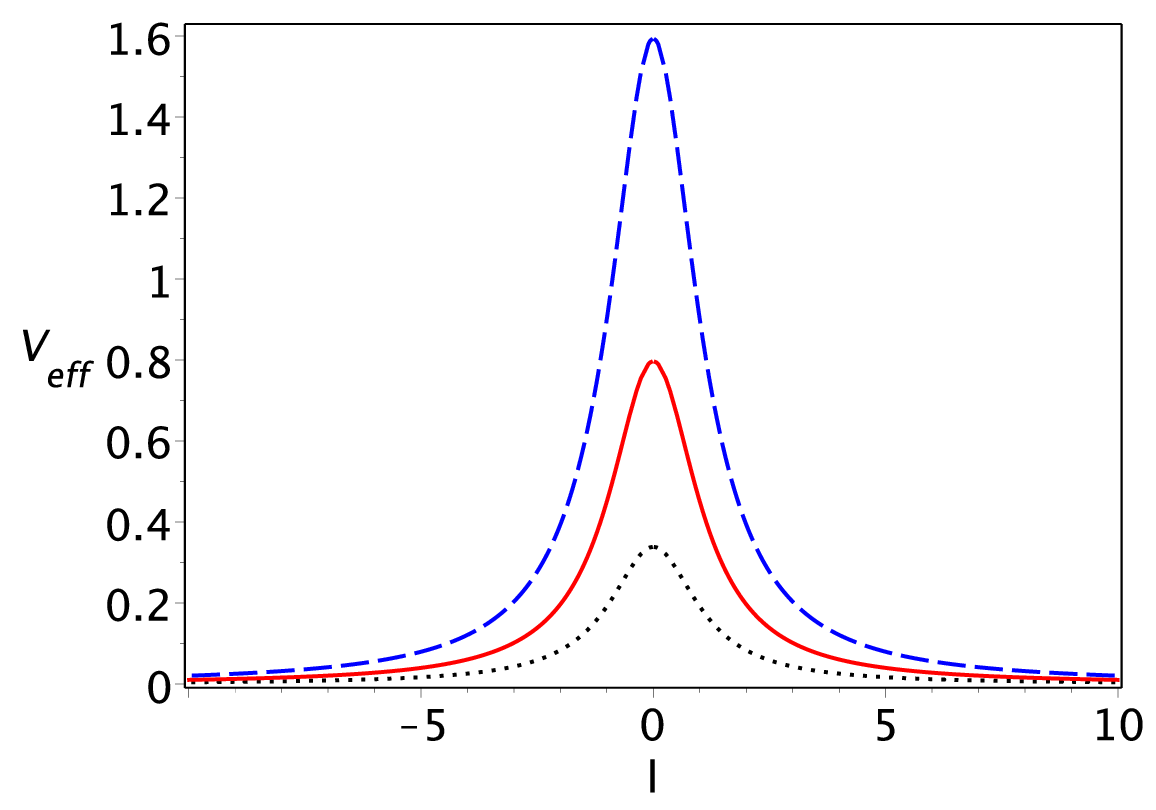}\label{nulleff2}
		\includegraphics[scale=0.348]{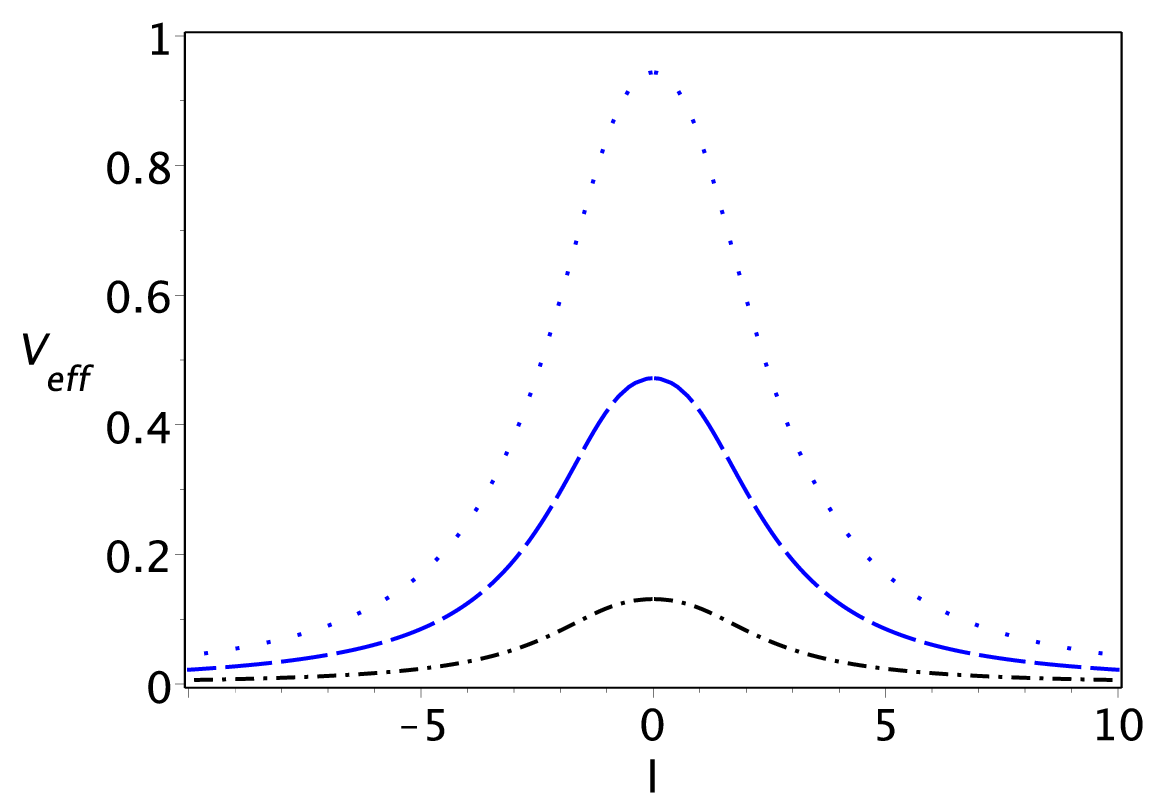}\label{nulleff3}
		\caption{Left panel: The behavior of effective potential for null geodesics against proper radial distance. For the model parameters we have set $r_0=1$, $\lambda=-0.7$, $L=0.5$ (black dotted curve), $L=1$ (red solid curve) and $L=2$ (blue dashed curve). Right panel: The behavior of effective potential for $r_0=1$, $\lambda=0.2$, $L=0.5$ (dot-dashed curve), $L=1$ (blue long-dashed curve) and $L=2$ (blue space-dot curve).}\label{veffw}
	\end{center}
\end{figure}

\begin{figure}
	\begin{center}
		\includegraphics[scale=0.43]{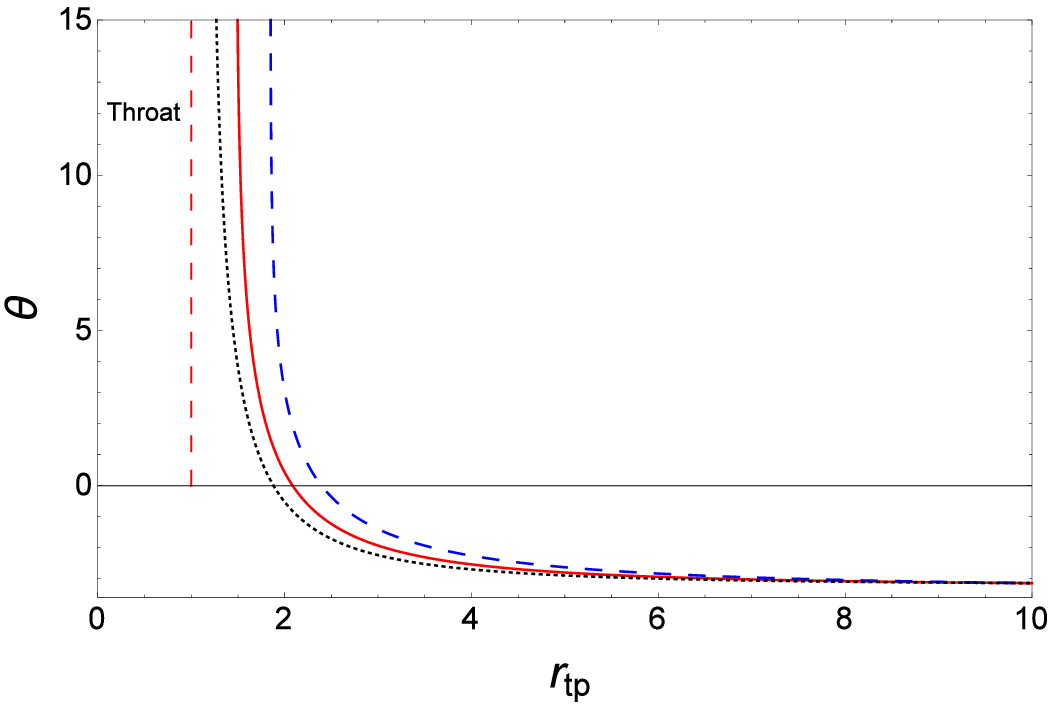}
		\includegraphics[scale=0.43]{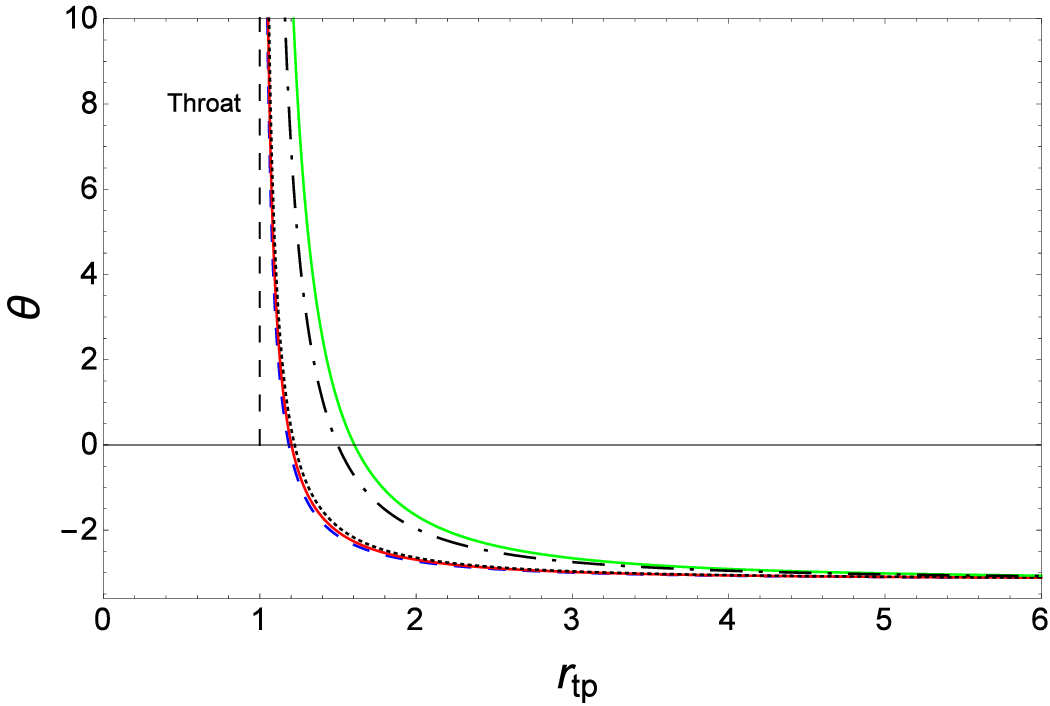}
		\caption{Left panel: The behavior of deflection angle for $r_0=1$, $L=0.5$, $\lambda=0.7$ (blue dashed curve), $\lambda=0.6$ (red solid curve) and $\lambda=0.5$ (black dotted curve). For each value of the parameter $\lambda$, the asymptotes of the curves are found as $r_{\rm tp}=r_2=1.25,1.40,1.55$, respectively. Right panel: The behavior of deflection angle for $r_0=1$, $L=0.5$, $\lambda=-0.7$ (blue dashed curve), $\lambda=-0.6$ (red curve) and $\lambda=-0.5$ (black dotted curve), $\lambda=0.2$ (green curve) and $\lambda=0.1$ (dot-dashed curve).}\label{lensplot}
	\end{center}
\end{figure}

In this section we investigate geodesic equations in wormhole spacetime described by the metric~(\ref{evw}), using the Lagrangian formalism~\cite{lagf}. Due to the spherical symmetry of the wormhole spacetime, without loss of generality we can
restrict our analysis to planar motion in the equatorial plane $\theta=\pi/2$. The corresponding Lagrangian for metric~(\ref{evw}) is then found as
\begin{equation}
	\mathcal{L} = g_{\mu\nu} \dot{x}^\mu \dot{x}^\nu= -f(r)\dot{t}^2+\frac{\dot{r}^2}{g(r)}+r^2\dot{\phi}^2,
	\label{lag}
\end{equation}    
where a dot denotes derivative with respect to the affine parameter $\eta$. The above Lagrangian is constant along a geodesic, hence, we can set $\mathcal{L}(x^{\mu},\dot{x}^\mu)={\epsilon}$ so that time-like and null geodesics correspond to ${\epsilon}=-1$ and ${\epsilon}=0$, respectively. Using the Euler-Lagrange equation
\begin{equation}\label{lag2}
	\frac{d}{d\eta} \frac{\partial{\cal	L}}{\partial\dot{x}^{\mu}}-\frac{\partial{\cal L}}{\partial
		x^{\mu}}=0,
\end{equation}
one can easily find two constants of motion given as
\begin{equation}\label{lag5}
\dot{t}=\frac{E}{f(r)},~~~~~~~~~~~ r^2\dot{\phi}=L,
\end{equation}
where $E$ and $L$ are the energy and angular momentum of the test particle, respectively. Inserting these constants of motion into (\ref{lag}) we obtain
\begin{align}\label{radc}
	\dot{r}^{2}&=g(r)\left[\frac{E^2}{f(r)}-\frac{L^{2}}{r^{2}}+\epsilon\right].
\end{align}
It is convenient to rewrite the above equation in terms of the proper radial distance which is defined as
\begin{equation}\label{mt2}
l(r)=\pm\int_{r_0}^r\frac{dr}{\sqrt{g(r)}}.
\end{equation}
The proper radial distance is finite for all finite values of $r$ throughout the spacetime. We note that the extension of spacetime in terms of proper radial distance is in such a way that, $l$ monotonically increases from $-\infty$ in the lower universe to $l=0$ at the throat and then from zero to $+\infty$ in the upper universe. Using the proper radial distance, Eq.~(\ref{radc}) takes the simple form 
\begin{align}
	\label{eq:geodesics}
	\dot{l}^{2}=f(r)^{-1}[E^{2}-V_{{\rm eff}}(L,l)],
\end{align}
where the effective potential is defined as
\begin{align}
	\label{eq:potential}
	V_{{\rm eff}}(L,l)=f\left(r(l)\right)\left[\frac{L^{2}}{r(l)^{2}}-\epsilon\right].
\end{align}
In what follows, we discuss particle trajectories in the wormhole spacetime, using the above form for the effective potential. Indeed, geodesic equation~(\ref{eq:geodesics}) can be viewed as a classical scattering problem with the potential barrier $V_{{\rm eff}}(L,l)$. Moreover, using Eq. (\ref{lag5}) we can rewrite Eq. (\ref{eq:geodesics}) as an ordinary differential equation for orbital motion 
\begin{equation}\label{lag7}
	\left(\frac{dl}{d\phi}\right)^2=\frac{{\dot{l}}^2}{\dot{{\phi}}^2}=\frac{{r(l)}^4}{f(r)L^2}\left[E^2- V_{{\rm eff}}(L,l)\right].
\end{equation} 
We note that, in traversable wormhole spacetimes, particles can travel through the throat of the wormhole from one asymptotically flat part of the universe to other one. Then, a geodesic can pass through the throat into the other universe if 
\begin{align}
	E^{2}>V_{{\rm eff}}(L,0).
\end{align} 
Similarly, for those geodesics that get reflected back to the same universe by the potential barrier, we have $ E^{2}<V_{{\rm eff}}(L,0)$. In this case, there is a turning point at $l=l_{{\rm tp}}$ which is obtained by solving the following equation 
\begin{align}
	E^{2}=V_{{\rm eff}}(L,l_{{\rm tp}}).
\end{align}
From Eq.~(\ref{eq:potential}), it is easy to verify that
\bea
\frac{dV_{{\rm eff}}}{dl}\!\!\!\!&=&\!\!\!\!\sqrt{g(r)}\bigg[\left(\frac{L^2}{r^2}-\epsilon\right) f^{\prime}(r)-\frac{2L^2 f(r)}{r^3}\bigg],\\
\frac{d^{2}V_{{\rm eff}}}{dl^{2}}\!\!\!\!&=&\!\!\!\!\frac{L^2}{r^4}f(r)\left[6g(r)-rg^{\prime}(r)\right]+\frac{f^{\prime}(r)}{2r^4}\left[\left(L^2-\epsilon r^2\right)r^2g^{\prime}(r)-8L^2 r g(r)\right]+\frac{f^{\prime\prime}(r)}{r^2}\left(L^2-\epsilon r^2\right)g(r).\nn
\eea
A generic feature of this effective potential in the case $f(r)=constant$  is that it possesses a global maximum at the throat
\begin{align}
\frac{dV_{{\rm eff}}}{dl}\Big|_{l=0}=0,\quad\quad \quad \frac{d^{2}V_{{\rm eff}}}{dl^{2}}\Big|_{l=0}=-\frac{L^2f(r_0)g^{\prime}(r_0)}{r_0^3}.
\end{align}
From the second part of the above equation, we find that the flaring out condition leads to $\f{d^2V_{{\rm eff}}}{dl^2}<0$ at the throat. This clearly provides an unstable orbit since it occurs at the maximum of the potential for $E^{2}=V_{{\rm eff}}(L,l_{0})$. We note that these conditions are independent of whether the geodesic is null or timelike.
We now consider the wormhole solutions presented in subsection (\ref{WHS2}) and restrict ourselves to the class of wormholes with $m=4$. Substituting the shape function Eq.~(\ref{b4}) into Eq.~(\ref{mt2}) we find
\begin{align}
\label{eq:embedding}
l(r)=\frac{\pm 1}{2 r_0}\left[(r_0^{2}+\lambda)\ln\left(\frac{-r_0^{2}-\lambda+2r{r_0}+2\sqrt{r_0\left(r-r_0\right) \left(rr_0-\lambda \right)}}{r_0^{2}-\lambda}\right)+2\sqrt{r_0\left(r-r_0\right)\left(rr_0-\lambda\right)}\right].
\end{align}
We can substitute the redshift function (\ref{red4}) into equation (\ref{eq:potential}) to get the effective potential as
\begin{align}
V_{{\rm eff}}(L,l)=\left(1-\frac{\lambda}{r_0 r} \right)^{\frac{r_0^{2}+\lambda}{\lambda}}\left[\frac{L^{2}}{r(l)^{2}}-\epsilon\right].
\end{align}
Also, we calculate the derivatives of the above potential as
\bea
V^{\prime}_{\rm eff}(L,l(r))&=&\frac{\left({r_0}-\lambda \right)^{{\frac{r_0^2}{\lambda}}}\sqrt{{\frac {\left(r-r_0\right)\left( {r_0}r-\lambda\right)}{{r_0}r}}}}{{r}^{5}{{r_0}}^{{\frac {r_0^2}{\lambda}}}{r}^{{\frac {r_0^2}{\lambda}}}{r_0}
}\left[r^2\epsilon\left(r_0^2+\lambda \right)+2{L}^{2}{r_0}r-\left(r_0^2+3\lambda \right)L^2\right],\label{d1veff}\\
V^{\prime \prime}_{\rm eff}(L,l(r))&=& \frac{(rr_0-\lambda)^{\f{r_0^2}{\lambda}}}{r^{\f{7\lambda+r_0^2}{\lambda}}r_0^{\f{2\lambda+r_0^2}{\lambda}}}\Big[L^2 r_0^2 \left(3 r-2 r_0\right) \left(4 r^2-6 r_0 r+r_0^2\right)-2 \lambda  L^2 r_0 \left(19 r^2-29 r_0 r+8 r_0^2\right)\nn&+&3 \lambda ^2 L^2 \left(9 r-10 r_0\right)+r^2 \epsilon  \left(\lambda +r_0^2\right) \left(r_0 \left(6 \lambda +4 r^2-7 r_0 r+2 r_0^2\right)-5 \lambda  r\right)\Big].\label{d2veff}
\eea
Next, we proceed to study null geodesics ($\epsilon=0$) for the class of wormhole solutions with $m=4$ and nonzero redshift function. From Eq.~(\ref{d1veff}) we can find two roots that satisfy equation ${V^{\prime}_{\rm eff}}=0$, these roots are given by $r_{1}=r_0$ and $r_{2}=\frac{r_0}{2}+\frac{3 \lambda}{2 r_0}$. The condition $r_{2}>r_{0}$ leads to the inequality $3\lambda>{r_0}^2$, that  by using the second derivative the effective potential (\ref{d2veff}) for positive $\lambda$ we have  $V^{\prime \prime}_{\rm eff}|_{r=r_2}<0$, i.e., a local maximum. For this case we have a photon sphere located outside the throat, see the left panel in Fig.~(\ref{nuleff1}) where we have sketched the behavior of effective potential as a function of proper radial distance for different values of $\lambda$ parameter. We further observe that the effective potential admits a local minimum at the throat, i.e., $V^{\prime \prime}_{\rm eff}(L,0)>0$. In this case the wormhole throat acts as an anti-photon sphere~\cite{taporbrata2019,Shaikh2019}. For $\lambda<0$ or $0<\lambda<r_0^2/3$ the effective potential assumes a maximum value at the throat, i.e. ${V^{\prime \prime}_{\rm eff}}|_{r=r_0}<0$, hence, the throat acts as a photon sphere, see the right panel in Fig.~(\ref{nuleff1}). Fig.~(\ref{veffw}) shows the changes in effective potential with respect to angular momentum where we observe that the height of potential barrier increases with increasing the angular momentum. Now, in order to discuss the photon orbits we may eliminate $d\eta$ between the second part of Eq.~(\ref{lag5}) and Eq.~(\ref{radc}) to find 
\begin{align}\label{defdifangle}
\left(\f{dr}{d\phi}\right)^2=\left[1-\f{b(r)}{r}\right]\left[\f{r^4}{\mu^2}{\rm e}^{-2\phi(r)}-r^2\right],
\end{align}
where $\mu=L/E$ is the impact parameter and $\left(dr/d\phi\right)|_{r=r_{\rm tp}}=0$ hence $\mu=r_{\rm tp}{\rm e}^{-\phi(r_{\rm tp})}$. For a photon that comes from the polar coordinate $\lim_{r\rightarrow\infty}(r,-\pi/2-\theta/2)$ and passes through the turning point located at $(r_{tp}, 0)$ before reaching the point $\lim_{r\rightarrow\infty}(r,\pi/2+\theta/2)$ one can define the deflection angle $\theta(r_{tp})$ as~\cite{Weinberg,defangle}
\begin{align}\label{defdifangleint}
\theta(r_{\rm tp})=-\pi+2\int_{r_{\rm tp}}^{\infty}\f{dr}{r\left[\left(1-\f{b(r)}{r}\right)\left(\f{r^2}{\mu^2}{\rm e}^{-2\phi(r)}-1\right)\right]^{\f{1}{2}}}.
\end{align}
It is possible that the light rays get trapped in a sphere of constant radius and consequently may not reach asymptotically $\lim_{r\rightarrow\infty}(r,\pi/2+\theta/2)$. In such a scenario the above integral diverges and such a sphere is called photon sphere. As mentioned before, the location of photon sphere can be determined through the behavior of effective potential. The left panel in Fig.~(\ref{lensplot}) shows the behavior of deflection angle as a function of turning point for $\lambda>0$. As we observe from the left panel of Fig.~(\ref{nuleff1}), the maxima of the effective potential represent the location of photon spheres for each value of $\lambda$ parameter. These points correspond to the asymptotic values of $r_{\rm tp}=r_2>r_0$ at which the integral (\ref{defdifangleint}) diverges. Hence, for each value of the parameter $\lambda>r_0^2/3$, we have a photon sphere outside the throat. In the right panel of Fig.~(\ref{lensplot}) we have sketched the behavior of deflection angle for $\lambda<0$ and $\lambda<r_0^2/3$. These value of the parameter $\lambda$ determine the behavior of effective potential as shown in the right panel of Fig.~(\ref{nuleff1}), where we observe that the effective potential assumes a maximum at the throat, i.e., $l(r)|_{r_0=1}=0$. This maximum indicates existence of a photon sphere at the throat where, the deflection angle diverges in the limit $r_{\rm tp}\rightarrow r_0$.
\par
\begin{figure}
	\begin{center}
		\includegraphics[scale=0.371]{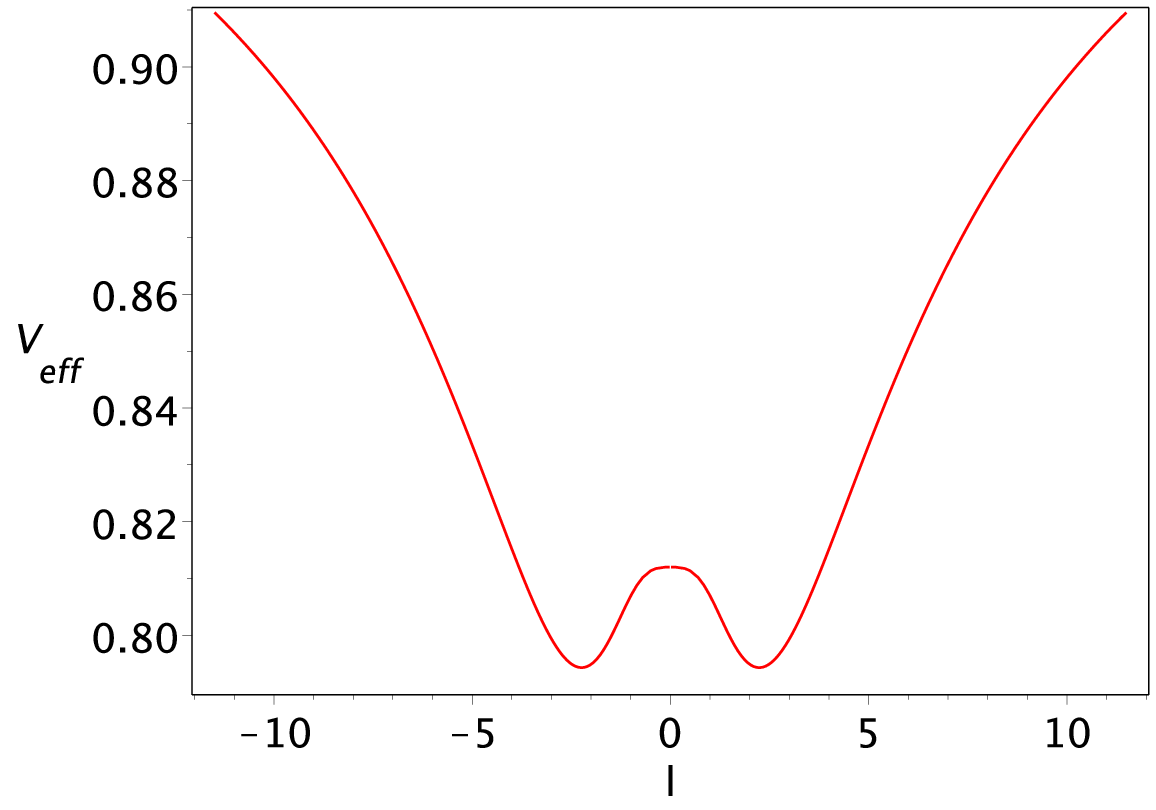}\label{timeff4}
		\includegraphics[scale=0.368]{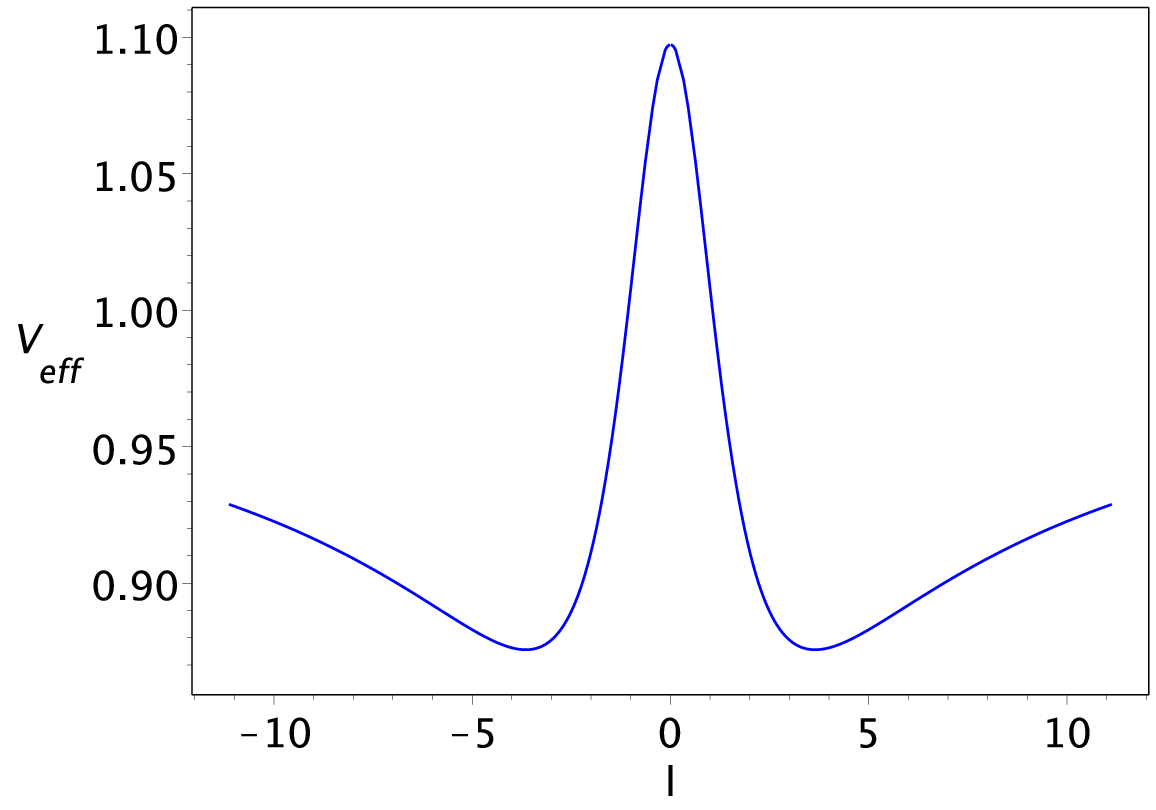}\label{timeeff5}
		\caption{Behavior of Effective potential for timelike geodesics versus proper radial distance. In the left panel we have set $\lambda=0.1$ and in the right panel $\lambda=-0.1$. In these plots we have considered angular momentum and throat radius as $L=1.26$ and $r_0=1$.}\label{vefftime11}
	\end{center}
\end{figure}
\begin{figure}
	\begin{center}
		\includegraphics[scale=0.368]{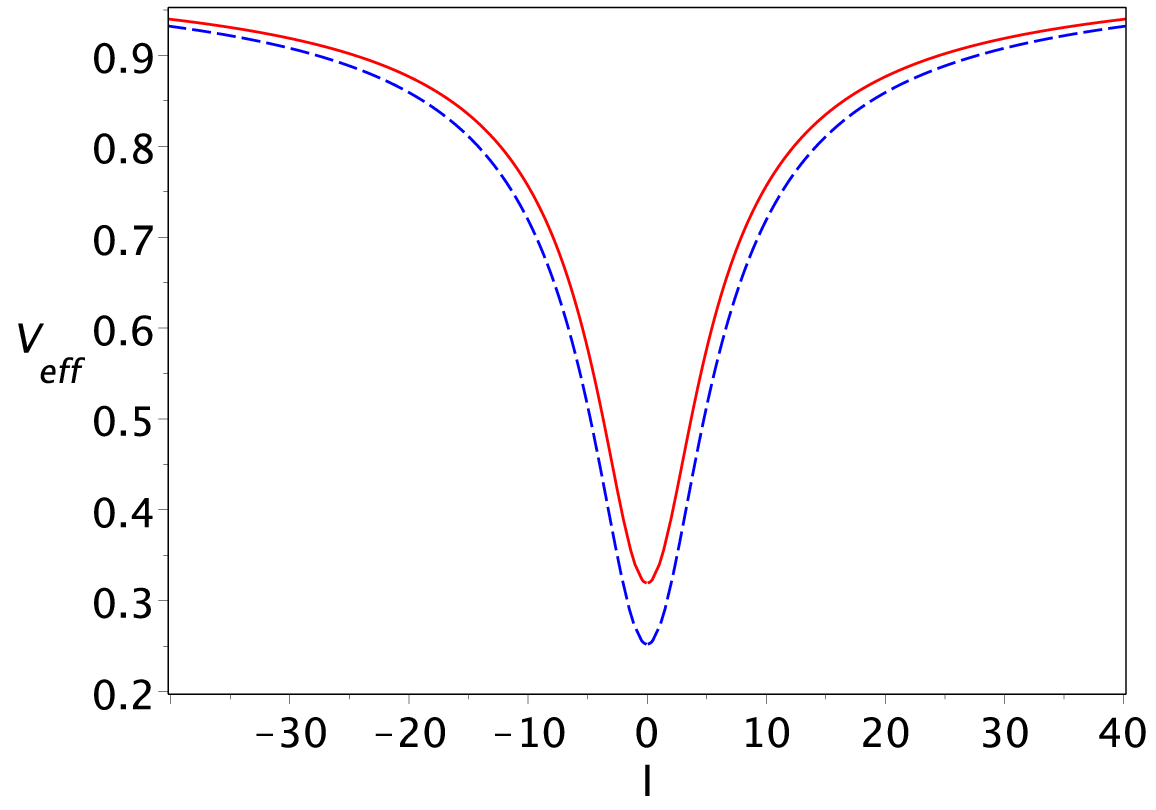}\label{timeff6}
		\includegraphics[scale=0.368]{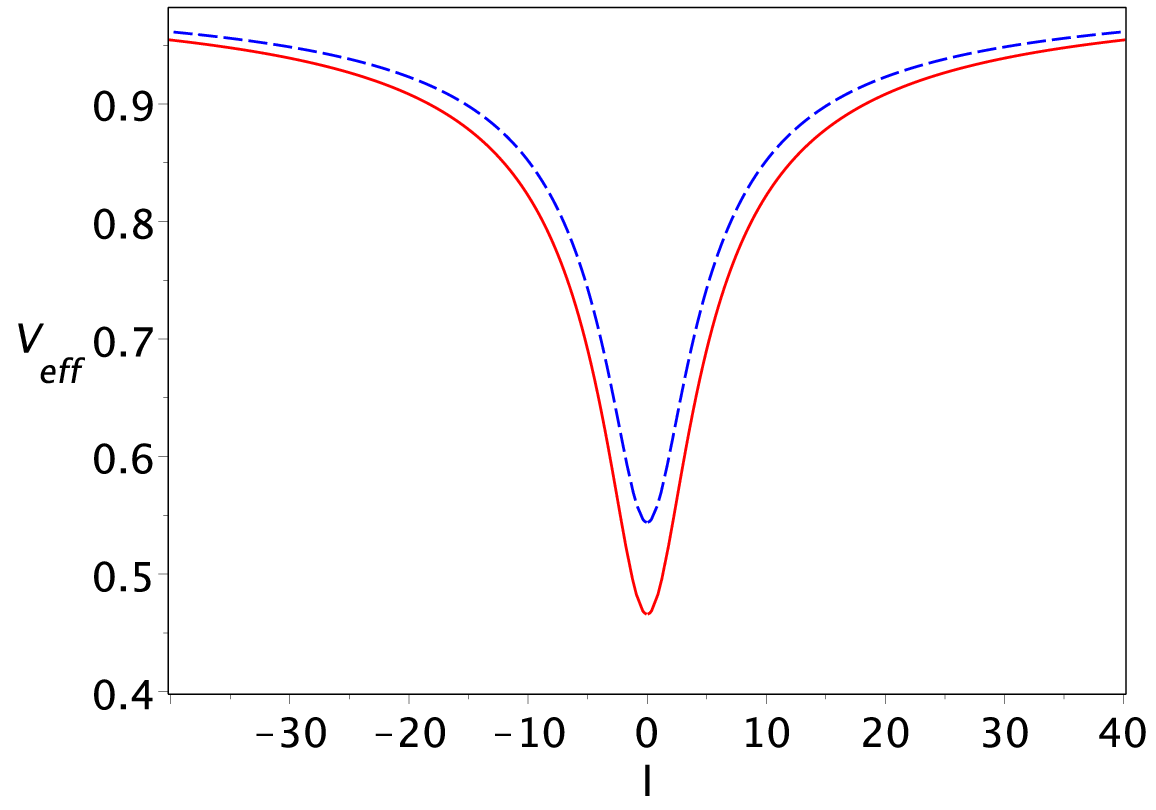}\label{timeeff7}
		\caption{The behavior of effective potential for timelike geodesics. In the left panel we have set $\lambda=1$ (blue curve) and $\lambda=0.5$ (red curve) and in the right panel $\lambda=-1$ (blue curve) and $\lambda=-0.5$ (red curve). In these plots the angular momentum and throat radius have been taken as $L=0.5$ and $r_0=2$.}\label{vefftime12}
	\end{center}
\end{figure}
\begin{figure}
	\begin{center}
		\includegraphics[scale=0.491]{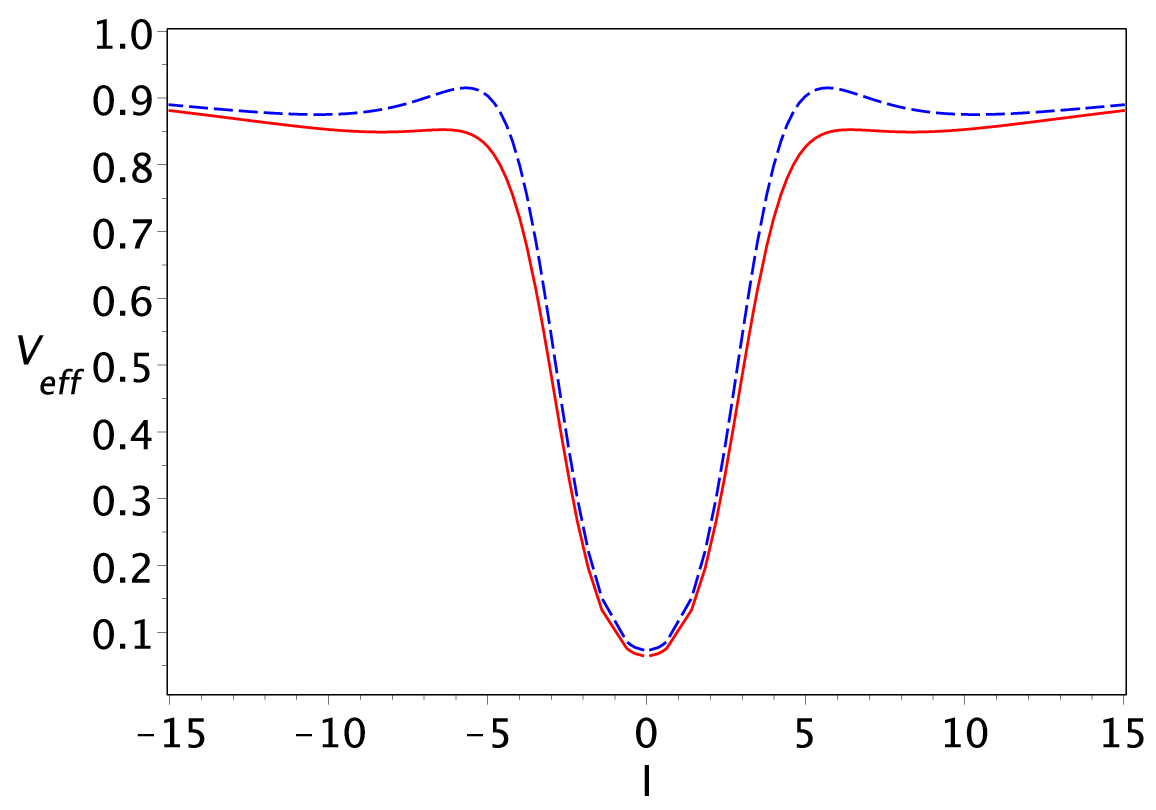}
		\caption{The behavior of effective potential for timelike geodesics. In this diagram we have set $r_0=1$, $\lambda=0.9$, $L=2.9$ (blue curve) and $L=2.7$ (red curve).}\label{timeeff13}
	\end{center}
\end{figure}

In the case of timelike geodesics, we can use Eq.~(\ref{d1veff}) with $\epsilon=-1$ to get the first derivative of effective potential as
\begin{align}
V^{\prime}_{\rm eff}(L,r(l))=\frac{\sqrt{\left(r-r_0\right)\left(r_0r-\lambda\right)}\left(r_0r-\lambda \right)^{{\frac{r_0^2}{\lambda}}}}{r^5r_0^{{\frac{r_0^2}{\lambda}}}{r}^{{\frac{r_0^2}{\lambda}}}{r_0^{\frac{3}{2}}}}\left[\left( {r}^{2}+{L}^{2} \right)r_0^2-2{L}^{2}{r_0}r+3\lambda{L}^{2}+\lambda{r}^{2}
\right],
\end{align}
whereby we find the following three roots for equation $V^{\prime}_{\rm eff}(L,r_{c})=0$ as  
\begin{align}
r_{\rm c}=r_0,~~r_{c,+}=\frac{r_0L^2+L\sqrt {r_0^2L^2-r_0^4-4\lambda r_0^2-3\lambda^2}}{r_0^2+\lambda},~~~~ r_{c,-}=\frac{r_0L^2-L\sqrt{ r_0^2L^2-r_0^4-4\lambda r_0^2-3\lambda^2}}{r_0^2+\lambda}.
\end{align}
With the help of Eq.~(\ref{d2veff}), we can calculate the second derivative of effective potential at the wormhole throat as  
\begin{align}
V^{\prime \prime}_{\rm eff}(L,0)=\frac{\left(r_0^{2}-\lambda \right)^{{\frac{r_0^{2}}{\lambda}}+1}}{2r_0^{{\frac{8\lambda+2r_0^2}{\lambda}}}}\left[r_0^{4}+ \left( \lambda-{L}^{2} \right) r_0^{2}+3{L}^{2}\lambda\right],
\end{align}
whence we find that there exists a critical value for the angular momentum at which $V^{\prime \prime}_{\rm eff}(L,0)$ changes its sign
\begin{align}\label{angucrit}
L_{\rm cr}=\sqrt{\frac{r_0^2(r_0^2+\lambda)}{(r_0^2-3\lambda)}}.
\end{align}
For $L>L_{\rm cr}$ ($L<L_{\rm cr}$) the effective potential admits a local maximum (local minimum) at wormhole throat. Then, unstable circular orbits can occur due to the existence of a maximum for the effective potential and bound orbits are possible when the effective potential gets a minimum at  the throat. In order that the roots ($r_{c,\pm}$) assume real values the angular momentum must obey the condition $L>L_{p}$ where $L_{p}$ is given by
\begin{align}\label{angup}
L_{p}=\frac{\sqrt{r_0^{4}+4\lambda r_0^2+3\lambda^2}}{r_0}.
\end{align}
To check the stability of orbits, we may solve equation $V^{\prime}_{\rm eff}(L,r_{c})=0$ for the square of angular momentum and substitute the result into Eq.~(\ref{d2veff}) to obtain
\begin{align}\label{Vdiff2c}
V^{\prime\prime}_{\rm eff}(\lambda,r_c)=\frac{2\left(r_0r_c-\lambda\right)^{{\frac{r_0^2}{\lambda}}}\left(r_0^2+\lambda \right)\left( r_0-r_c\right)\left(r_0^2-r_0r_c+3\lambda\right)\left(r_0r_c-\lambda \right)}{r_c^5\left(r_0^2-2r_0r_c+3\lambda\right)r_0^{{\frac{r_0^2}{\lambda}}}r_c^{{\frac {r_0^2}{\lambda}}}{r_0}^{2}}.
\end{align}
For this class of wormhole solutions, we can obtain the behavior of effective potential as a function of the proper distance for timelike geodesics. Generally, for circular orbits we require that $r=constant$ and so $\dot{r}=\ddot{r}=0$. In this situation, the only possible position of the particle will be a circle for which the conserved total energy of the particle is equal to the extremum of the effective potential. More precisely, if the energy corresponds to a maximum or a saddle point of the effective potential, then the particle moves on an unstable orbit. If the energy corresponds to a minimum of the effective potential then the trajectory of the particle will be a stable orbit. In Fig.~(\ref{vefftime11}) we have plotted the behavior of effective potential against proper radial distance for $L>L_p$. It is therefore seen that for $r_0=1$ and $\lambda=\pm0.1$ we have $r_{c,\pm}\in\mathbb{R}^+$ but only one of the roots is larger than $r_0$. Using then Eq.~(\ref{Vdiff2c}) one can recognize that $V^{\prime\prime}_{\rm eff}(\lambda,r_c)>0$ which corresponds to a minimum of the effective potential. In Fig.~(\ref{vefftime12}) we choose the values of parameters $r_0$ and $\lambda$ in such a way that $L<L_p$, therefore the two roots $r_{c,\pm}$ are no longer real. For this case there exists only a local minimum at wormhole throat for both $\lambda>0$ and $\lambda<0$. Finally, in Fig.~(\ref{timeeff13}) we taken the values of $r_0$ and $\lambda$ so that $L>L_p$. Then, there exist two real roots with values greater than the throat radius. One of these roots corresponds to the maximum ($V^{\prime\prime}_{\rm eff}(\lambda,r_c)<0$) and the other one corresponds to the minimum ($V^{\prime\prime}_{\rm eff}(\lambda,r_c)>0$) of the effective potential. We also note that for $L<L_{p}$, we have only a local maximum or a local minimum at the throat. The former occurs for $L>L_{\rm cr}$ and the latter for $L<L_{\rm cr}$.

\section{Quasi-normal modes}\label{QNMO}
\begin{figure}
	\begin{center}
		\includegraphics[scale=0.37]{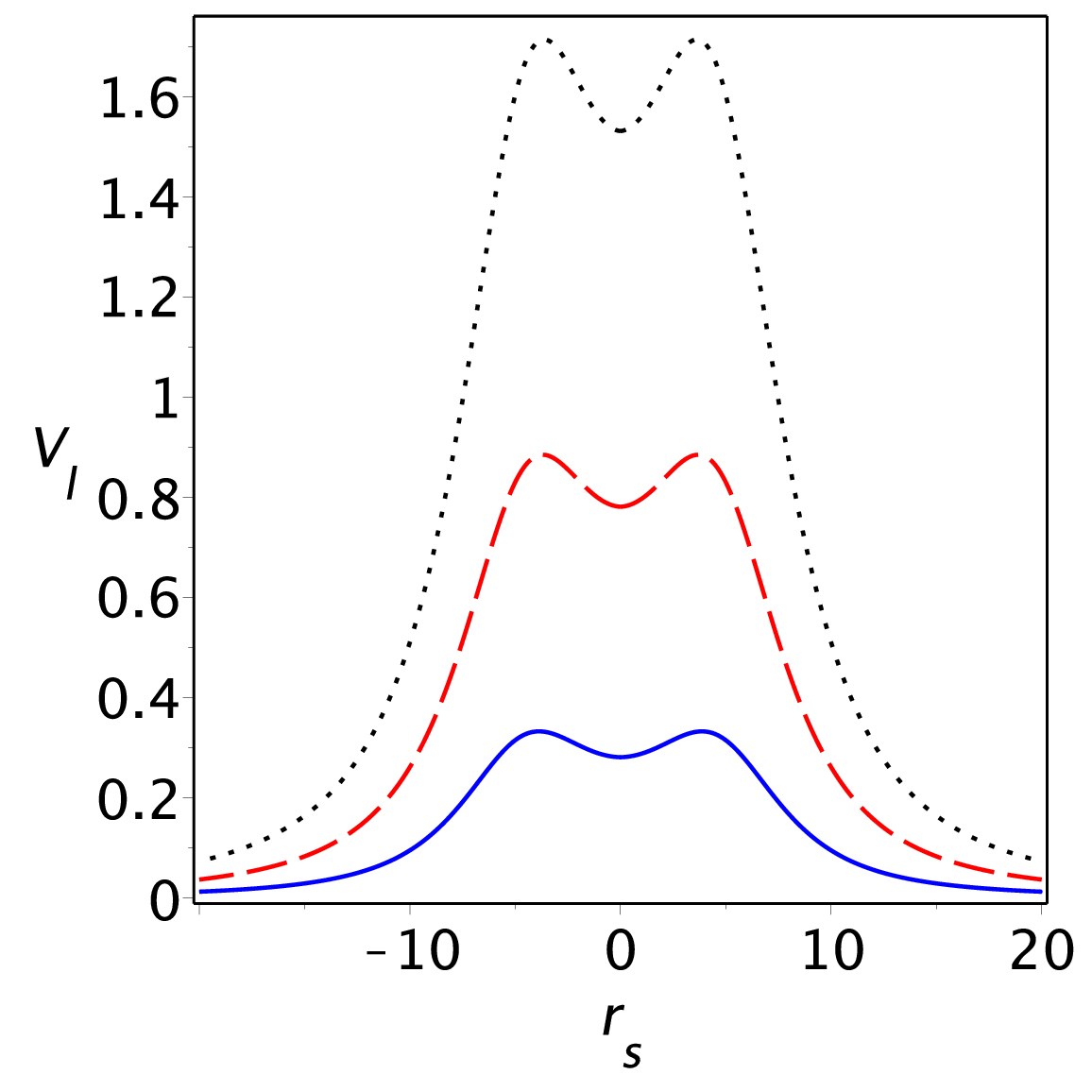}
		\includegraphics[scale=0.37]{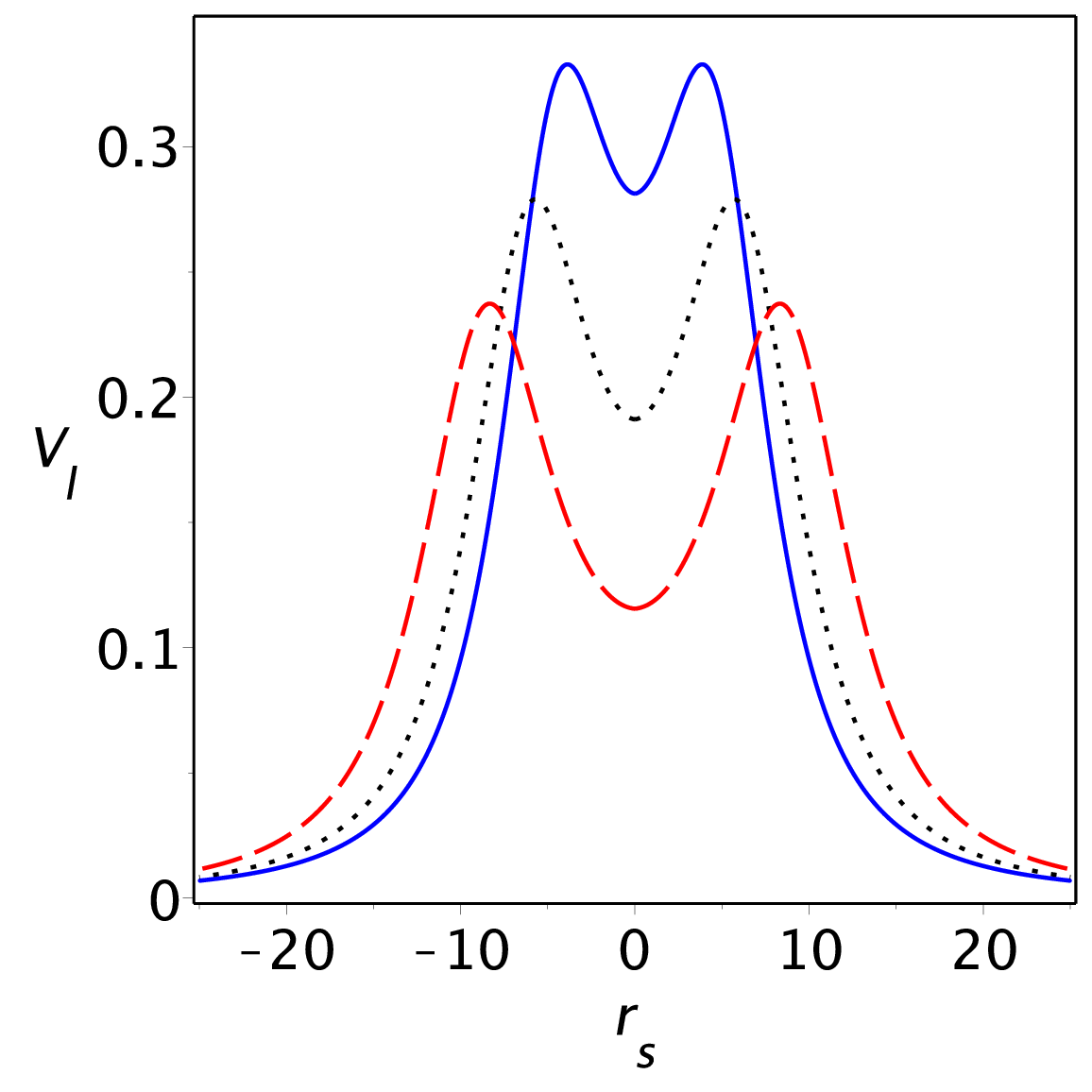}
		\caption{Left panel: The behavior of effective potential against tortoise coordinate for $m=4$, $r_0=1$, $\lambda=0.5$, $\ell=1$ (blue curve), $\ell=2$ (red curve) and $\ell=3$ (dotted curve). Right panel: The behavior of effective potential against tortoise coordinate for $m=4$, $r_0=1$, $\ell=1$, $\lambda=0.7$ (red curve), $\lambda=0.6$ (black curve) and $\lambda=0.5$ (blue curve).}\label{VSL}
	\end{center}
\end{figure}

\begin{figure}
	\begin{center}
		\includegraphics[scale=0.37]{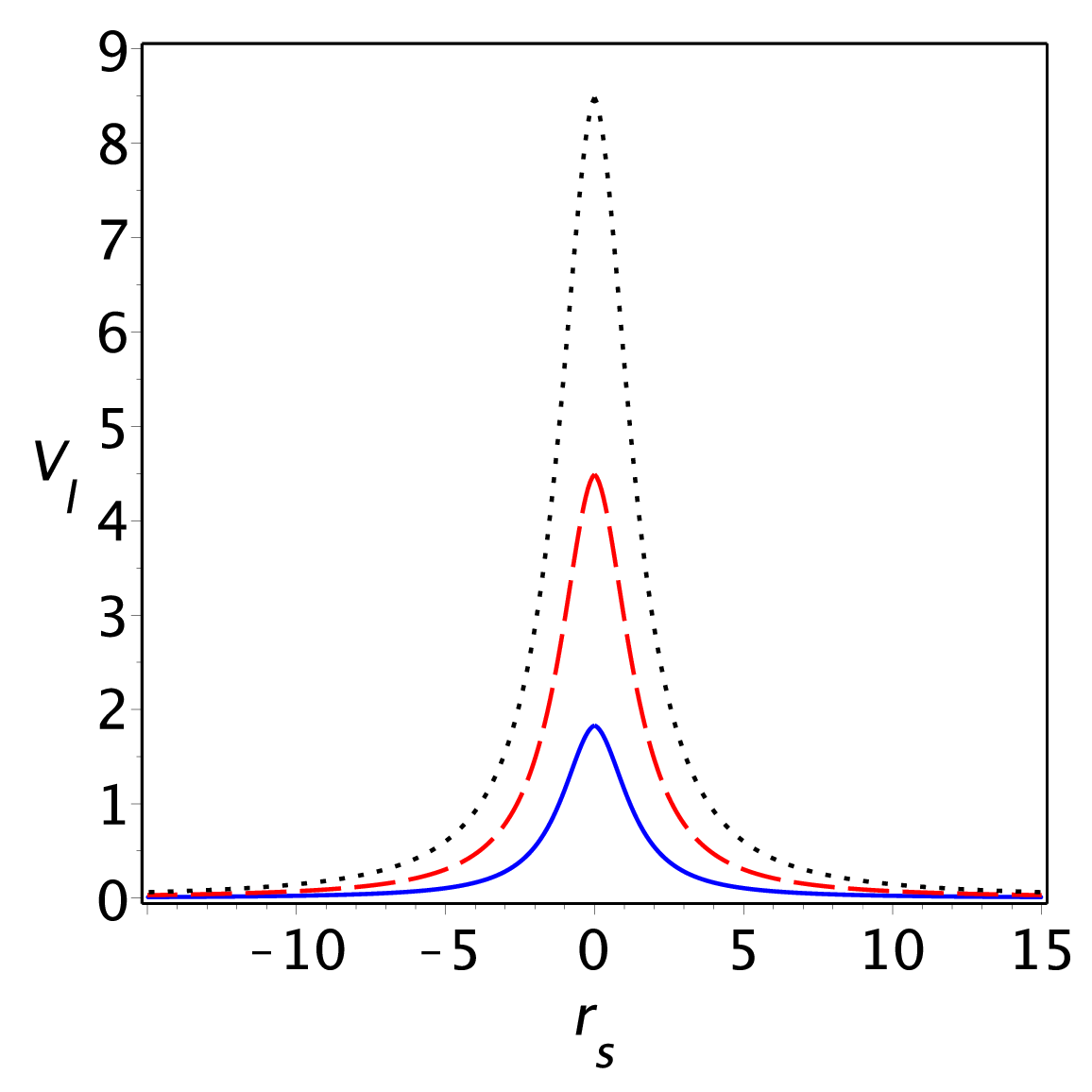}
		\includegraphics[scale=0.369]{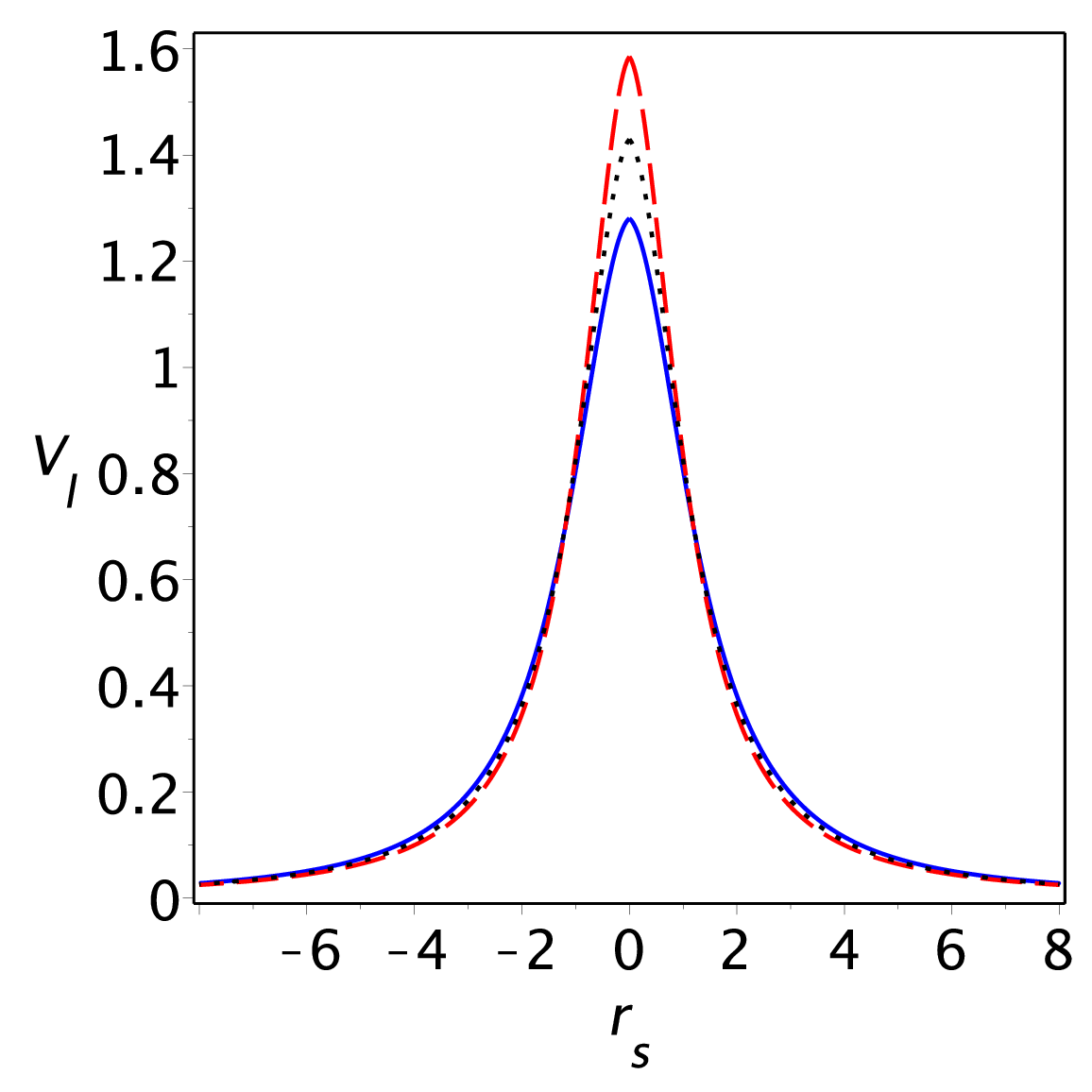}
		\caption{Left panel: The behavior of effective potential against tortoise coordinate for $m=4$, $r_0=1$, $\lambda=-0.5$, $\ell=1$ (blue curve), $\ell=2$ (red curve) and $\ell=3$ (dotted curve). Right panel: The behavior of effective potential against tortoise coordinate for $m=4$, $r_0=1$, $\ell=1$, $\lambda=-0.7$ (red curve), $\lambda=-0.6$ (dotted curve) and $\lambda=-0.5$ (blue curve).}\label{VSLAM}
	\end{center}
\end{figure}

{In the present section we would like study the spectrum of QNMs for a scalar field as another example of observable quantity within the spacetimes of traversable wormholes. These modes are characteristic frequencies of compact objects such as black holes or wormholes which are independent of the initial conditions of perturbations and are fully determined by the parameters of the object~\cite{qnm1,qnm2}, see e.g.~\cite{qnm3,qnm4,qnm5,qnm6,qnm7,qnm8} for calculation of QNMs of various wormhole configurations and also~\cite{qnm000,qnm110,qnm120,qnm130,qnm140,qnm150} for the case of black holes. The final stages of merger of two exotic compact objects are dominated by superposition of quasinormal modes that the spectrum of which provides useful information about the spacetime geometry near the photon orbits~\cite{qnm1}. The boundary conditions and reflection of modes from the maxima of potential barriers determine the spectrum of these modes. The initial ring-down signals in wormholes are caused by oscillatory modes produced by photon spheres. At the location potential maxima there exist unstable orbits (due to the positive Lyapunov exponent) from which massless particles will deviate with minimal perturbation producing thus rapid damping QNMs. Such initial modes are very similar to the oscillatory modes of black holes whose potential is close to that of a wormhole~\cite{qnm9}. At late times, the spectrum of modes is corrected by the presence of trapped modes and this leads to echos in the propagated modes. The time delay between two echos is almost equal to the travel time of light between two symmetric potential barriers relative to the wormhole throat. Such resonant modes are characteristics of wormhole spectra which, in the Eikonal limit, are determined by physical properties of photon spheres~\cite{qnm10}. In the following, we try to briefly review the results of theoretical research related to understanding of different spectra caused by wormholes. Let us now consider the master equation for scalar perturbations on a curved spacetime. The general form for perturbation equation for a massless scalar field $\psi$ is given by
\begin{eqnarray}
\frac{1}{\sqrt{-g}}\partial_\mu(\sqrt{-g}g^{\mu\nu}\partial_\nu)\psi=0.\label{eq1}
\end{eqnarray}
Introducing the solution $\psi=e^{-i\omega t}Y_{\ell,m}(\theta,\varphi)u_\ell(r)/r$ with $\omega$ being the unknown frequency (to be determined after imposing the concrete boundary conditions), and $Y_{\ell,m}(\theta,\varphi)$ denotes the spherical harmonics with $\ell=0, 1, 2, 3 . . .$ is the multipole number which arises from the separation of angular variables by expansion into spherical harmonics. Using the metric (\ref{evw}) and after separation of variables the perturbation equation takes the following wave-like form for the radial function $u_\ell(r)$}
{\begin{eqnarray}
\frac{d^2 u_{\ell}(r_s)}{dr_s^2}+\left[\omega^2-V_\ell(r_s)\right] u_{\ell}(r_s)=0,\label{e3}
\end{eqnarray}
where the tortoise coordinate $r_s$ is defined by the relation 
\begin{eqnarray}
r_s(r)=\int_{r_0}^r \frac{e^{-\phi(r^\prime)}dr'}{\sqrt{1-b(r')/r'}}.
\end{eqnarray}
We Note that the tortoise coordinate $r_s$ is defined in the interval $(-\infty, \infty)$ in such way that the spatial infinity at both sides of the wormhole corresponds to $r_s=\pm\infty$ and the wormhole throat is located at $r_s=0$. The effective potential is given by
\begin{eqnarray}\label{veff}
V_\ell(r)= e^{2\phi} \bigg[\frac{\ell(\ell+1)}{r^2}-\frac{r b' -b}{2r^3} +\frac{\phi'}{r} \bigg(1-\frac{b}{r} \bigg) \bigg],\label{vqs}
\end{eqnarray}}
{Therefore, we have transformed the problem into the well-known one-dimensional wave equation with energy $\omega^2$ and an effective potential $V_\ell(r)$. Now, we are interested to find QNMs for wormhole solutions presented in subsection (\ref{WHS2}) and restrict ourselves
to the class of wormholes with $m=4$. Substituting the shape function Eq.~(\ref{b4}) and the redshift function Eq.~(\ref{red4}) into Eq.~(\ref{vqs}) we find}
\begin{eqnarray}\label{veff1}
V_\ell (r)=\frac{\left(r_0r-\lambda\right)^{\frac{r_0^2}{\lambda}+1}\ell\left(\ell+1\right)}{r_0^{\frac{r_0^2}{\lambda}+1}{r}^{\frac{r_0^2}{\lambda}+3}} +\frac{\big(2rr_0^2-r_0^3-3\lambda r_0+2r\lambda\big)\left(r_0r-\lambda\right)^{{\frac{r_0^2}{\lambda}}+1}}{2r_0^{\frac{r_0^2}{\lambda}+2}{r}^{{\frac {r_0^2}{\lambda}}+5}}.\label{vqs2}
\end{eqnarray} 
{It is obvious that the effective potential depends only on the value of parameters $\ell$ and $\lambda$. In Figs.~(\ref{VSL}) and (\ref{VSLAM}) we have plotted the behavior of effective potential for different values of model parameters. As the left panel in Fig.~(\ref{VSL}) shows, the effective potential is symmetric about the throat of wormhole and takes the form of a well at the throat and a double barrier near it. For a specified value of $\lambda>0$ the height of potential barriers and depth of potential well grow as the multipole number increases. The QNMs are strongly dependent on the shape of the effective potential so that the height of potential peaks can decide their behavior. In the the right panel we find that increasing $\lambda$ widens the potential but the height of the potential peaks becomes shorter. As a result, for larger values of $\lambda$, the echo signal of these class of wormholes assume longer time-delay and utilizing the pattern of the echoes one may determine the value of this parameter. Furthermore, the amplitude of echoes decreases with increasing the height of potential peaks, consequently, the trapped modes will have less chance to escape the potential barriers and the lifetime of them increases since the photon spheres are approximately located at the maxima of the potential~\cite{qnm00}. From Fig.~(\ref{VSLAM}) it is seen that the effective potential assumes a single barrier at the throat of wormhole for $\lambda<0$. As the left panel shows, the potential peak increases for larger values of multipole number. For a specified value of the parameter $\ell$, the potential peak grows as $\lambda$ increases in negative direction, see the right panel. It is worth to conclude this section with the following comments: In Figs.~(\ref{VSL}) and (\ref{VSLAM}), it is seen that the height of maxima grows with the increase of multipole number, consequently the growth in the peaks of effective potential leads to the modes with higher frequency. In the left panel of Fig.~(\ref{VSLAM}), for attractive Casimir energy with a single potential barrier at the throat, the oscillating modes will be very similar to those of regular black holes~\cite{qnm3,qnm11,qnm12}, however, we need more information to distinguish between them~\cite{qnm13}. In the left panel of Fig.~(\ref{VSL}), for a repulsive Casimir effect, the trapped resonant modes between two maxima with long-life play an important role in understanding their behavior.}
\section{Concluding Remarks}\label{concluding}
In this work, we studied static spherically symmetric wormhole spacetimes that sustained by the Casimir energy as the source. In the first case, under zero tidal force condition, we obtained the shape function by imposing a general form for the Casimir energy density. In this case, we showed that the wormhole solutions are asymptotically flat or AdS and dS so that each situation depends on the value of model parameters, i.e $\lambda$ and power of $m$ in the Casimir energy density. Then, we checked the WEC and SEC and showed that the energy conditions are violated at the wormhole throat. We calculated radial and tangential EoS parameters as the ratio of the respective pressures to the energy density. In the second case, by using a linear EoS between the radial pressure and energy density we derived a general form for the red shift function. In addition, in order to be a traversable wormhole, the curvature singularity at the wormhole throat must be absent. In this regard, we obtained a general condition for the state parameter that meets this condition and avoids formation of an event horizon at the throat. Hence, by choosing specific state parameter as $w=-r_0^{m-2}/\lambda$, all curvature invariants such as Ricci and Kretshmann scalars assume finite values in the range $r_0\leq r$, providing then traversable wormhole solutions. Furthermore, we studied three geometric configurations of the Casimir effect in details. For the case of Casimir effect between two parallel plates we obtained the shape and red shift functions of the wormhole metric and investigated the WEC and SEC for positive and negative values of $\lambda$ parameter against the radial distance from the wormhole throat. Also, for the case of parallel cylinders and spherical shells, we obtained the shape function and numerically solved the differential equation for the red shift function for both $\lambda>0$ and $\lambda<0$. In addition, we checked the WEC and SEC at the wormhole throat and showed that in general the classical energy conditions are violated for three geometric configurations, as expected. We further investigated the stability of wormhole solutions utilizing TOV equation and found that our obtained wormhole solutions are stable. Also, we analyzed trajectory of null and timelike particles for a class of wormhole solutions with $m=4$ and nonzero redshift function. We found that for $\lambda>0$ (repulsive Casimir force) there exists a photon sphere located outside the throat and an anti-photon sphere at the throat. For $\lambda<0$ (attractive Casimir force) or $0<\lambda<r_0^2/3$ there exists only a photon sphere at the wormhole throat. We also found that the deflection angle vanishes at a critical value of turning point, namely $r_{\rm tp}^\star$, see Fig~(\ref{lensplot}). For $r_{\rm tp}>r_{\rm tp}^\star$ the deflection angle is negative and for $r_{\rm tp}<r_{\rm tp}^\star$ it is positive. In the former case the negatively deflected light rays are deflected away from the wormhole and in the latter case the positively deflected light rays get attracted by the wormhole lens to form stable or unstable photon orbits. It is worth mentioning that negative deflection angle has been also reported in gravitational lensing by naked singularities, see e.g.~\cite{negdefang} and references therein. In the particular case for which $r_{\rm tp}=r_{\rm tp}^\star$, light rays are unaffected by the gravitating object. For timelike geodesics, we observed that there exists a certain value for particle angular momentum, Eq.~(\ref{angup}), with the help of which one can determine whether the effective potential admits any extrema outside the throat. Hence, depending on the values of $\lambda$ parameter and angular momentum, the effective potential could assume a local minimum outside the throat allowing thus stable circular orbits for timelike particles. Moreover, there exists a critical value for angular momentum, Eq.~(\ref{angucrit}), that decides the extrema of the effective potential at the throat. As it is shown in Fig.~(\ref{vefftime12}), stable circular orbits are possible at the wormhole throat. Depending on model parameters, the effective potential could assume a minimum as well as a maximum value outside the throat, thus, stable (unstable) and bound orbits are possible for this situation. {The study of QNMs for a scalar field was performed in the last section of the article and physical properties of them along with the behavior of effective potential for these modes were discussed for both attractive and repulsive Casimir effects. Finally, to close this article, we would like to mention some points: $i)$ It is well known that in GR framework, the violation of NEC, as a consequence of exotic matter, is a fundamental ingredient of static traversable wormholes. The exotic matter possesses the negative energy that is required to sustain the wormhole. One practical example of such matters can be found in Casimir effect~\cite{bordag,miltonbook}. However, the violation of NEC is a signal that there are unstable fields such as tachyons or ghosts~\cite{tachgost} which may cause instabilities in the wormhole configuration. In this sense, one may argue that the stability analysis of the obtained wormhole solutions via TOV equation may not be enough and the stability of the wormhole configuration against several types of perturbation is always an important issue. For example, the study of stability of thin-shell wormholes, constructed using the cut-and-paste technique, by considering specific equations of state~\cite{stcp,stcp1,stcp2,stcp3,stcp4,stcp5}, or by applying a linearized radial perturbation around a stable solution~\cite{linradp,linradp1,pvis,linradp3,linradp4,linradp5}. $ii)$ It is also shown that there is a relation between violation of NEC and instability of the wormhole~\cite{necinstab}. In static 4-dimensional wormhole spacetimes, the NEC is violated (at least at the throat) owing to the flare-out condition. However, in case such a violation leads to instability in the wormhole configuration one may employ electromagnetic fields to preserve the stability~\cite{casgup2,stabem}. $iii)$ Since the original work of~\cite{gara}, many authors have investigated Casimir wormholes under different conditions and within alternative gravity theories~\cite{casi3,casid,altcas,altcas1,altcas2,casgup,casgup2,casgup3,casgup4,yukcas,yukcas1,yukcas2,othercas,othercas2,othercas3,othercas4}. In the present work we tried to find suitable wormhole solutions for the general form of Casimir energy~(\ref{rh10}). A traversable wormhole must not contain a spacetime singularity at its throat, hence we obtained the EoS parameter in such a way that the singularity is avoided. In this regard, some authors have obtained the EoS parameter by fixing the metric potentials and studied its behavior for a Casimir wormhole, see e.g.,~\cite{altcas2,casgup2,casgup3}. In other studies~\cite{gara,yukcas,othercas3,casoli,garatin2023}, assuming non-homogeneity in the EoS parameter, the authors have tried to establish the best correspondence between a traversable wormhole and the Casimir EMT~\cite{book visser}. $iv)$ An important and interesting feature of the Casimir effect involves the preservation or violation of the Lorentz symmetry and its possible effects on the values of Casimir energy~\cite{Lorsymcas,Lorsymcas1,Lorsymcas2}. As high energies disrupt Lorentz symmetry, the study of Lorentz symmetry breakdown has attracted a lot of attention in the scientific community. Investigations in this arena have been focused on the interplay of Casimir energy with various directions of Lorentz-violating quantum fields, some valuable results on this issue have been reported in~\cite{Lorsymcas3,Lorsymcas4,Lorsymcas5,Lorsymcas6,Lorsymcas7,Lorsymcas8,Lorsymcas9}. Having tried to preserve Lorentz symmetry, the author of~\cite{qassab} calculated the Casimir EMT of a scalar field in the spacetime of a long-throated traversable wormhole and examined whether this exotic matter is sufficient to maintain the stability of the wormhole.} 
\section{Acknowledgements} 
The authors would like to appreciate the anonymous referees for providing useful and constructive comments that helped us to improve the original version of our manuscript.

\end{document}